\documentclass[12pt,a4paper,twoside]{article}
\usepackage[dvips]{graphics}
\usepackage{cite}

\long\def\ifundef#1#2#3{\expandafter\ifx\csname
  #1\endcsname\relax#2\else#3\fi}
\ifundef{ensuremath}{\newcommand{\ensuremath}[1] {\relax\ifmmode {#1}\else {$#1$} \fi}}{}
\ifundef{mathcal}{\newcommand{\mathcal}[1]{{\cal #1}}}{}
\ifundef{mathrm}{\newcommand{\mathrm}[1]{\rm #1}}{}
\ifundef{resizebox}{\newcommand{\resizebox}[3] 
{#3 
}}{}
\ifundef{includegraphics}{\newcommand{\includegraphics}[1] 
{$#1$ 
\typeout{PostScript File #1 not included!}
}}{}

\newcommand{\epem}   {\ensuremath{\mathrm{e^+e^-}}}

\newcommand{\lmsbf}  {\ensuremath{\Lambda^{(5)}_{\mathrm{\overline{MS}}}}}

\newcommand{\as}     {\ensuremath{\alpha_s}}

\newcommand{\asq}    {\ensuremath{\alpha_s(Q)}}
\newcommand{\asmz}   {\ensuremath{\alpha_s(M_{\mathrm{Z^0}})}}

\newcommand{\oaa}    {\ensuremath{\mathcal{O}(\alpha_s^2)}}
\newcommand{\oaaa}   {\ensuremath{\mathcal{O}(\alpha_s^3)}}

\newcommand{\znull}  {\ensuremath{\mathrm{Z^0}}}
\newcommand{\zzero}  {\ensuremath{\mathrm{Z^0}}}

\newcommand{\mz}     {\ensuremath{M_{\mathrm{Z^0}}}}

\newcommand{\bt}     {\ensuremath{B_T}}
\newcommand{\bw}     {\ensuremath{B_W}}
\newcommand{\mh}     {\ensuremath{M_H}}
\newcommand{\thr}    {\ensuremath{1-T}}

\newcommand{\chisq}  {\ensuremath{\chi^2}}
\newcommand{\chisqd} {\ensuremath{\chi^2/\mathrm{d.o.f.}}}
\newcommand{\xmu}    {\ensuremath{x_{\mu}}}

\ifundef{pt}{ }
            { }

\newcommand{\ycut}   {\ensuremath{y_{\mathrm{cut}}}}

\newcommand{\dtwo}   {\ensuremath{D_2}}

\newcommand{\bm}[1]  {\mbox{\boldmath\ensuremath{#1}}}

\newcounter{hours}
\newcounter{minutes}

\newcommand{\Printtime}{%
  \setcounter{hours}{\time/60}%
  \setcounter{minutes}{\time-\value{hours}*60}%
  \ifthenelse{\value{hours}<10}{0}{}\thehours:%
  \ifthenelse{\value{minutes}<10}{0}{}\theminutes}

\begin{document}
%
\begin{titlepage}
\vspace*{-10mm}
\hbox to \textwidth{ \hsize=\textwidth
\hspace*{0pt\hfill} 
\vbox{ \hsize=58mm
{
\hbox{ PITHA 97/27\hss}
\hbox{ August 25, 1997\hss } 
\hbox{rev. October 29, 1997\hss } 
}
}
}

\bigskip\bigskip\bigskip
\begin{center}
{\Huge\bf
A Study of Event Shapes and \\[1.5mm]
Determinations of {\ensuremath{\alpha_s}} \\[1.5mm]
using data of ${\ensuremath{\mathrm{e^+e^-}}}$ Annihilations \\[1.5mm]
at $\sqrt{s} = 22$ to $44$ GeV
}

\end{center}
\bigskip\bigskip
\begin{center}
P.A.~Movilla~Fern\'andez$^{(1)}$, O.~Biebel$^{(1)}$, S.~Bethke$^{(1)}$, 
S.~Kluth$^{(2)}$, 
P.~Pfeifenschneider$^{(1)}$\\
 and the JADE Collaboration$^{(3)}$
\end{center}
\bigskip

\begin{abstract}
\noindent 
Data recorded by the JADE experiment at the PETRA 
{\ensuremath{\mathrm{e^+e^-}}}\ collider 
were used to measure the event shape observables thrust, heavy
jet mass, wide  and total jet broadening and the differential 2-jet rate
in the Durham scheme.
For the latter three observables, no experimental 
results have previously been presented at
these energies.
The distributions were compared with resummed QCD calulations 
({\ensuremath{\mathcal{O}(\alpha_s^2)}}+NLLA),
and the strong coupling constant {\ensuremath{\alpha_s(Q)}}\ was 
determined at different energy
scales $Q=\sqrt{s}$. The results,
\begin{displaymath}
{\ensuremath{\alpha_s}}(22\ \mathrm{GeV}) = 0.161 ^{+0.016}_{-0.011}\ ,\ %
{\ensuremath{\alpha_s}}(35\ \mathrm{GeV}) = 0.143 ^{+0.011}_{-0.007}\ ,\ %
{\ensuremath{\alpha_s}}(44\ \mathrm{GeV}) = 0.137 ^{+0.010}_{-0.007}\ ,
\end{displaymath}
are in agreement with previous combined results of PETRA albeit 
with smaller uncertainties.
Together with corresponding data from LEP, the energy
dependence of {\ensuremath{\alpha_s}}\ is significantly tested and 
is found to be in good
agreement with the QCD expectation.
Similarly, mean values of the observables were compared to analytic QCD
predictions where hadronisation effects are absorbed in
calculable power corrections.
\end{abstract}

\vspace*{0pt\vfill}
\vfill
\bigskip\bigskip\bigskip\bigskip

{
\small
\noindent
$^{(1)}$ 
\begin{minipage}[t]{155mm} 
III. Physikalisches Institut der RWTH Aachen,
D-52056 Aachen, Germany \\
contact e-mail: Otmar.Biebel@Physik.RWTH-Aachen.DE
\end{minipage}\\
$^{(2)}$ 
\begin{minipage}[t]{155mm} 
CERN, European Organisation for Particle Physics, CH-1211 Geneva 23, Switzerland
\end{minipage}
$^{(3)}$ 
\begin{minipage}[t]{155mm} 
for a full list of members of the JADE Collaboration see Reference \cite{bib-naroska}
\end{minipage}
\hspace*{0pt\hfill}
}

\end{titlepage}
%
%
\newpage
\section{ Introduction }
Summaries of measurements of \as\ from various processes and at
different energy scales $Q$
demonstrate \cite{bib-world-alphas-sb,bib-world-alphas-ms} that the energy
dependence of \asq\ is
in good agreement with the prediction of Quantum Chromodynamics (QCD). 
The uncertainties of these measurements, 
both experimental and theoretical, are different and their correlations 
are, in general, not known \cite{bib-world-alphas-sb,bib-world-alphas-ms}.
More quantitative studies of the running of \as\ therefore require the
existence of consistent measurements over large ranges of the
energy scale $Q$, for the same process, using identical experimental
techniques and theoretical calculations in order to minimise
point-to-point systematic uncertainties.

Significant progress has been made in perturbative QCD calculations since
1992. Observables have been proposed for which perturbative predictions 
are extended beyond the next-to-leading-order (\oaa ) \cite{bib-ERT}, 
through the inclusion of leading and next-to-leading logarithms which 
are summed to all orders of \as\ (NLLA) 
\cite{bib-NLLA-1,bib-NLLA-2,bib-NLLA-3}.
These calculations exhibit a better stability to contributions of 
unknown higher order corrections, which are usually estimated by 
variations of the renormalisation scale $\mu$.

The experiments at LEP and SLC provided a number of significant 
determinations of \as\ from hadronic event shapes and jet 
production, based on \oaa+NLLA calculations,
at centre-of-mass energies $\sqrt{s}$ at and above 91~GeV, the mass 
of the \zzero\ boson.
Detailed studies of the high statistics data samples from
the LEP experiments provide a better understanding of the
phenomenology of the hadronisation process and thus of the modelling
of hadronic final states with Monte Carlo programs.
Determinations of \as\ 
at LEP and 
SLC~\cite{bib-ALEPHalphas,bib-DELPHIalphas,bib-L3alphas,bib-OPALresummed,bib-SLDalphas}
therefore have smaller uncertainties than those which are available from 
previous measurements at lower \epem\ centre-of-mass 
energies~\cite{bib-naroska,bib-sb-budapest}.
There are only a few recent measurements of \as\ at lower energies. 
These either 
employed only some of the observables 
which are now available~\cite{bib-TPC2gamma,bib-TOPAZ} or they
were based on limited
samples of \znull\ decays with final state 
photon radiation~\cite{bib-L3-FSR}. 
Therefore equivalent studies with data in the 
centre-of-mass energy range from 
$\sqrt{s} = 14$ to $46.7$~GeV, taken at the PETRA collider which was
shut down in 1986, are desirable.

In this paper we present an \oaa+NLLA determination of \as\ at $\sqrt{s} =
22$,  $35$, and
$44$ GeV using data from the JADE experiment~\cite{bib-naroska,bib-JADEdet} at 
PETRA. 
The selection of the JADE data and Monte Carlo event samples 
are described in Section~\ref{sec-data}. 
The measurement of event shape distributions, the
corrections for detector imperfections and for initial state
photon radiation as well as the estimate of the experimental uncertainties
are outlined in Section~\ref{sec-procedure}. 
The corrected event shape distributions and the determination
of the strong  coupling constant \asq\ 
are presented in Section~\ref{sec-alphas}.
A study of the energy dependence of mean values of event shape
distributions and their comparison with analytic QCD calculations comprising
power corrections to account for hadronisation effects
is presented in Section~\ref{sec-meanvalues}.
In Section~\ref{sec-conclusions} the results are summarised and the 
conclusions are drawn.
  
\section{ Data samples and Monte Carlo simulation }
\label{sec-data}
For the studies presented in this paper, we analysed 
data recorded with the JADE detector in 1981, 1984 to 1985, and 1986 at 
centre-of-mass 
energies of $22$~GeV, $39.5$-$46.7$~GeV, and around $35$~GeV,
respectively. 
The JADE detector was one of the five experiments at the 
PETRA electron-positron
collider. 
It was operated from 1979 until 1986 at centre-of-mass energies of
$\sqrt{s} = 12$ to $46.7$ GeV. 
A detailed description of the JADE detector can be found
in~\cite{bib-naroska,bib-JADEdet}. 
The main components of the detector were the
central jet chamber to measure charged particle tracks
and the lead glass calorimeter to measure energy depositions of
electromagnetic showers, which both
covered almost the whole
solid angle of $4\pi$. 

Multihadronic events were selected by the standard JADE selection 
cuts~\cite{bib-JADEtrigger} which were based on
minimum energy deposits 
in the calorimeter and a minimum number of tracks emanating
from the interaction region.
All charged particle tracks with a total momentum of $|\vec{p}| >
100$~MeV$/c$ were considered in the analysis. Energy clusters in the
electromagnetic calorimeter were considered if their energies
exceeded $150$ MeV after correction for 
energy deposited by associated tracks.  
Charged particle tracks were assumed to be pions 
while the photon hypothesis was assigned to electromagnetic
energy clusters.

In order to remove background from two-photon processes and 
$\tau$-pair events and from events which lost a substantial part of
their energy due to hard initial state photon radiation,
further constraints were imposed on the visible energy 
$E_{\mathrm{vis}} = \sum E_i$, 
the total missing momentum
$p_{\mathrm{miss}} = |\sum \vec{p}_i|$ ($\vec{p}_i$ and $E_i$
are the 3-momentum and the energy of the tracks and clusters), 
the longitudinal balance relative to the \epem\ beam axis of momenta 
$p_{\mathrm{bal}} = |\sum p^z_i/E_{\mathrm{vis}}|$
and the polar angle of the
thrust axis, $\theta_{T}$:
\begin{itemize}
\item $E_{\mathrm{vis}} > \sqrt{s}/2$~;
\item $p_{\mathrm{miss}} < 0.3\cdot \sqrt{s}$~;
\item $p_{\mathrm{bal}} < 0.4$~;
\item $|\cos\theta_{T}| < 0.8$~.
\end{itemize}
With these cuts, the backgrounds from 
$\gamma\gamma$ and $\tau$-pair events were reduced to less than 
$0.1\%$ and $1\%$, respectively \cite{bib-JADEeventsel}.
The final numbers of events which were retained for this analysis
are listed in Table~\ref{tab-eventnumbers}.

\begin{table}[b]
\begin{center}
\begin{tabular}{|c|c||c|c|}
\hline
year & $\sqrt{s}\ [\mathrm{GeV}]$ & data & MC \\ 
\hline\hline
   1981 &  $22$      &             $\ 1404$ &               ---   \\ \hline
1984/85 &  $40$-$48$ &             $\ 6158$ & $14\thinspace 497$  \\ \hline
   1986 &  $35$      &   $20\thinspace 926$ & $25\thinspace 123$  \\
\hline
\end{tabular}
\end{center}
\caption{\label{tab-eventnumbers}
Number of events in data and in Monte Carlo detector simulation retained after 
application of the multihadron selection cuts described in the text. 
}
\end{table}

The retrieval of data files eleven years after shutdown of 
the experiment was difficult and turned out to be
incomplete at this stage of the analysis. 
Comparisons of the numbers given in Table~\ref{tab-eventnumbers} with
previous JADE publications \cite{bib-JADEeventsel,bib-jadjet1,bib-JADEtune} 
revealed that we were missing data sets of about 250
events around 22~GeV and about 450 events around 44~GeV.
In addition, the original files containing 
information about the luminosity of different
running periods could not be retrieved, so that 
only approximate values of integrated
luminosities corresponding to our final number of events 
can be given\footnote{It is not 
expected that the missing events alter the measured distributions
in a systematic manner. 
Also, detailed knowledge of luminosities is not required for 
the following studies
of {\em normalised} event shape distributions.}: 
the data samples shown in
Tab.~\ref{tab-eventnumbers} correspond to about $2.4$~pb$^{-1}$, 
$80$~pb$^{-1}$ and
$40$~pb$^{-1}$ at 22~GeV, 35~GeV and 44~GeV centre-of-mass energy, 
respectively.

In order to verify the compatibility of this study with results which were
previously published by JADE, we repeated a determination of the relative
2-, 3- and 4-jet event production rates as published in
\cite{bib-JADEeventsel}, using the original JADE jet finder with a
resolution parameter of $\ycut = 0.08$.
The results are presented in Table~\ref{tab-oldjetrates}.
Considering the fact that our present data samples at 22~GeV and 44~GeV 
lack about 10\% of the original ones and that the samples around 35~GeV
are from different running periods (1986 for this analysis, 1984-1985 for 
Reference~\protect\cite{bib-JADEeventsel}), the agreement between the old
and this new study is very good.  This demonstrates that we are able to
perform detailed studies of event properties in a consistent way.

\begin{table}[t]
\begin{center}
\begin{tabular}{|l|c|c|}
\hline
$\sqrt{s}$ & Ref. \cite{bib-JADEeventsel} & this analysis \\ 
\hline\hline
\ 22 GeV & & \\
{\hfill $R_2$} & $ 72.5 \pm 1.2$ & $ 72.7 \pm 1.2$ \\
{\hfill $R_3$} & $ 27.1 \pm 1.2$ & $ 27.0 \pm 1.2$ \\
{\hfill $R_4$} & $ 0.42 \pm 0.16$ & $ 0.28 \pm 0.14$ \\ \hline
\ 35 GeV & & \\
{\hfill $R_2$} & $ 77.7 \pm 0.4$ & $ 78.2 \pm 0.3$ \\
{\hfill $R_3$} & $ 22.0 \pm 0.4$ & $ 21.6 \pm 0.3$ \\
{\hfill $R_4$} & $ 0.31 \pm 0.05$ & $ 0.24 \pm 0.03$ \\ \hline
\ 44 GeV & & \\
{\hfill $R_2$} & $ 79.8 \pm 0.5$ & $ 79.2 \pm 0.5$ \\
{\hfill $R_3$} & $ 20.1 \pm 0.5$ & $ 20.7 \pm 0.5$ \\
{\hfill $R_4$} & $ 0.14 \pm 0.05$ & $ 0.15 \pm 0.05$ \\
\hline
\end{tabular}
\end{center}
\caption{\label{tab-oldjetrates}
A comparison of relative $n$-jet production
rates $R_n$, as
percentages of all hadronic events, using the JADE jet finding
algorithm with $\ycut = 0.08$ \protect\cite{bib-JADEeventsel}.
No corrections are applied; the errors are statistical only.
}
\end{table}
 
Corresponding Monte Carlo detector simulation data were 
retrieved for $35$ and $44$~GeV.
They were generated using the QCD parton shower event generator
JETSET 6.3~\cite{bib-JETSET}.
The Monte Carlo events at 
$35$~GeV were generated using the coherent branching for the parton 
shower while the $44$~GeV events had non-coherent branching
\footnote{The different treatment of coherence in these samples of
simulated data has no visible influence on the results of this study; see
also Fig.~1, Figs.~3-5 and the discussions in Section~\ref{sec-procedure}.}.
The main parameters used for event
generation are given in Section~\ref{subsec-systematics}.
Both samples included a simulation of the acceptance and resolution
of the JADE detector.

\begin{figure}[!htb]
\vspace*{-7mm}
\begin{center}
\resizebox{79mm}{!}{\includegraphics{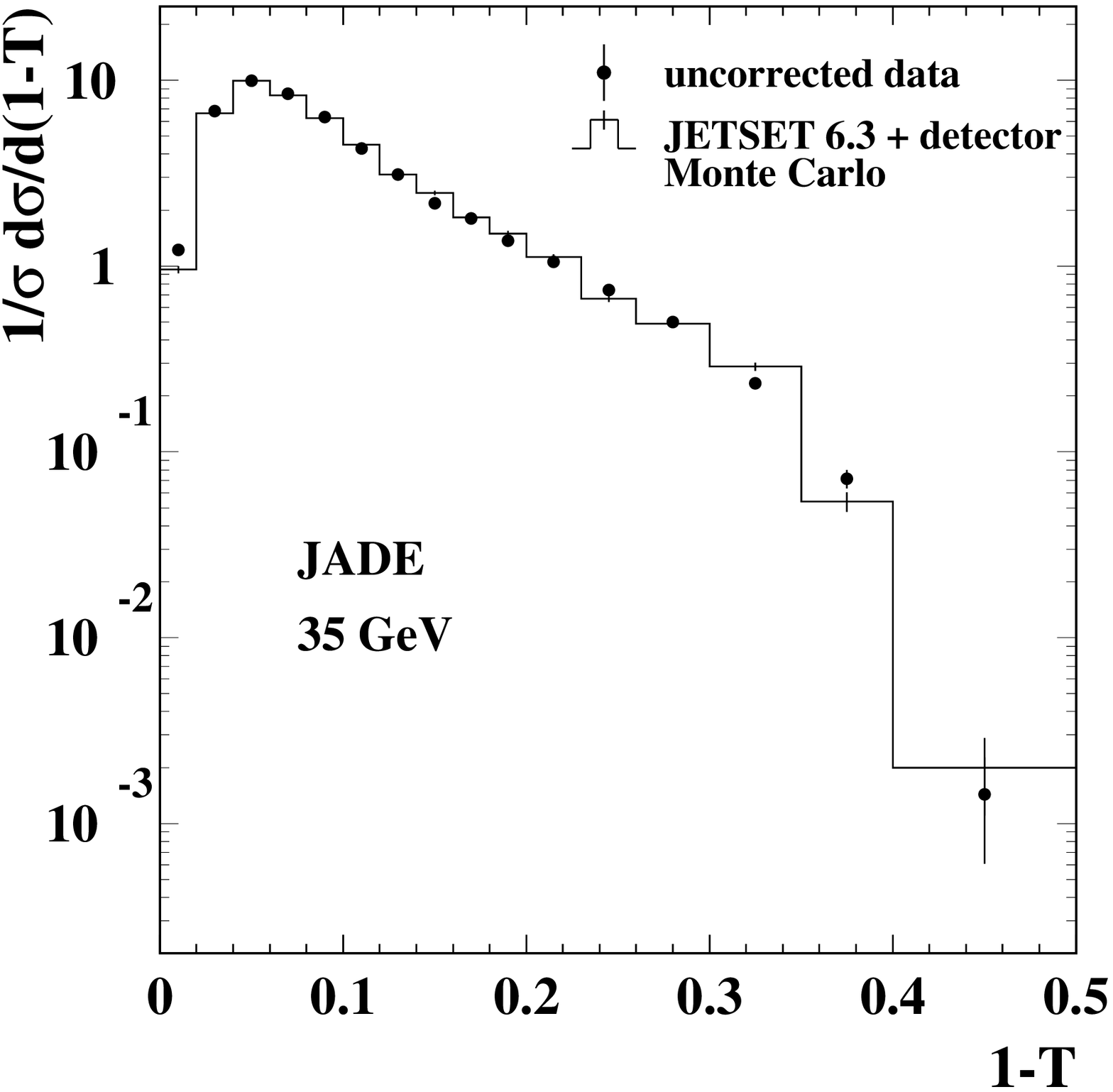}}
\resizebox{79mm}{!}{\includegraphics{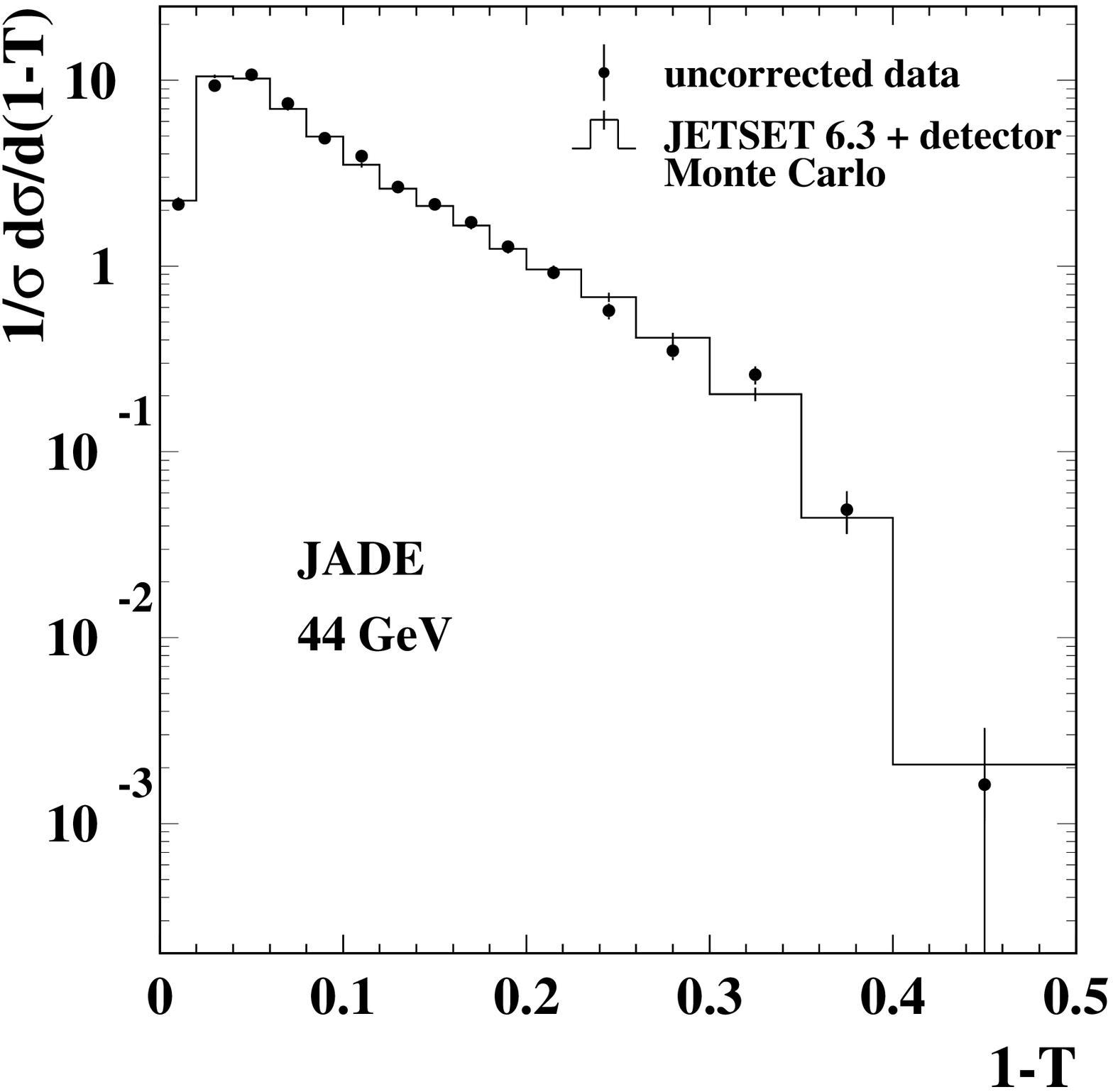}}\\
\resizebox{79mm}{!}{\includegraphics{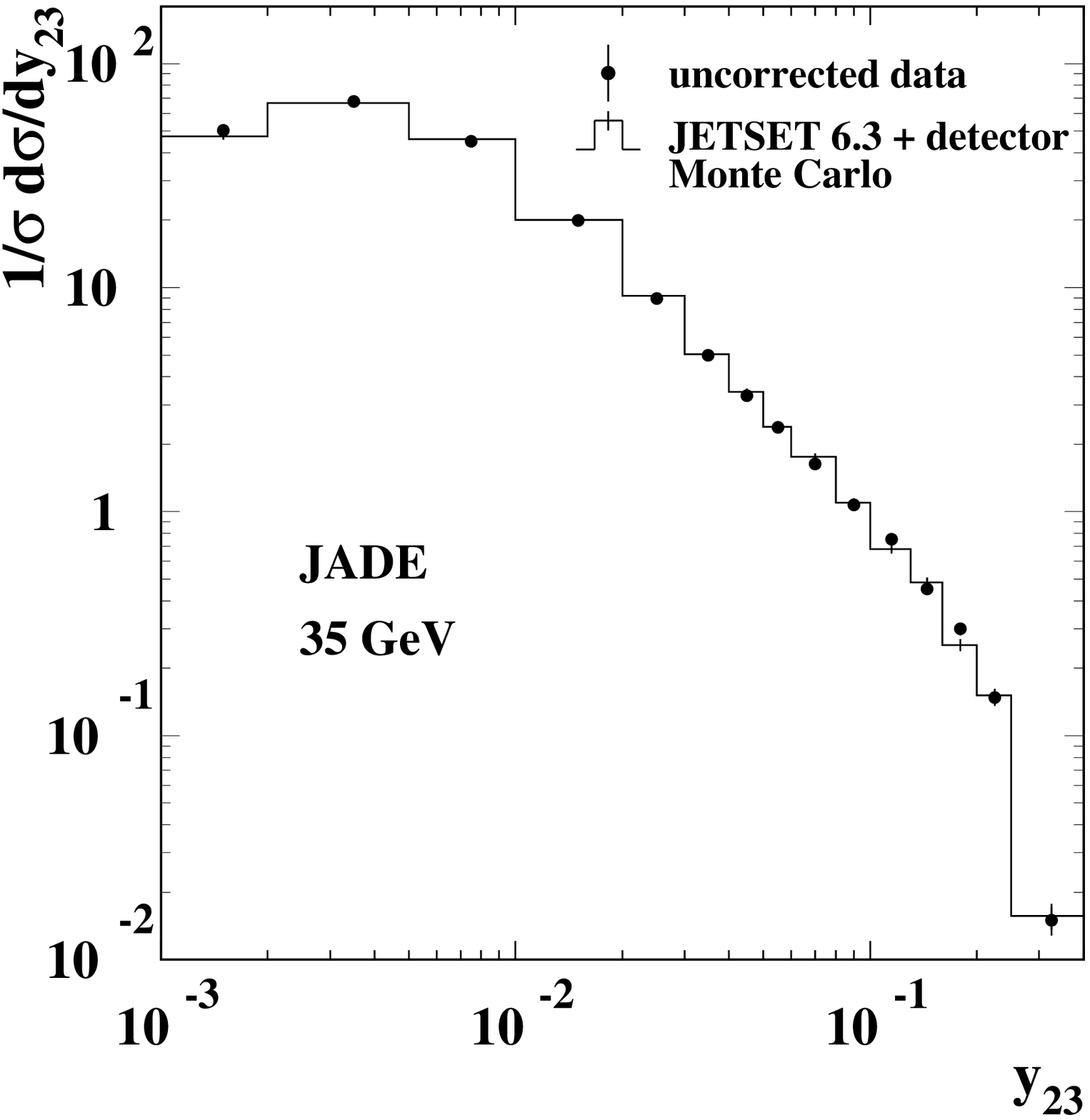}}
\resizebox{79mm}{!}{\includegraphics{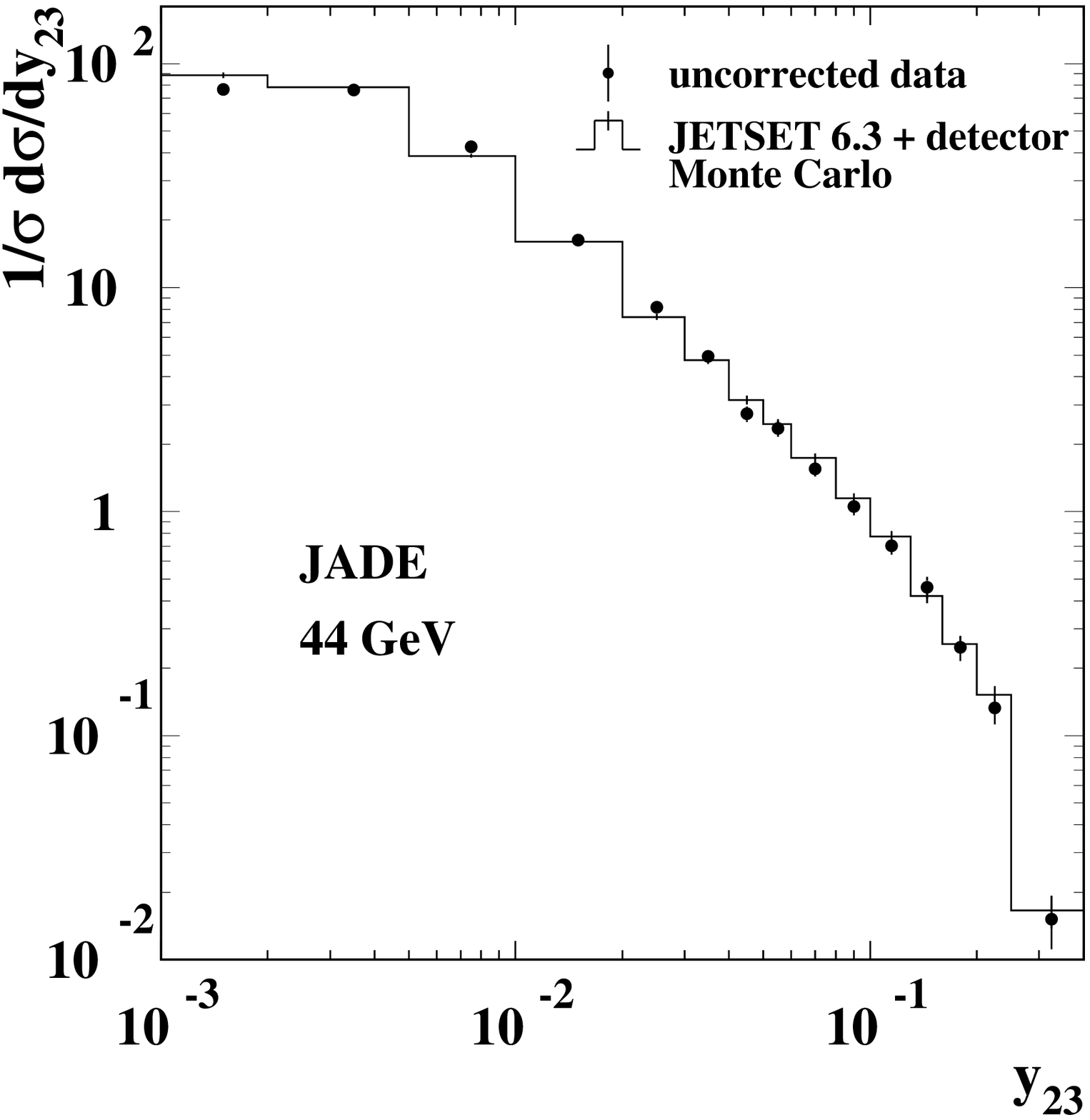}}
\end{center}
\caption{\label{fig-thrust-uncorr} 
Measured  and uncorrected distributions of the thrust observable $1-T$ (top)
and of the differential 2-jet rate $D_2$ (bottom) at $35$ (left) 
and $44$~GeV (right). The simulated data are overlayed as a 
solid line histogram.
Only statistical errors are shown by the error bars.
}
\end{figure}

Comparisons of the measured and simulated distributions of visible 
energy, momentum balance, missing momentum and other quantities showed 
that the Monte 
Carlo simulation gave a reasonable description of the measurements. 
The simulated data can thus be used to correct for 
detector effects in the measured data.
As an example we show in Figure~\ref{fig-thrust-uncorr} 
the distributions of the thrust observable $1-T$ and of the 
differential two-jet rates $D_2$, measured at $35$ and at
$44$~GeV. The definitions of these observables are given in
Section~\ref{subsec-eventshapes}.
In general, we found a good agreement of the detector
simulation with data for all event shape distributions studied here,
irrespective of coherent or non-coherent parton branching.

\section{Experimental procedure}
\label{sec-procedure}

\subsection{Event shapes and differential 2-jet rate}
\label{subsec-eventshapes}
From the data samples described in the previous section, the
event shape distributions of thrust, the heavy jet
mass, the total and wide jet broadening and the
differential 2-jet event rate using the Durham jet finder were
determined. For convenience we list the definitions of these observables.
\begin{description}
\item[{\bf Thrust} \bm{T}:]  \hspace*{0pt\hfill} \\
     The thrust value of a hadronic event is defined by the
     expression~\cite{bib-thrust-def}
     \begin{displaymath}
            T = \max_{\vec{n}}\left( \frac{\sum_i | \vec{p}_i \cdot \vec{n} | }
                                          {\sum_i | \vec{p}_i | } \right)\ \ .
     \end{displaymath}
     The vector $\vec{n}$ which maximises the expression in parentheses 
     is the thrust axis $\vec{n}_T$. It is used to divide an event into
     two hemispheres $H_1$ and $H_2$ by a plane through the origin and
     perpendicular to the thrust axis.
\item[{\bf Heavy Jet Mass} \bm{M_H}:]  \hspace*{0pt\hfill} \\
     From the particles in each of the two hemispheres defined by the 
     thrust axis an invariant mass is calculated. The heavy jet mass 
     $M_H$~\cite{bib-jetmass-def,bib-LEP1report} is defined by the 
     larger of the two
     masses. This analysis used the 
     measured                                              
     heavy jet mass scaled by the 
     visible energy $E_{\mathrm{vis}}$ 
     which is, after correction for detector resolution,   
     acceptance, and for initial state radiation, equal to 
     $M_H/\sqrt{s}$.                                       
\item[{\bf Jet Broadening} \bm{B}:]  \hspace*{0pt\hfill} \\
     The jet broadening measures are calculated 
     by the expression~\cite{bib-NLLA-1}:
     \begin{displaymath}
          B_k = \left( \frac{\sum_{i\in H_k} | \vec{p}_i \times \vec{n}_T | }
                                         {2\sum_i | \vec{p}_i | } \right)
     \end{displaymath}
     for each of the two hemispheres, $H_k$, defined above.
     The total jet broadening is given by $B_T = B_1 + B_2$. The wide
     jet broadening is defined by $B_W = \max(B_1, B_2)$.
\item[{\bf Durham differential 2-jet rate} \bm{D_2}:] \hspace*{0pt\hfill} \\
     Jets are reconstructed by a standard recombination algorithm: For any 
     combination of two particles $i$ and $j$ in an event a measure of 
     distance, $y_{ij}$, is calculated according to the Durham 
     recombination scheme~\cite{bib-durham-def} 
     \begin{displaymath}
          y_{ij} = 
            \frac{2 \cdot \min( E_i^2, E_j^2) \cdot (1 - \cos\theta_{ij})}
                              {E_{\mathrm{vis}}^2}\ ,
     \end{displaymath}
     where $E_i$ and $E_j$ are the energies of the particles and 
     $\cos\theta_{ij}$ is the angle between their 3-momentum vectors. 
     The pair $i$, $j$ of particles with the smallest value of $y_{ij}$ 
     is replaced by a pseudoparticle $k$ with 4-momentum $p_k = p_i + p_j$.
     This procedure is repeated 
     until exactly three pseudoparticles remain which are called jets. The 
     smallest $y_{ij}$ corresponding to these three jets is indicated 
     by $y_{23}$ throughout the paper. At this particular value the 
     number of reconstructed jets changes from $3$ to $2$. 
     $D_2$ is the normalised differential cross-section as a function of 
     $y_{23}$~\cite{bib-OPAL89-MARKII}.
\end{description}
In the following we use the symbols $T$, $M_H$, $B_T$, $B_W$ and $D_2$
to denote thrust, heavy jet mass, total and wide jet broadening, and the
differential 2-jet rate, respectively.

\subsection{Correction procedure}
\label{subsec-correction}
The event shape data were corrected for the limited acceptance and 
resolution of the detector and for initial state photon radiation effects 
by applying a bin-by-bin correction procedure.
Correction factors were 
defined by the ratio of the distribution calculated from events
generated by JETSET 6.3
at {\em hadron level} over the same distribution at {\em detector
level}.
The {\em hadron level} distributions were obtained from JETSET 6.3
generator runs without detector simulation and without initial state
radiation, using all particles with lifetimes $\tau > 3\cdot
10^{-10}$~s.
The model events at {\em detector level} contained initial state photon
radiation and a detailed simulation of the detector response and 
were processed in the same way as the data.

In a second step, the data distributions were further corrected for
hadronisation effects. 
This was done by applying bin-by-bin correction factors derived from
the ratio of the distribution at {\em parton level} over the same
distribution at {\em hadron level}, which were 
calculated from JETSET generated
events before and after hadronisation, respectively. 
The data distributions, thus corrected to the {\em parton level}, 
can be compared to analytic QCD calculations.

\subsection{Systematic uncertainties}
\label{subsec-systematics}
To study systematic uncertainties of the corrected data distributions
we modified details of the event selection and of the correction procedure.
For each variation the whole analysis was repeated and 
any deviation from the main result was considered a systematic
error. 
In general, the maximum deviation from the main result for each 
kind of variation was regarded as symmetric systematic 
uncertainty.
The main result was obtained using the default selection and 
correction procedure as described above.

We restricted the measurement of the event shape 
distributions to rely either on tracks or on clusters only. We varied the cut
on $\cos\theta_T$ by $\pm 0.1$. The cut on $p_{\mathrm{miss}}$ was
either removed or tightened to $p_{\mathrm{miss}} < 0.25 \cdot \sqrt{s}$. 
Similarly, the momentum balance requirement was either restricted to 
$p_{\mathrm{bal}} < 0.3$ or dropped. We also varied the cut
for the visible energy $E_{\mathrm{vis}}$ by $\pm 0.05 \cdot \sqrt{s}$.
In order to check the residual contributions from $\tau$-pair events we
also required at least seven well-measured charged tracks.

To study the impact of the hadronisation model of the JETSET 6.3
generator, the values of several significant model parameters were
varied around their tuned default values used for 
our main result. We took the tuned values 
from Reference~\cite{bib-JADEtune},
where the QCD parameter was $\Lambda_{\mathrm{LLA}} = 400$~MeV, the cut-off 
for the parton shower development was $Q_0 = 1$~GeV, and the width of the 
transverse momentum distribution
of the hadrons with respect to the direction of the quark
was $\sigma_0 = 300$~MeV. According to this Reference we
chose the LUND symmetric fragmentation function with $a = 0.5$ and $b = 0.9$
for the fragmentation of the light u, d, and s quarks. The heavy c and b
quark were fragmented applying the Peterson et al.~\cite{bib-Peterson}
fragmentation function using $\epsilon_{\mathrm{c}}=0.05$ and 
$\epsilon_{\mathrm{b}}=0.01$~\cite{bib-JADEtune}.

Different sets of correction factors to correct the data from {\em
hadron level} to {\em parton level} were generated by varying single
parameters of the JETSET generator.
The variations were chosen to be similar to the one 
standard deviation percentage limits
obtained by the OPAL Collaboration from a parameter 
tuning of JETSET at $\sqrt{s} = \mz$~\cite{bib-OPALtune}. 

In particular, we investigated the effects due to 
parton shower, hadronisation parameters, and quark masses. 
The amount of gluon radiation during the parton shower development was modified by varying 
$\Lambda_{\mathrm{LLA}}$ by $\pm 50$~MeV. 
To vary the onset of hadronisation, we altered the parton shower cut-off 
parameter $Q_0$ by $\pm 0.5$~GeV. We 
used            
the 
full            
observed variation of 
\as\ to reflect a variation of $Q_0$ between $0$ and $2$~GeV.
The width of the transverse momentum distribution 
in the
hadronisation process was varied by $\pm 30$~MeV.
The LUND symmetric fragmentation function was varied by changing the 
$a$ parameter by $\pm 0.225$ whereas the $b$ parameter was 
kept fixed. As a systematic variation we used the LUND
fragmentation function also for charm and bottom quarks.
The effects due to the bottom quark mass were studied by 
restricting the model calculations which were used to determine the
correction factors
to up, down, 
strange, and charm quarks (udsc) only. 
In this case, any deviation 
from our main 
result was treated as asymmetric error.  

\begin{figure}[!htb]
\resizebox{\textwidth}{!}{\includegraphics{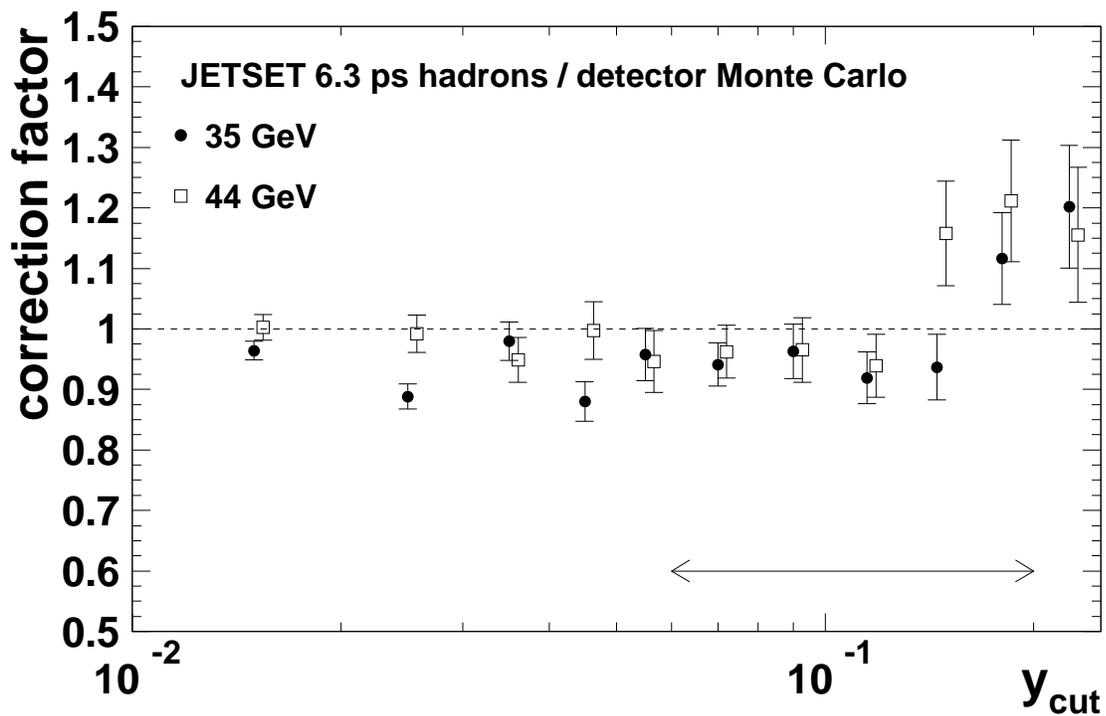}}
\caption{\label{fig-D2-correction} 
The detector correction factors at $35$ (points) and at $44$~GeV (open squares)
are shown for the differential 2-jet rate, $D_2$, in the Durham jet finder 
scheme.  
The error bars represent the statistical error. The arrow indicates the 
range of data considered to determine \as\ at 22~GeV.
}
\end{figure}

No {\em detector level}
Monte Carlo simulation data
were available
for the $22$~GeV data. In order to obtain consistent detector corrections also
at this energy, 
we studied the energy dependence of the
detector correction from the $35$ and $44$~GeV Monte Carlo samples. 
Here 
we considered only the 
differential 2-jet rate, $D_2$, of the Durham jet finder scheme 
because it is known to depend to a lesser extent on hadronisation 
and detector effects. 
In Figure~\ref{fig-D2-correction} the detector correction factors as
obtained at $35$~GeV and 
at $44$~GeV are displayed.
In general, the corrections are small, and 
their size is about the same at 
both centre-of-mass energies. 
There is no apparent energy 
dependence of the detector correction within the range, indicated by the
arrow, which was considered for the fit of \as\ at $22$~GeV.
We therefore applied the $35$~GeV detector correction 
to the differential
2-jet rate measured from the $22$~GeV data, and studied only the 
well known dominating sources of systematic uncertainties.
The correction of {\em hadronisation} effects was then determined from
JETSET generator runs at 22~GeV centre-of-mass energies, exactly as for
the data at 35 and 44~GeV.

\section{ Determination of \as }
\label{sec-alphas}
\subsection{ Corrected event shape distributions}
\begin{figure}[!htb]
\vspace*{-7mm}
\begin{center}
\resizebox{79mm}{!}{\includegraphics{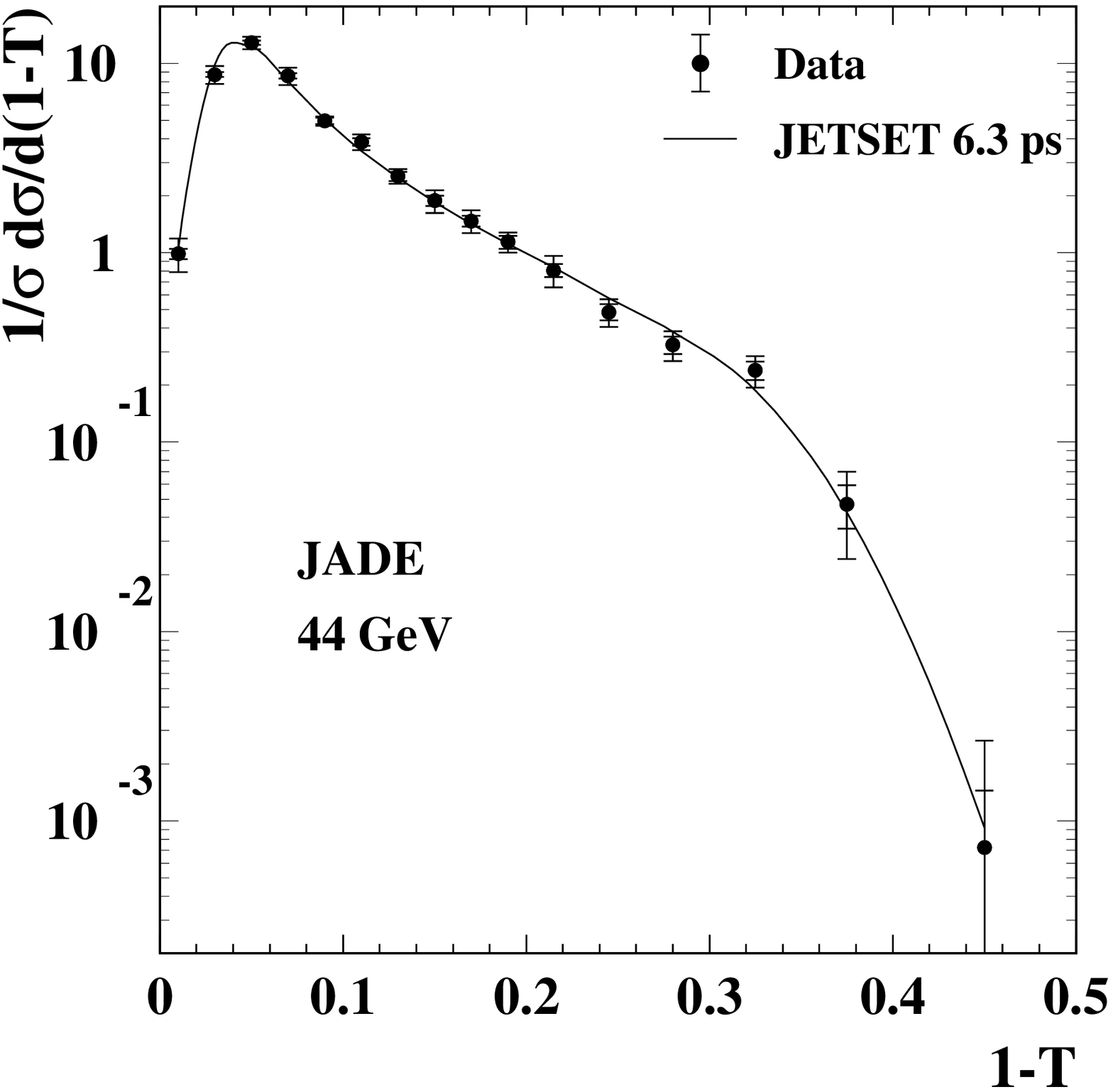}}
\resizebox{79mm}{!}{\includegraphics{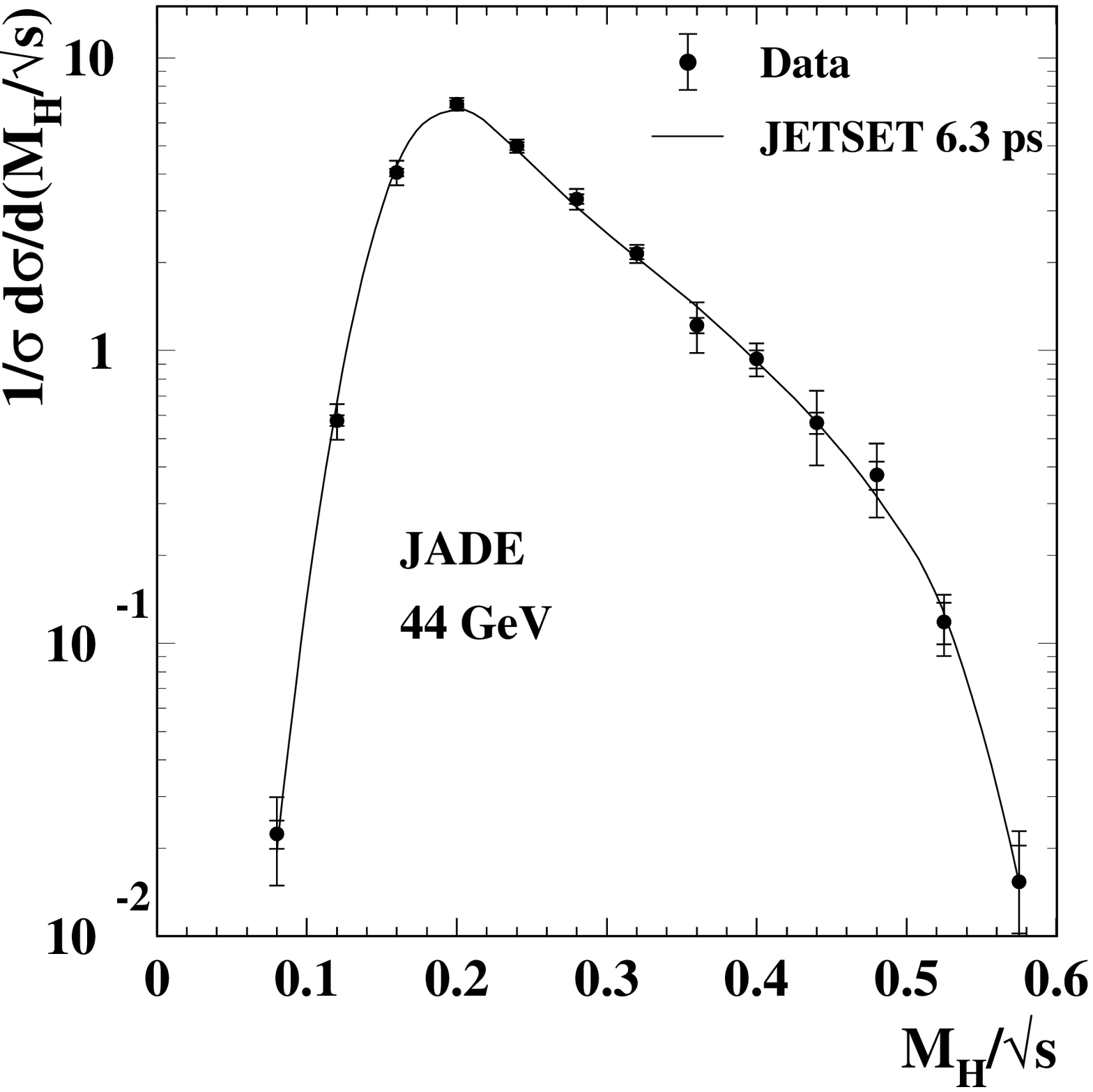}}\\
\resizebox{79mm}{!}{\includegraphics{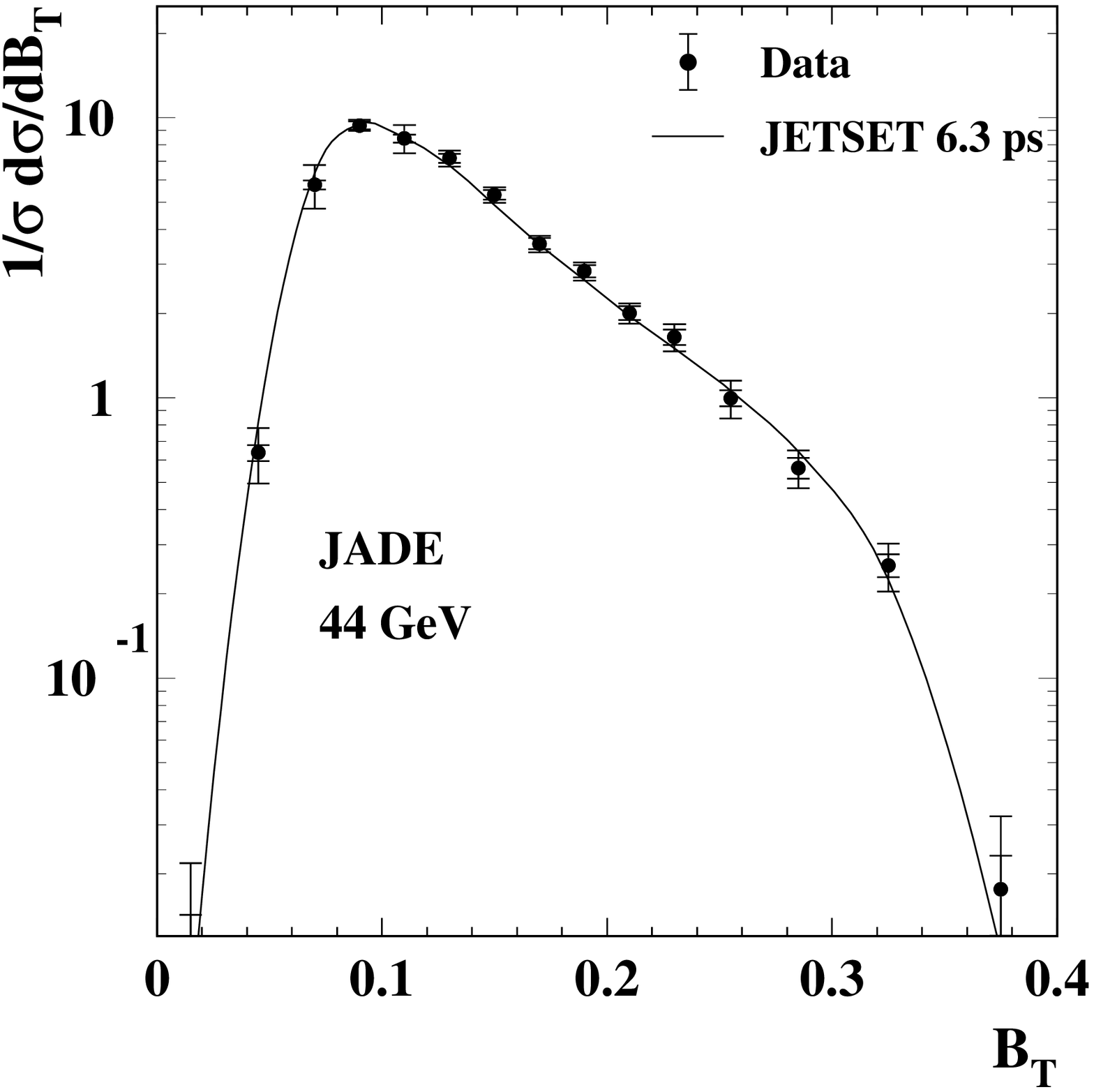}}
\resizebox{79mm}{!}{\includegraphics{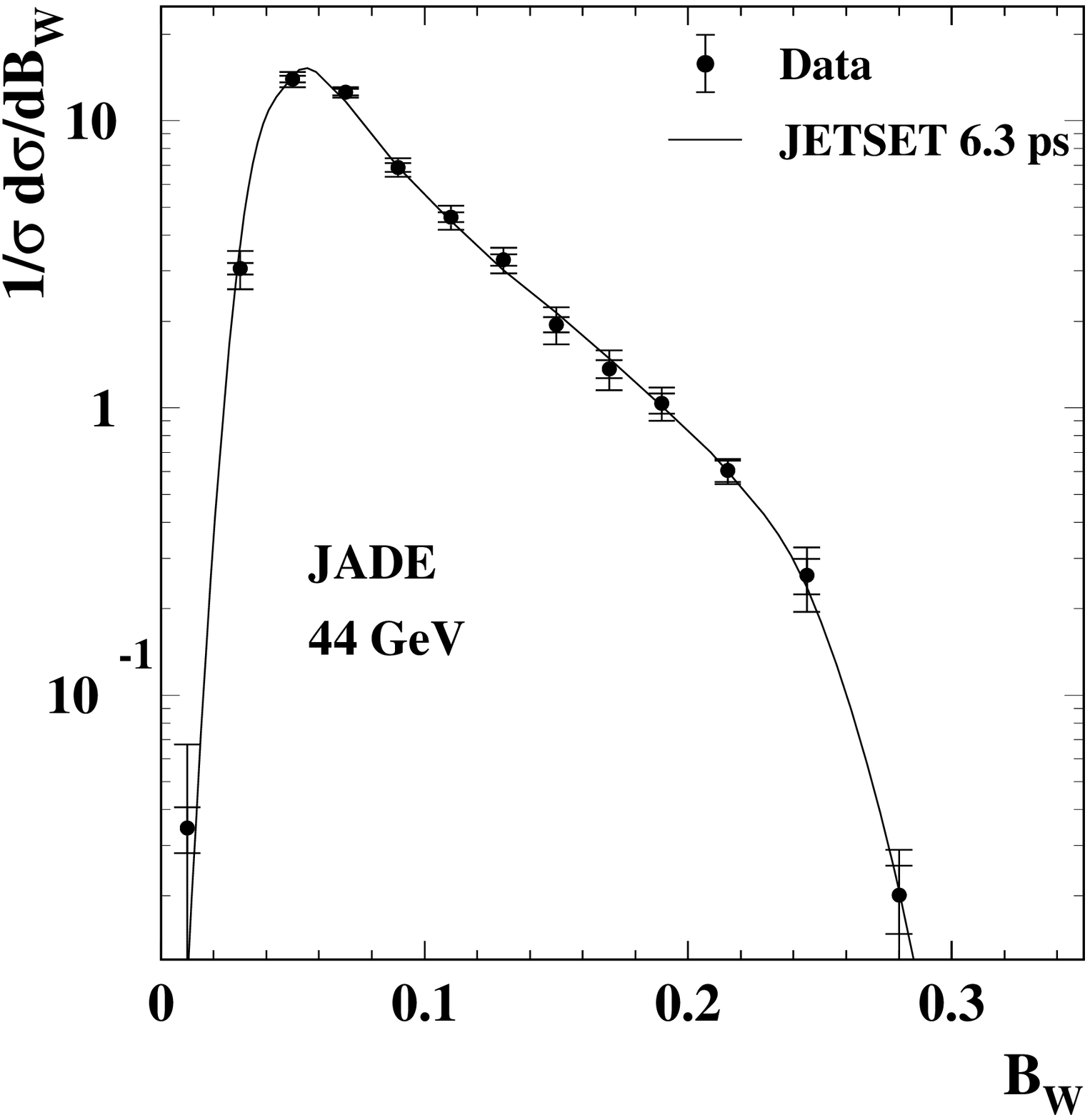}}
\end{center}
\caption{\label{fig-eventshapes-44GeV}
Event shape distributions at $\protect\sqrt{s} = 44$~GeV corrected to the 
hadron level are 
shown for 
Thrust ($T$), heavy jet mass ($M_H$), total ($B_T$) and wide jet broadening 
($B_W$).
The error bars show the statistical 
error (inner tick marks) and the total error obtained by adding the
statistical and experimental systematic error in quadrature.
The solid line represents the JETSET 6.3 parton shower model prediction.
}
\end{figure}

\begin{figure}[!htb]
\vspace*{-7mm}
\begin{center}
\resizebox{79mm}{!}{\includegraphics{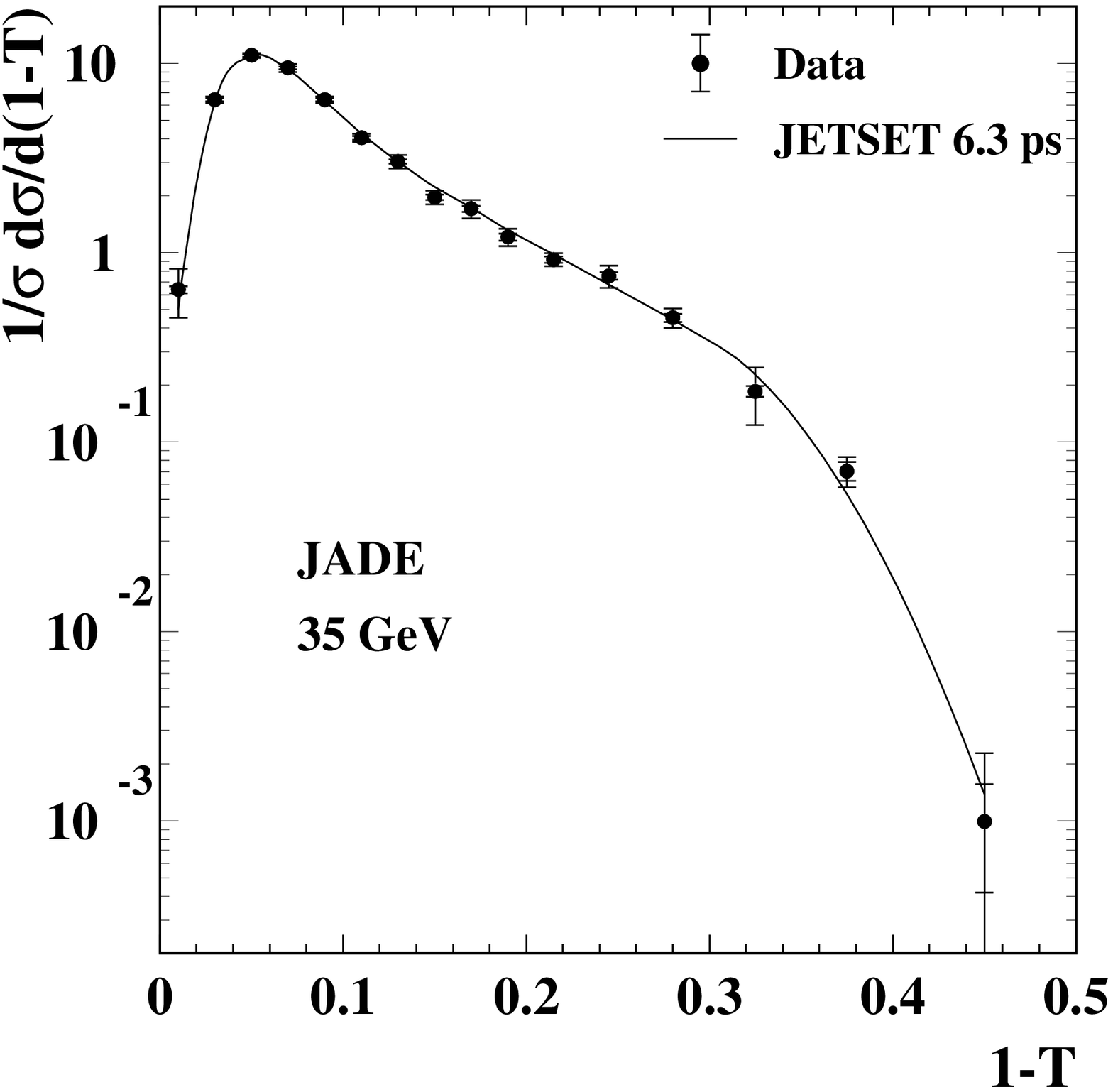}}
\resizebox{79mm}{!}{\includegraphics{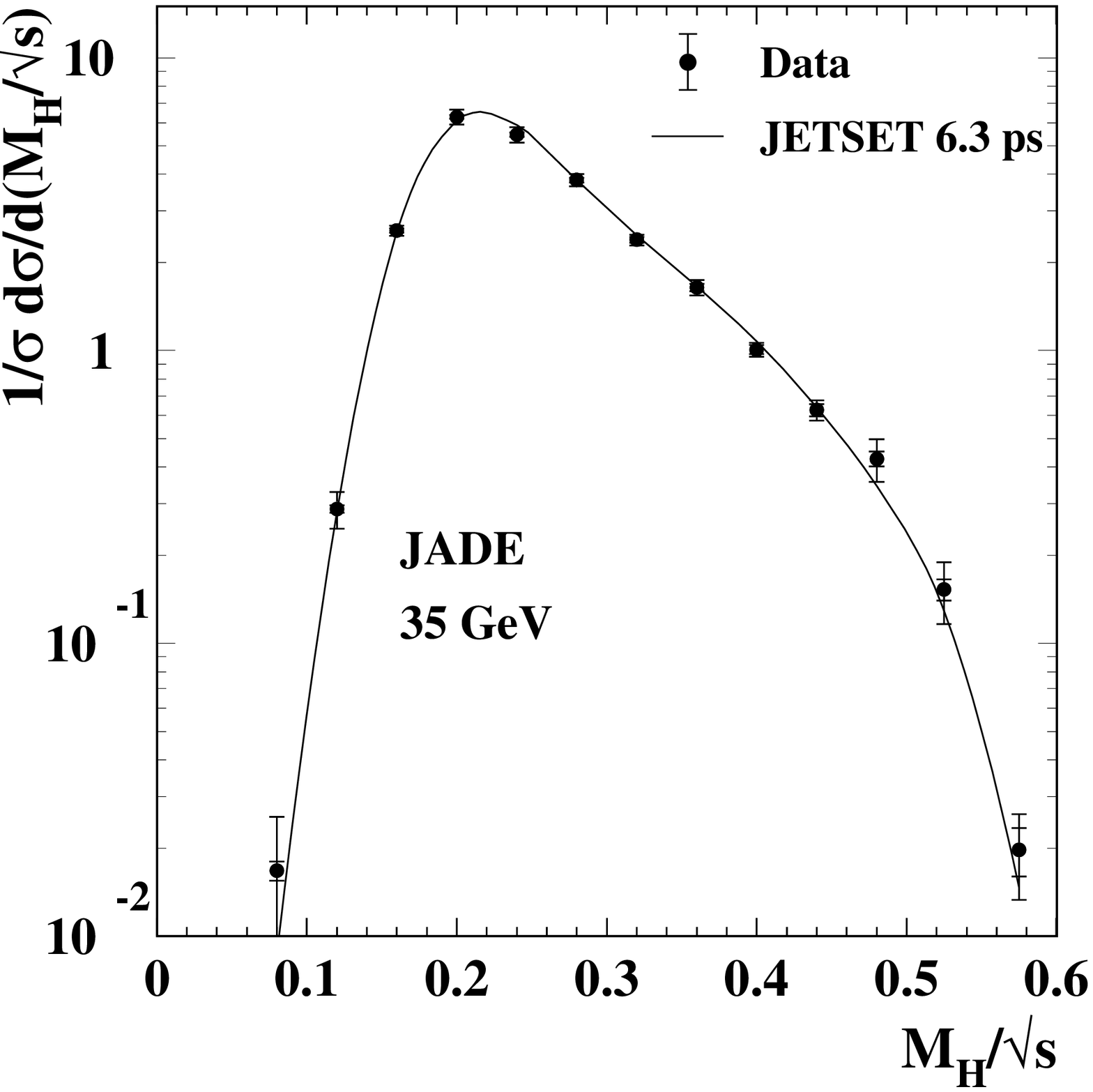}}\\
\resizebox{79mm}{!}{\includegraphics{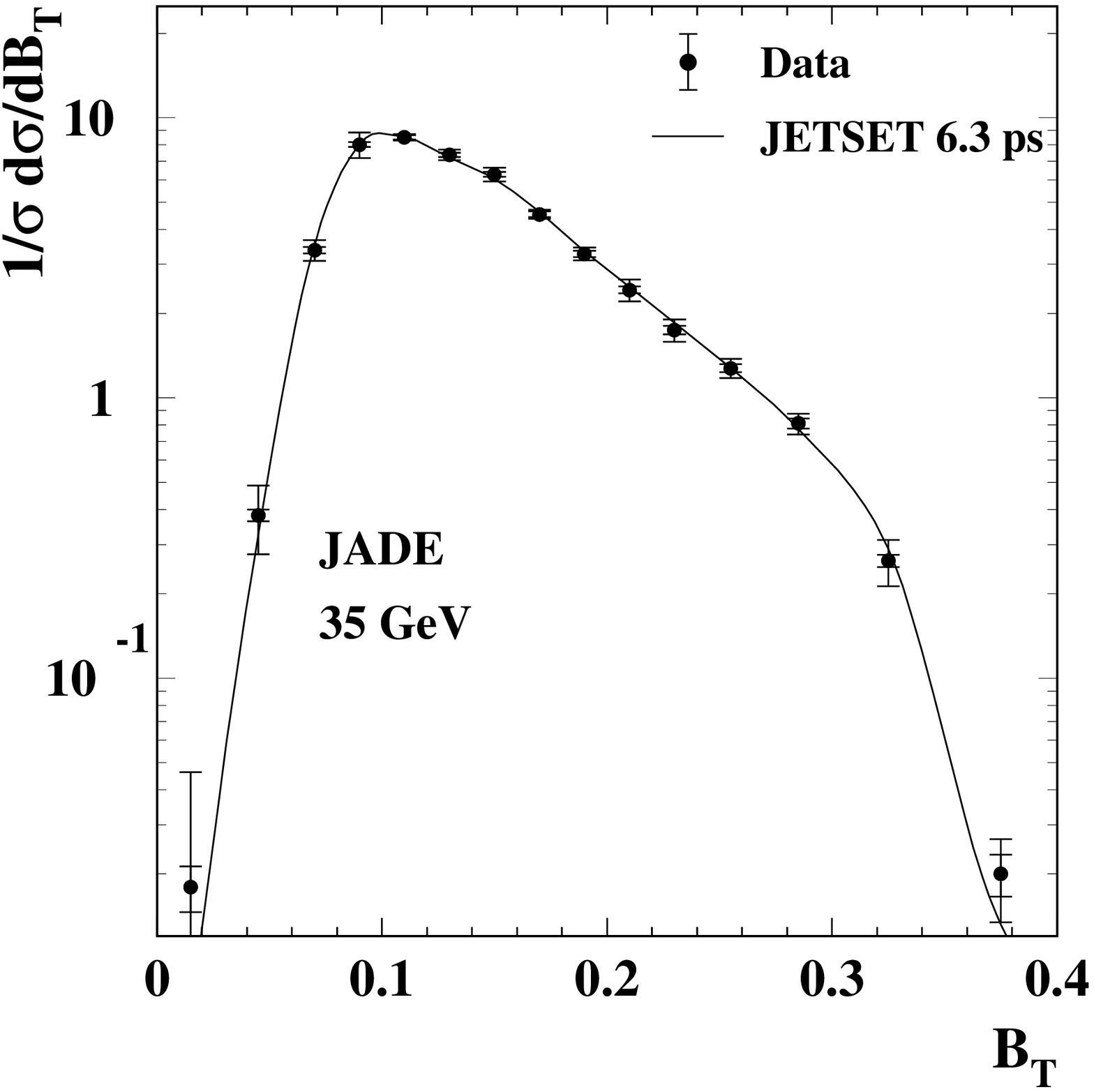}}
\resizebox{79mm}{!}{\includegraphics{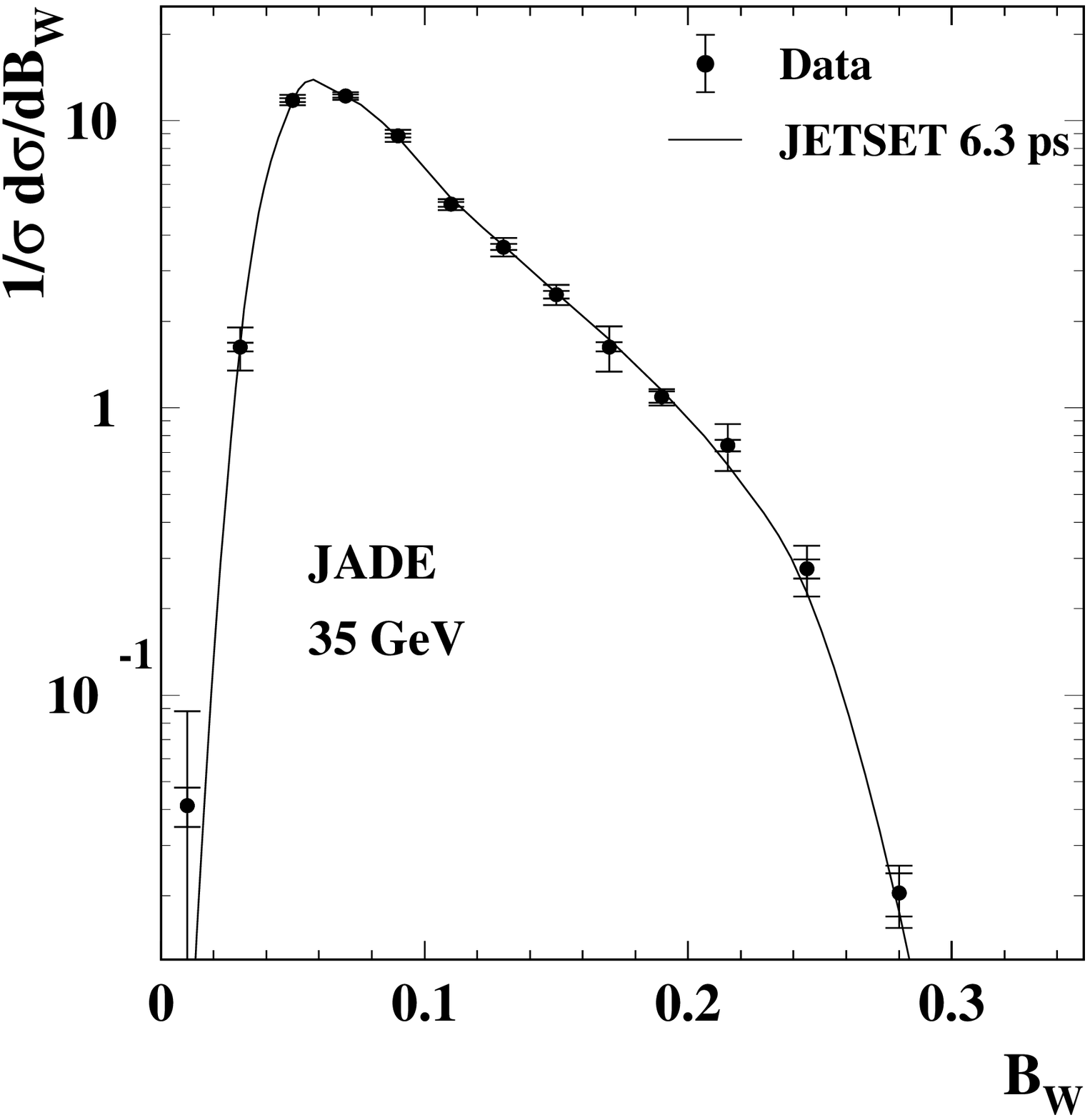}}
\end{center}
\caption{\label{fig-eventshapes-35GeV}
Event shape distributions corrected to the hadron level as for 
Figure~\protect\ref{fig-eventshapes-44GeV} 
but at $\protect\sqrt{s} = 35$~GeV. The solid 
line represents the JETSET 6.3 parton shower model prediction.
}
\end{figure}

\begin{figure}[!htb]
\vspace*{-7mm}
\begin{center}
\resizebox{79mm}{!}{\includegraphics{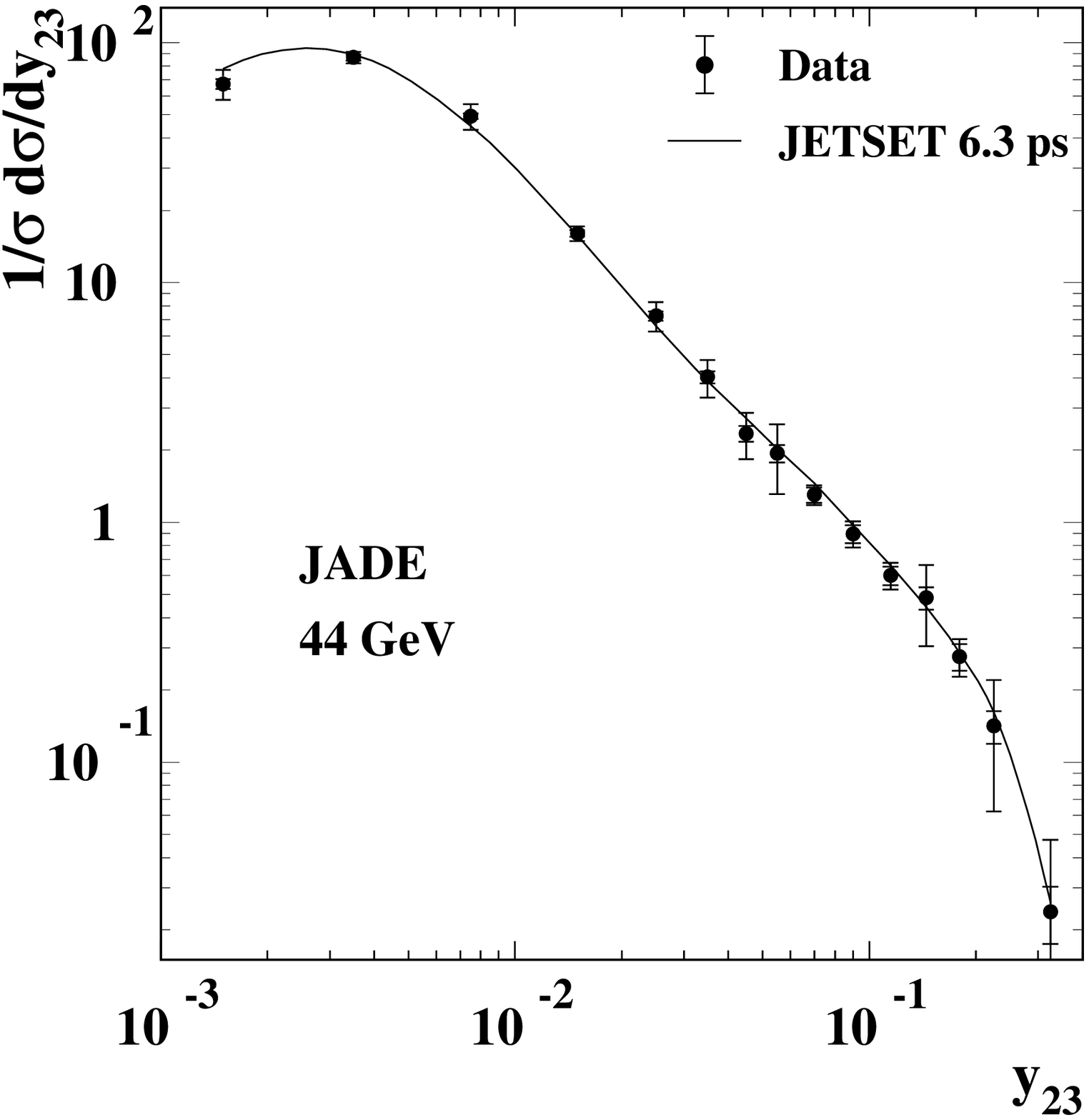}}
\resizebox{79mm}{!}{\includegraphics{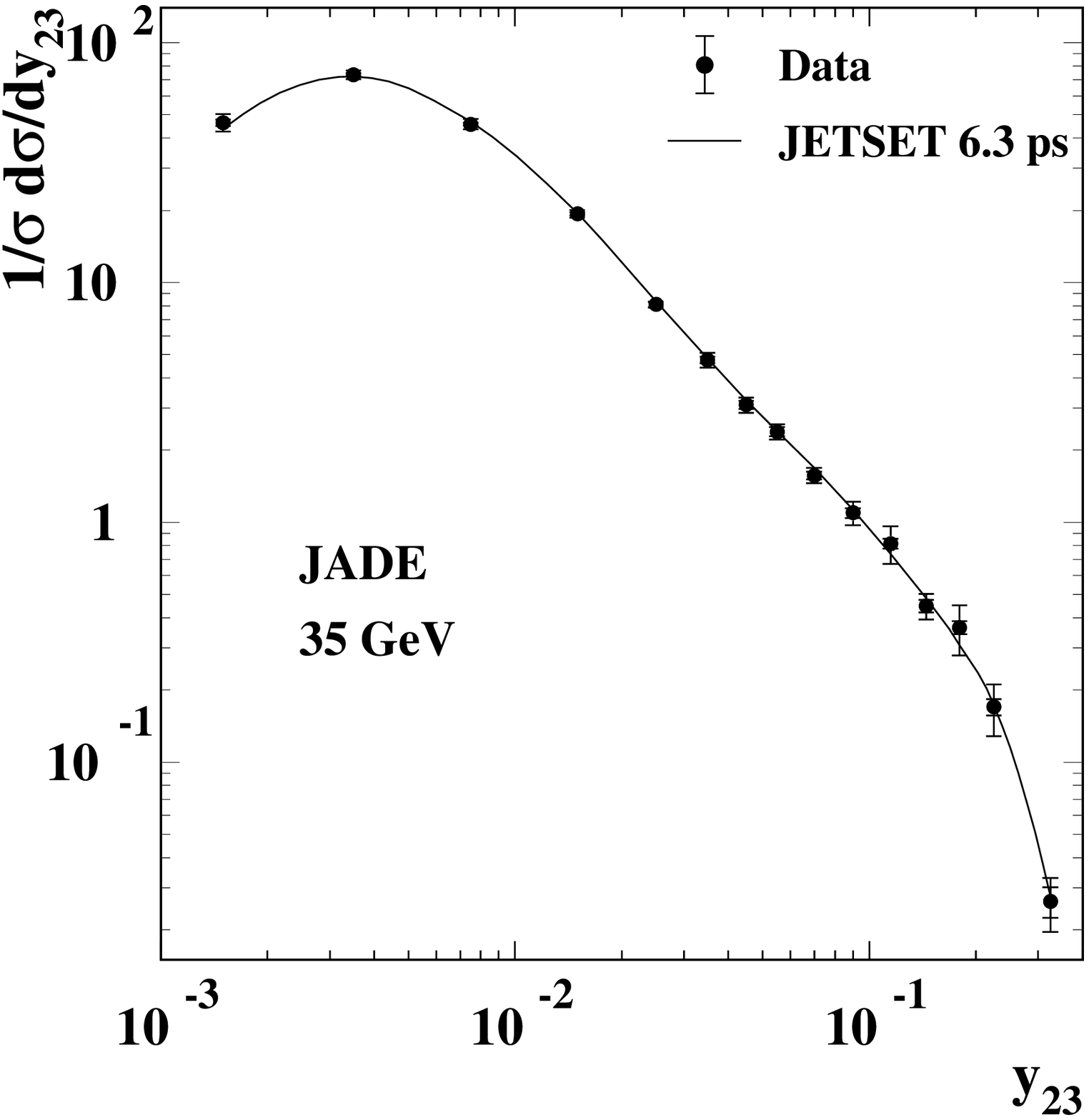}}\\
\resizebox{79mm}{!}{\includegraphics{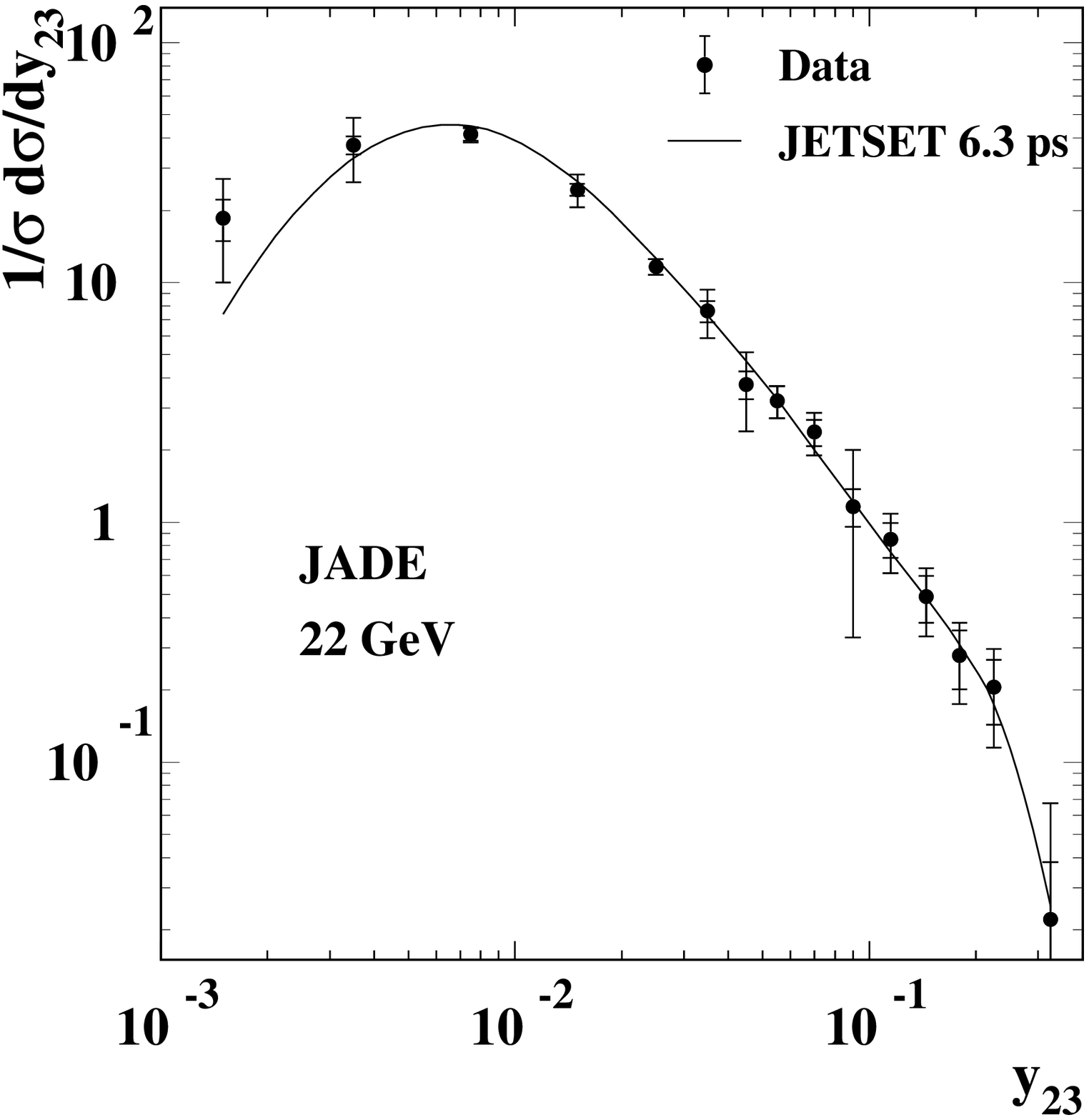}}
\end{center}
\caption{\label{fig-eventshapes-22GeV}
Event shape distributions corrected to the hadron level 
at $\protect\sqrt{s} = 44$, 
$35$, and $22$~GeV are shown for the differential 2-jet 
rate ($D_2$) in the Durham scheme. The error bars show the statistical 
error (inner tick marks) and the total error obtained by adding the
statistical and experimental systematic error in quadrature. The solid 
line represents the JETSET 6.3 parton shower model prediction.
}
\end{figure}

After applying the corrections for detector and for initial state radiation 
effects we obtained the event shape distributions at {\em hadron level}. 
These are shown in Figures~\ref{fig-eventshapes-44GeV}, 
\ref{fig-eventshapes-35GeV}, and \ref{fig-eventshapes-22GeV} 
for the $44$, $35$, and $22$~GeV data samples.
For comparison the respective 
distributions predicted by the JETSET 6.3 generator, at {\em hadron level}, 
are also shown.
There is excellent agreement between the data and the model over the
whole kinematic range of the observables.
In Tables~\ref{tab-eventshapes-44GeV}, \ref{tab-eventshapes-35GeV} and
\ref{tab-D2shapes} the corrected data values are listed with
statistical errors and experimental systematic uncertainties. The 
mean values of the distributions are also given.

\subsection{ QCD calculations for event shapes}
The distributions of the
event shape observables used in this analysis
are predicted in perturbative QCD by a 
combination of the \oaa~\cite{bib-ERT} and the
NLLA~\cite{bib-NLLA-1,bib-NLLA-2,bib-NLLA-3} calculations.
The \oaa\ calculation yields an
expression of the form 
\begin{displaymath}
    R_{\oaa}(y) = 1 + A(y)\left(\frac{\as}{2\pi}\right) 
                    + B(y)\left(\frac{\as}{2\pi}\right)^2,
\end{displaymath}
where 
$R(y) = \int_{0}^{y} {{\mathrm{d}} y}\ 1/\sigma_0 \cdot 
        {{\mathrm{d}}}\sigma/{{\mathrm{d}}y}$
is the cumulative cross-section of an event shape 
observable $y$
normalised to the lowest order Born cross-section $\sigma_0$. 
The NLLA calculations
give an expression for $R(y)$ in the form:
\begin{displaymath}
R_{\mathrm{NLLA}}(y) = \left(1 + C_1\left(\frac{\as}{2\pi}\right) 
                               + C_2\left(\frac{\as}{2\pi}\right) ^2\right) 
                       \cdot
                       \exp\left[L\, g_1\!\left(\frac{\as}{2\pi} L\right)
                                 +   g_2\!\left(\frac{\as}{2\pi} L\right)\right]
\end{displaymath}
where $L = {\mathrm{ln}}(1/y)$. The functions $g_1$ and $g_2$ are given by
the NLLA calculations. The coefficients $C_1$ and $C_2$ are known from the
\oaa\ matrix elements.

\subsection{ Determination of \as\ using \oaa+NLLA calculations}
We determined \as\ by \chisq\ fits to event 
shape distributions of $1-T$, $M_H$, $B_T$, $B_W$ and of $D_2$
corrected to the {\em parton level}.
For the sake of direct comparison to other published results we 
closely followed the procedures described in 
\cite{bib-OPALresummed,bib-eventshapes,bib-OPALNLLA}. 
We chose the so-called ln($R$)-matching
scheme to merge the \oaa\ with the NLLA calculations. 
The renormalisation scale factor, $\xmu \equiv \mu/\sqrt{s}$, was
set to $\xmu = 1$ for what we chose to be the main result. 
Here, the value of $\mu$ defines 
the energy scale at which the theory is renormalised. 

\begin{figure}[!htb]
\vspace*{-7mm}
\begin{center}
\resizebox{79mm}{!}{\includegraphics{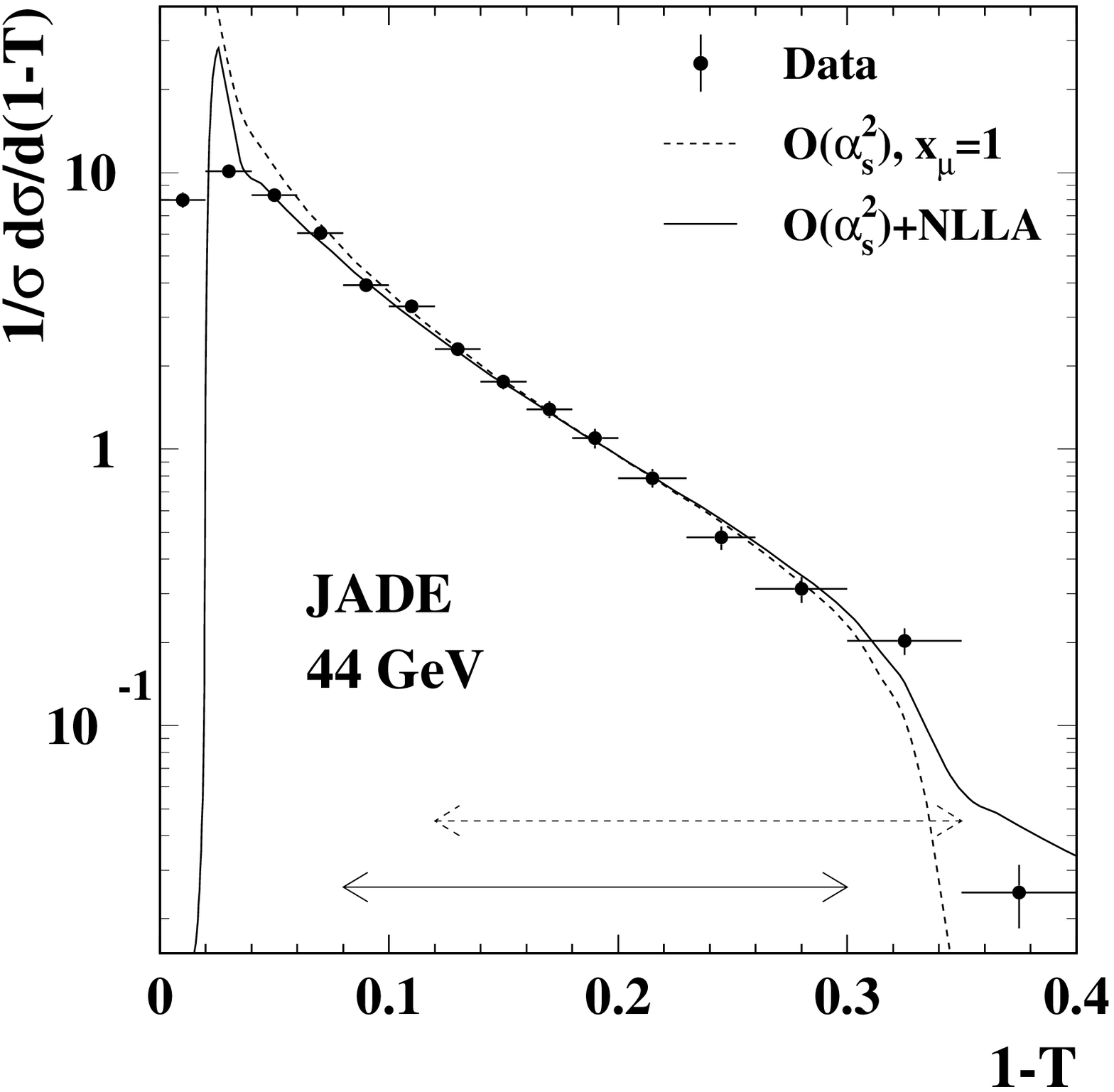}}
\resizebox{79mm}{!}{\includegraphics{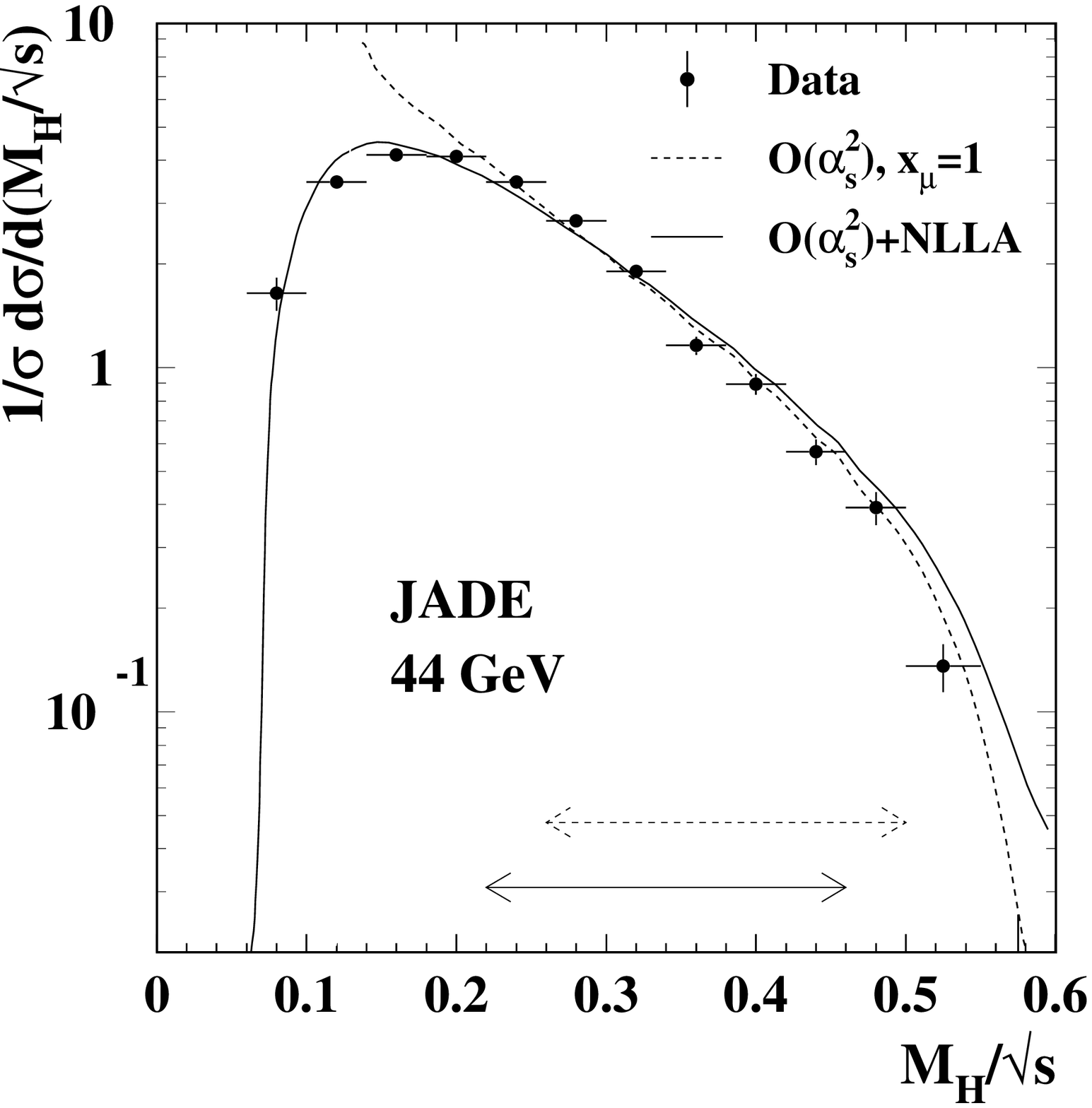}}\\
\resizebox{79mm}{!}{\includegraphics{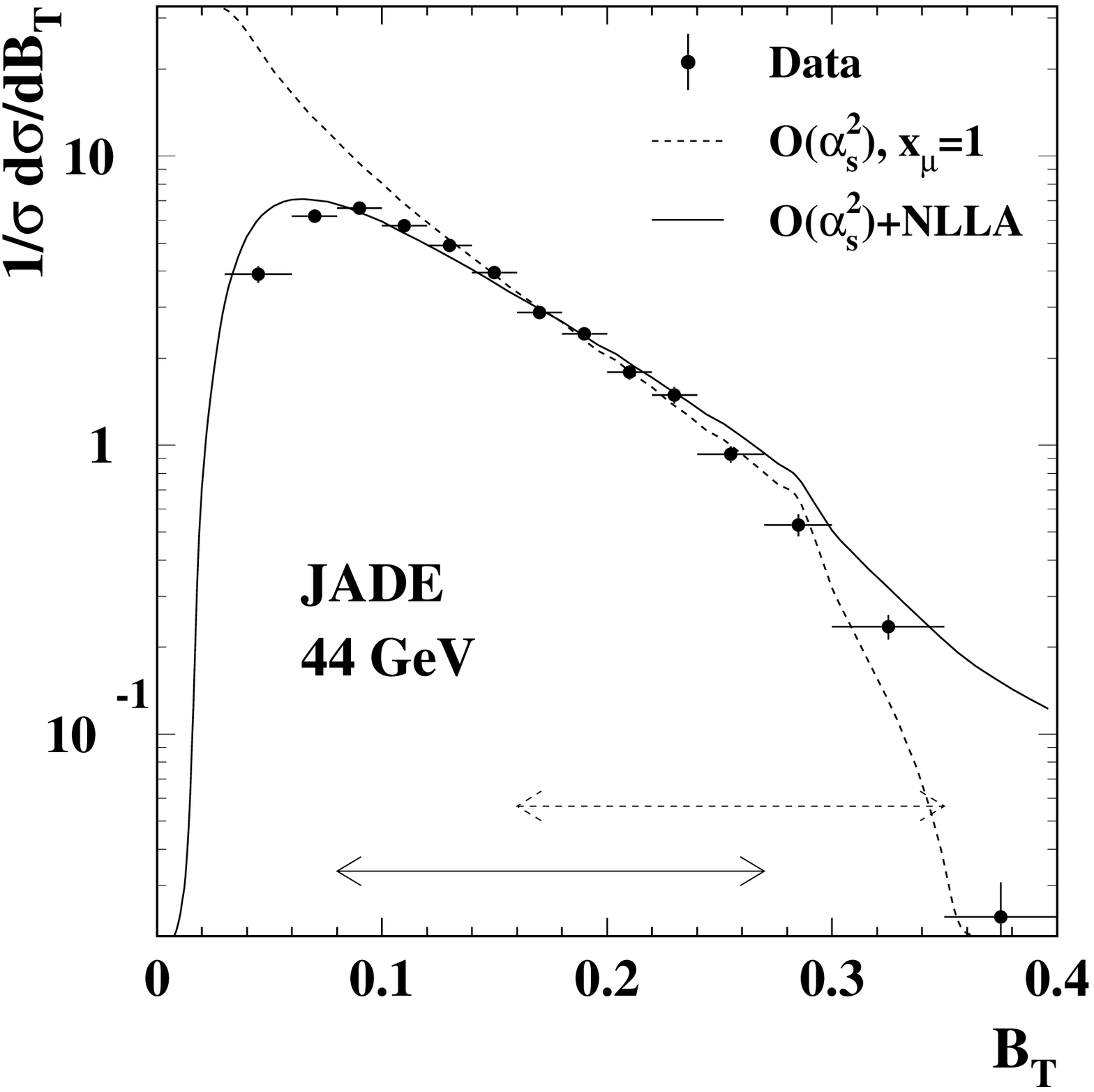}}
\resizebox{79mm}{!}{\includegraphics{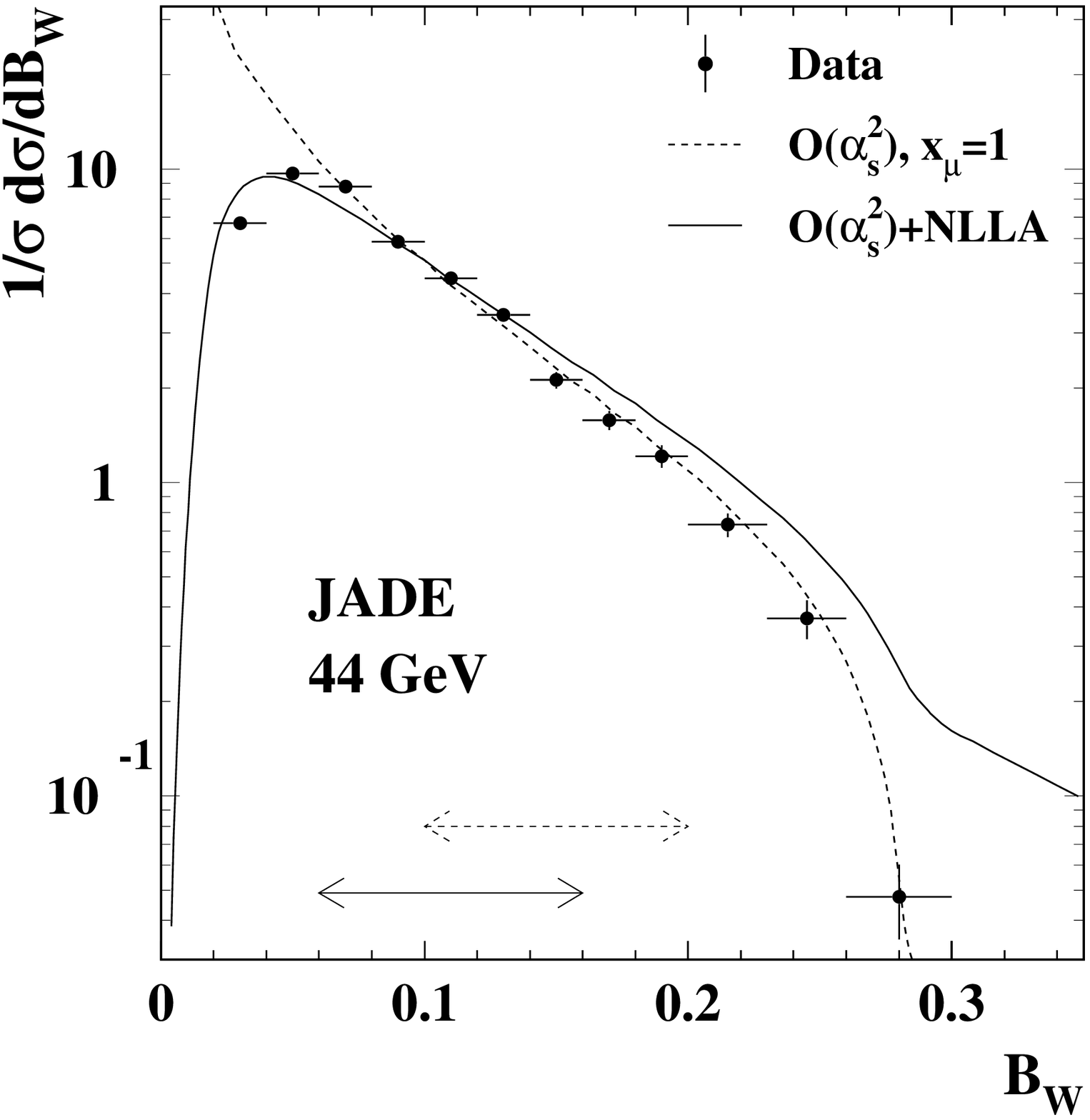}}
\end{center}
\caption{\label{fig-asresult-44GeV}
The distributions measured at $\protect\sqrt{s} = 44$~GeV and corrected 
to parton level are shown for thrust $T$, heavy jet mass $M_H$, 
total and wide jet broadening $B_T$ and $B_W$.
The fits of the \oaa+NLLA (solid line) and of the
\oaa($\xmu=1$) (dashed line) QCD predictions are overlayed and the fit 
ranges are indicated by the solid and dashed arrows. 
The error bars represent statistical errors only.
}
\end{figure}

\begin{figure}[!htb]
\vspace*{-7mm}
\begin{center}
\resizebox{79mm}{!}{\includegraphics{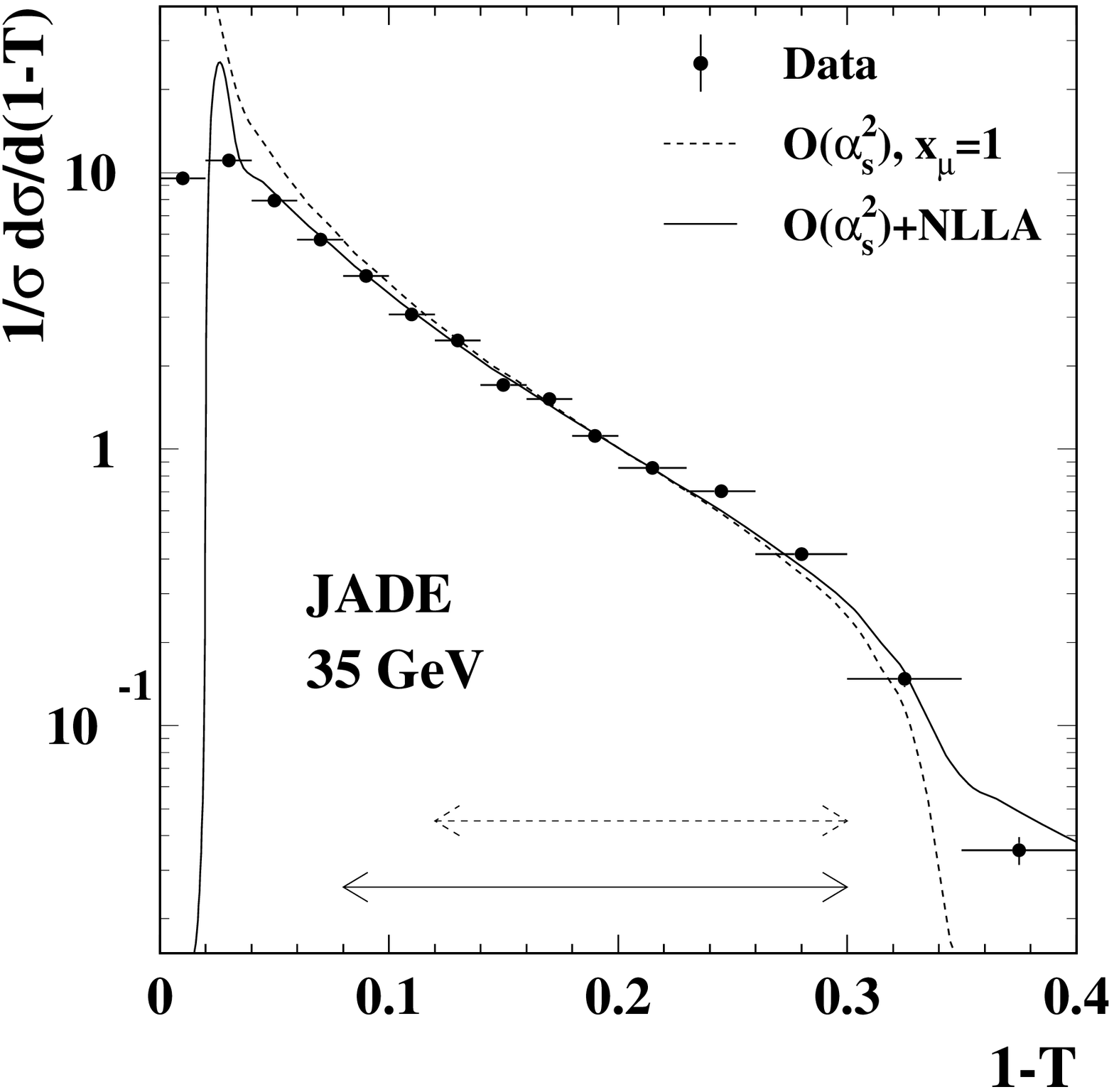}}
\resizebox{79mm}{!}{\includegraphics{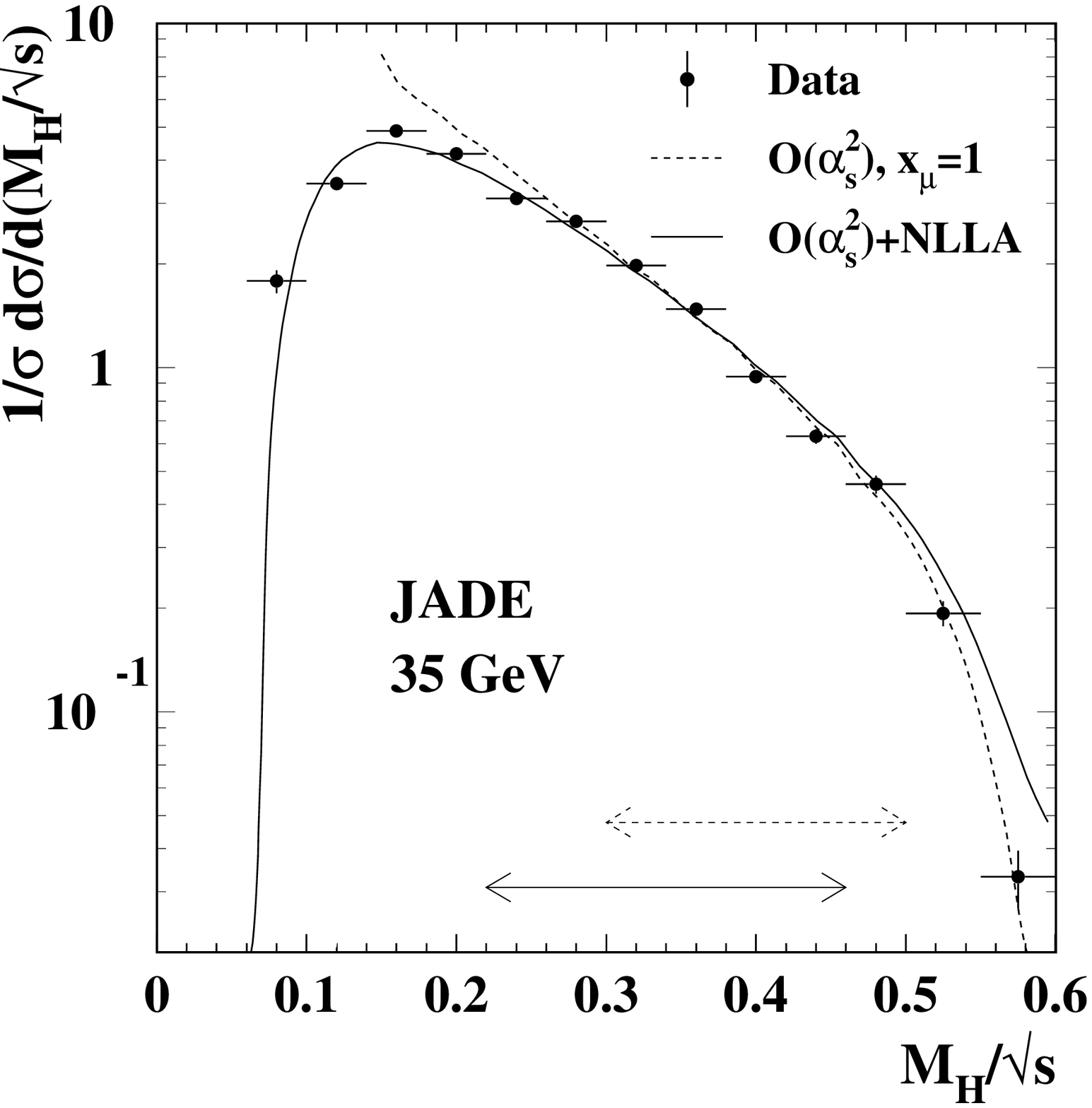}}\\
\resizebox{79mm}{!}{\includegraphics{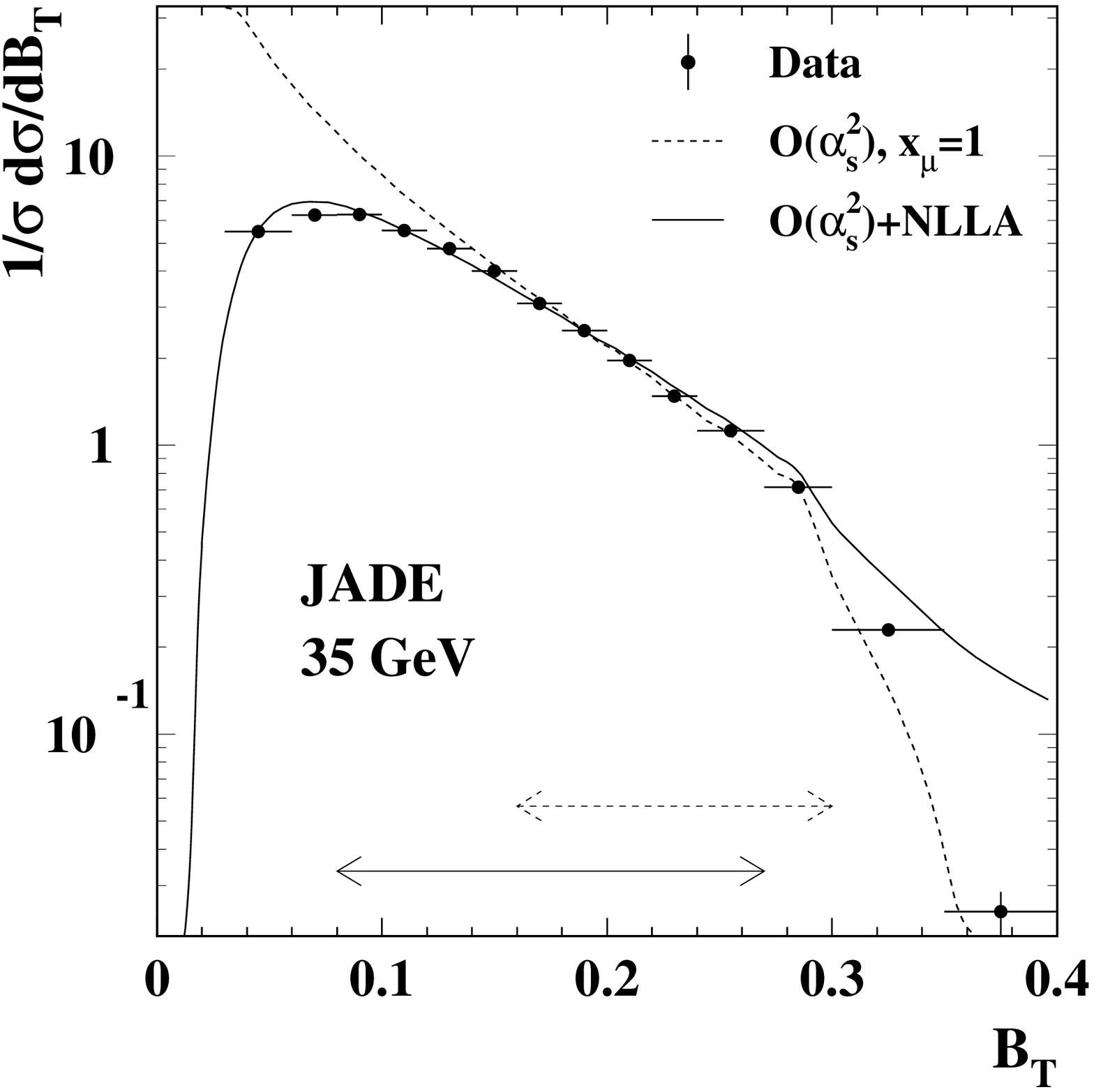}}
\resizebox{79mm}{!}{\includegraphics{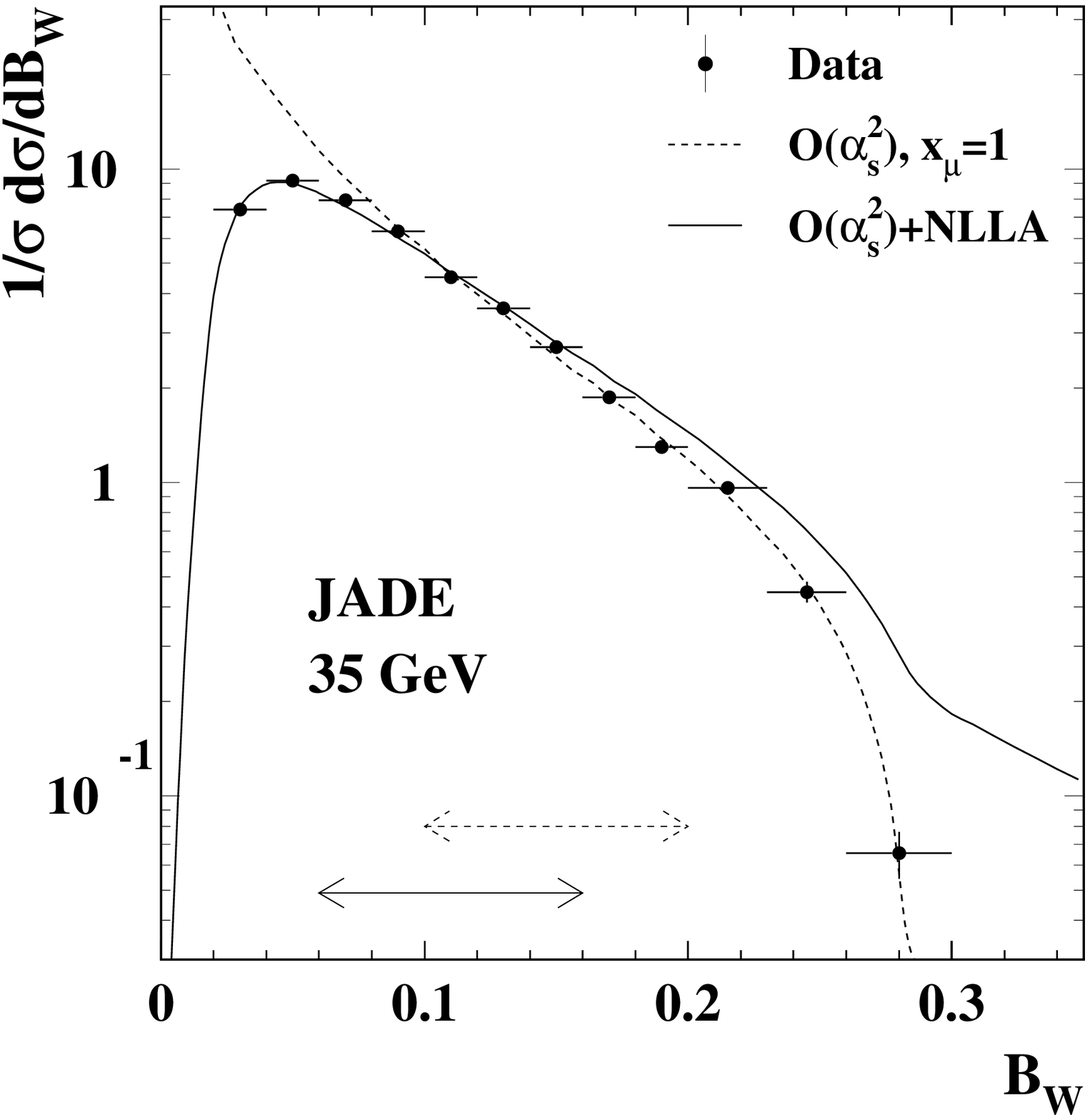}}
\end{center}
\caption{\label{fig-asresult-35GeV}
The same distributions as in Figure~\protect\ref{fig-asresult-44GeV} but for 
$\protect\sqrt{s} = 35$~GeV. 
}
\end{figure}

\begin{figure}[!htb]
\vspace*{-7mm}
\begin{center}
\resizebox{79mm}{!}{\includegraphics{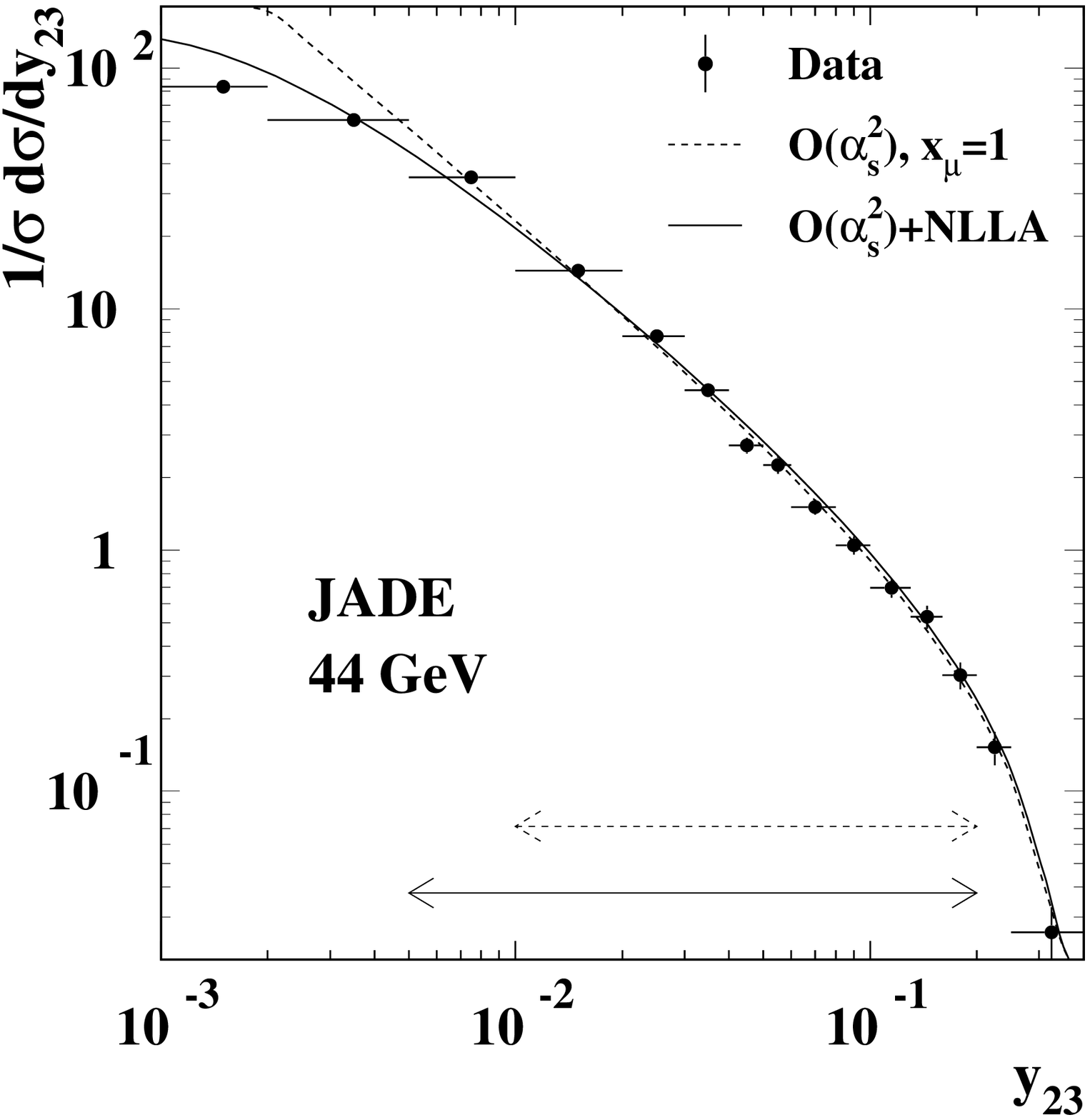}}
\resizebox{79mm}{!}{\includegraphics{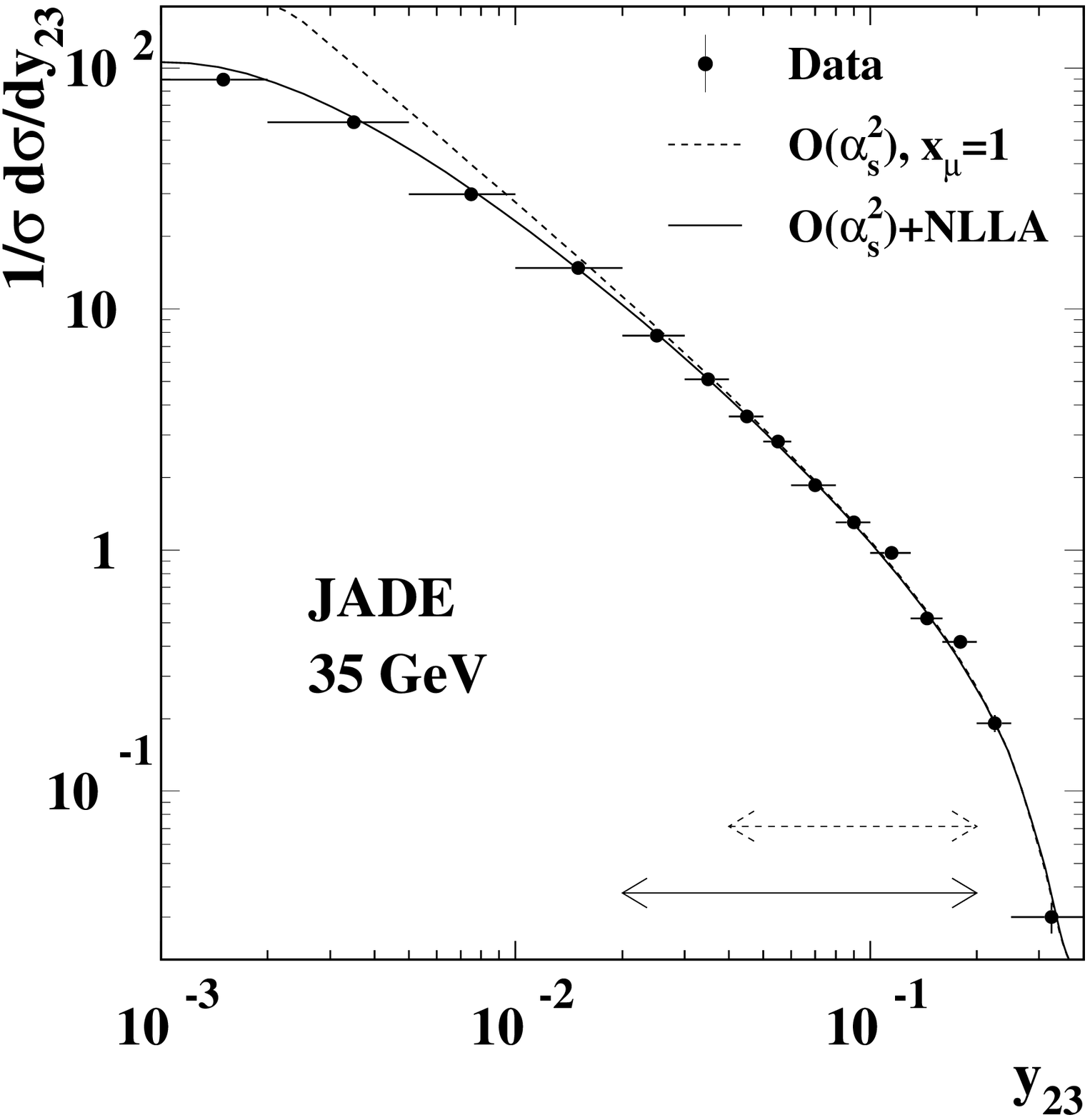}}\\
\resizebox{79mm}{!}{\includegraphics{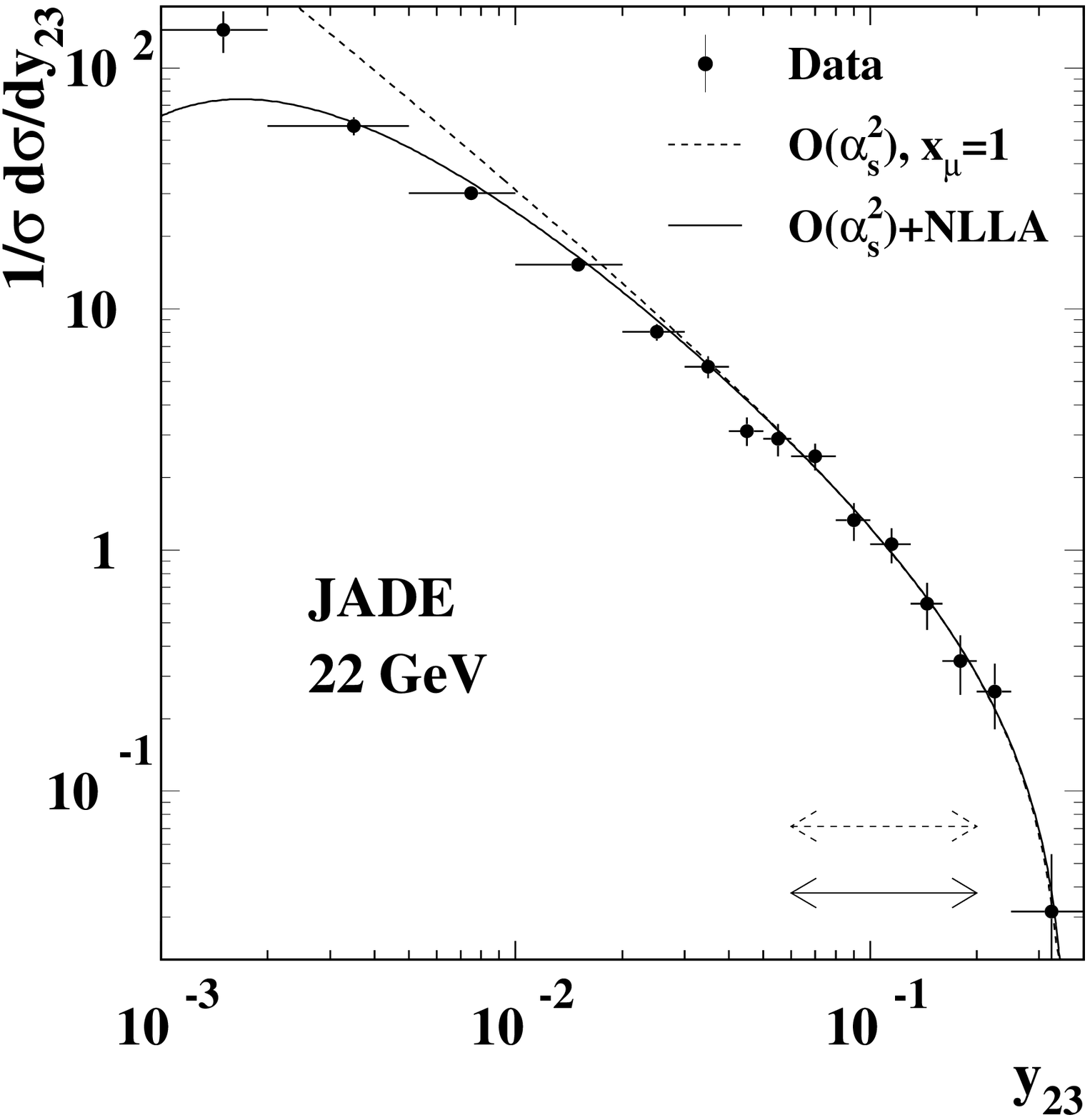}}
\end{center}
\caption{\label{fig-asresult-22GeV}
The distributions of the differential 2-jet rate, $D_2$, measured at 
$\protect\sqrt{s} = 44$, $35$, and $22$~GeV using the Durham 
scheme are shown after correction to the parton level. 
The solid and dashed lines correspond to the fit results as
in Figure~\protect\ref{fig-asresult-44GeV}.
}
\end{figure}

The fit ranges for each observable were determined   
by choosing the largest range for which the hadronisation
uncertainties remained 
below about $10$~\%, 
for which the \chisqd\ of the fits did not
exceed the minimum by more than a factor of two, 
and by 
aiming at results for \as\ that are independent of the fit range.        
The remaining 
changes when      
enlarging or reducing the fit range by one bin on either side   
were taken as systematic uncertainties.
Only statistical errors were considered in the fit thus resulting in   
\chisqd\ larger than unity.                                            
The finally selected fit ranges, the results of the \chisq\ fits and of 
the study of systematic uncertainties are 
tabulated in Tables~\ref{tab-asresult-44GeV}, \ref{tab-asresult-35GeV}, 
and \ref{tab-asresult-22GeV} and are shown in Figures~\ref{fig-asresult-44GeV},
\ref{fig-asresult-35GeV}, and \ref{fig-asresult-22GeV}. 

\begin{figure}[!htb]
\vspace*{-15mm}
\begin{center}
\resizebox{\textwidth}{!}{\includegraphics{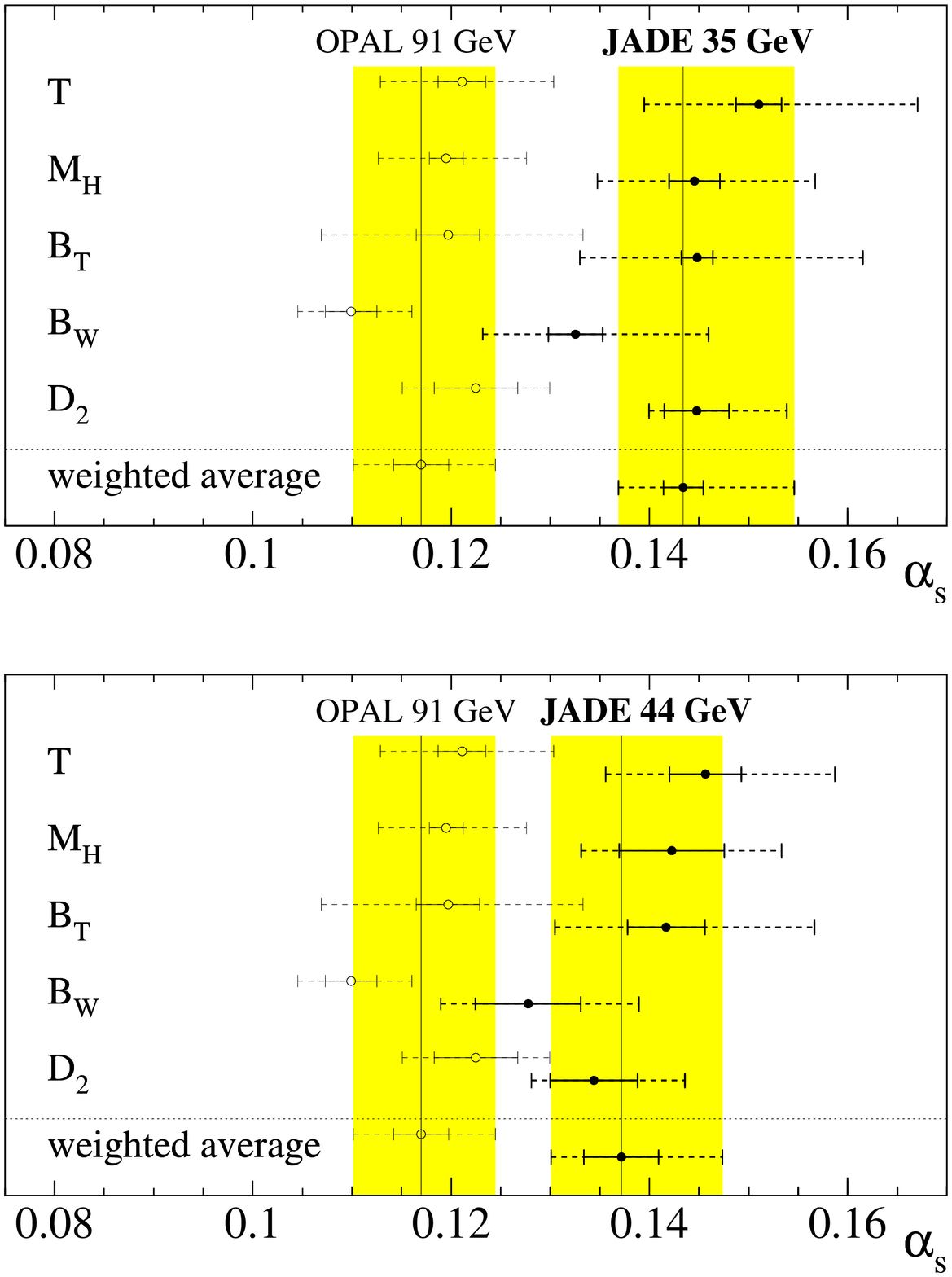}}
\end{center}
\vspace*{-5mm}
\caption{\label{fig-asresult-numbers}
Values of \as(35~GeV) and \as(44~GeV) derived from \oaa+NLLA fits to 
event shape distributions. The experimental and statistical uncertainties
are represented by the solid error bars. The dashed error bars show the
total error including hadronisation and higher order effects. The shaded
region shows the one standard deviation region around the weighted average
(see text). For comparison the \as\ values and errors measured by the 
OPAL Collaboration~\protect\cite{bib-OPALNLLA} for the same set of
observables are also shown. 
}
\end{figure}

The dependence of the 
fit result for \as\ on \xmu\ indicates the importance of higher order 
terms in the theory. 
We also changed the renormalisation
scale factor in the range of $\xmu = 0.5$ to $2.0$. 
We found variations of similar size as the
uncertainties from the detector correction and the 
hadronisation model dependence. The differential 2-jet rate, $D_2$,
in the Durham jet scheme exhibits the smallest 
renormalisation scale uncertainties, resulting in 
the smallest total error of all observables
considered in this analysis. The values of \as\ and the errors obtained
at $35$ and $44$~GeV are shown in Figure~\ref{fig-asresult-numbers}.
In these diagrams also the \as\ values measured by the OPAL
Collaboration at $\sqrt{s} = \mz$~\cite{bib-OPALresummed} are shown for 
comparison. 
The values of \as\ exhibit a similar scattering pattern at all energies.
This demonstrates the strong correlation of the systematic
uncertainties, which are dominated by theoretical and hadronisation
uncertainties. 

The individual results of the four event shape observables and the
differential 2-jet rate were combined into a single value following
the procedure described in 
References~\cite{bib-OPALresummed,bib-eventshapes,bib-globalalphas}.
This procedure accounts for correlations of the systematic uncertainties. 
At each energy, a weighted average of the five \as\ values was calculated
with the reciprocal of the square of the respective total error used
as a weight. 
In the case of asymmetric errors we took 
the average 
of the positive and negative error 
to determine the weight.
For each of the systematic checks, the mean of the \as\ values from all
considered observables was determined. 
Any deviation
of this mean from the weighted average of the main result was taken as
a  systematic uncertainty.

With this procedure we obtained as final results for \as
\begin{eqnarray*}
\as(44\ {\mathrm{GeV}}) & = & 0.1372 \pm 0.0017{\mathrm{(stat.)}}
                         \ ^{+0.0101}
                           _{-0.0069}{\mathrm{(syst.)}} \\
\as(35\ {\mathrm{GeV}}) & = & 0.1434 \pm 0.0010{\mathrm{(stat.)}}
                         \ ^{+0.0112}
                           _{-0.0065}{\mathrm{(syst.)}} \\
\as(22\ {\mathrm{GeV}}) & = & 0.1608 \pm 0.0083{\mathrm{(stat.)}}
                         \ ^{+0.0139} 
                           _{-0.0064}{\mathrm{(syst.)}}\ , 
\end{eqnarray*}
where the result at $22$~GeV is based on the differential 2-jet rate only.
The systematic errors at $44$, $35$, and $22$~GeV are 
the quadratic sums of
the experimental 
uncertainties ($\pm 0.0034$, $\pm 0.0018$, $\pm 0.0030$), the effects due
to the Monte Carlo modelling 
($^{+0.0049}_{-0.0027}$, $^{+0.0068}_{-0.0034}$, 
$^{+0.0119}_{-0.0056}$) 
and the contributions due to the variation of the renormalisation scale 
($^{+0.0082}_{-0.0054}$, $^{+0.0087}_{-0.0052}$, $^{+0.0066}_{-0.0001}$). 
It should be noted that the modelling uncertainties 
due to quark mass effects contribute significantly to the total error.

\subsection{ Determination of \as\ using \oaa\ calculations}
For comparison, we repeated the \as\ fits using fixed order \oaa\ %
calculations only. 
The fit ranges for each distribution had to be readjusted in order to
match the stability requirements given above\footnote{From corresponding
studies at LEP \cite{bib-OPALresummed,bib-globalalphas} it is known that
different fit ranges are required for the \oaa\ and for \oaa +NLLA
predictions. This is also supported by theoretical considerations, since
the inclusion of NLLA is supposed to extend the degree of reliability
especially in the 2-jet region of phase space, i.e. at small values of the
event shape observables used in this study.}.
All systematic checks were done as described 
above except for the variation of the renormalisation scale factor \xmu.
Instead, the \oaa-fits were performed once with $\xmu$ fixed to $1$ 
and once with \xmu\ as a free parameter of the fit. 
The fit ranges were the same in both fits except for $D_2$ where it had to    
be enlarged towards the lower end in order to obtain a stable fit with $\xmu$ 
as a free parameter.                                                          
The mean value of \as\ %
from the two fits was taken as the final result while half of the difference 
between the two was assigned as a systematic error due to the
unknown higher orders in perturbation theory. 
The results of the \oaa\ fits are summarised
in Tables~\ref{tab-oaa-results-44GeV} and \ref{tab-oaa-results-35GeV}.
The corresponding theoretical predictions 
were superimposed on the results of the \oaa+NLLA fits that are 
presented in Figures~\ref{fig-asresult-44GeV} and \ref{fig-asresult-35GeV}. 
All results at a given centre-of-mass energy agree with each other but
the \as\ values from the \oaa+NLLA fits are systematically lower.
Again, the pattern between these results and those obtained at
the higher energies of LEP \cite{bib-globalalphas} is very similar.

\section{Mean Values of Distributions and QCD Power Corrections}
\label{sec-meanvalues}
\subsection{ Power corrections} 
The value of \as\ can also be assessed by the energy dependence of 
mean values of 
event shape distributions. Presently, the mean values of the observables 
considered in this analysis are calculated up to \oaa. For 
an observable $\cal F$ the perturbative prediction is 
\begin{displaymath}
   \langle {\cal F}^{\mathrm{pert.}} \rangle = 
                              A_{\cal F}  \left( \frac{\as}{2\pi}\right)  +
              (B_{\cal F} - 2 A_{\cal F}) 
              \left( \frac{\as}{2\pi}\right)^2
\end{displaymath}
where the coefficients $A_{\cal F}$ and $B_{\cal F}$ were determined from the 
\oaa\ perturbative calculations~\cite{bib-ERT,bib-NLLA-1,bib-LEP1report,bib-EVENT2}. 
The term $-2 A_{\cal F}$ accounts for the difference between the total 
cross-section used in the measurement and the Born level cross-section used in 
the perturbative calculation. 
\begin{table}[b]
\begin{center}
\begin{tabular}{|c||c|r||c|c|c|}
\hline
Observable $\cal F$ & $A_{\cal F}$ & \multicolumn{1}{c|}{$B_{\cal F}$} 
                                                  & $a_{\cal F}$  & $p$ & $r$ \\
\hline\hline
$\langle T \rangle$       
                    & $2.103$      & $44.99$      & $-1$          & $1$ & $0$ \\
$\langle M_H^2/s \rangle$ 
                    & $2.103$      & $23.24$      & $1.0 \pm 0.5$ & $1$ & $0$ \\
$\langle B_T \rangle$ 
                    & $4.066$      & $64.24$      & $1.0 \pm 0.5$ & $1$ & $1$ \\
$\langle B_W \rangle$ 
                    & $4.066$      & $-9.53$      & $1.0 \pm 0.5$ & $1$ & $1$ \\
$\langle y_{23} \rangle$ 
                    & $0.895$      & $12.68$      & ---          & $2$ & --- \\
\hline
\end{tabular}
\end{center}
\caption{\label{tab-powcor}
Coefficients of the perturbative 
prediction~\protect\cite{bib-ERT,bib-NLLA-1,bib-LEP1report,bib-EVENT2} and coefficients and
parameters of the power corrections~\protect\cite{bib-webber} to the mean values of 
the event shape observables.
}
\end{table}
The 
numerical values of these coefficients are summarised in Table~\ref{tab-powcor}.

Instead of correcting for hadronisation effects with a Monte Carlo event
generator as we did for the \as\ determination presented in 
Section~\ref{sec-alphas}, we considered additive power-suppressed corrections 
($1/(\sqrt{s})^p$)
to the perturbative predictions of the 
mean values of the event shape observables.
Such corrections are expected on general grounds for hadronisation and 
other non-perturbative effects,
for example renormalons~\cite{bib-renormalons}. 
The non-perturbative effects are due to the emission of very low energetic 
gluons which can not be treated perturbatively due to the divergence of the 
perturbative expressions for \as\ at low scales. 
In the calculations of Reference~\cite{bib-webber} which we used in this
analysis a non-perturbative
parameter
\begin{displaymath}
\bar{\alpha}_p(\mu_I) = 
   \frac{p+1}{\mu_I^{p+1}} 
   \int_0^{\mu_I} {\mathrm{d}}k\ \ \as(k)\cdot k^{p}
\end{displaymath}
was introduced to replace the divergent portion of the perturbative
expression for $\as(\sqrt{s})$ below an infrared matching scale $\mu_I$.  
The general form of the power 
correction to the mean value of an observable $\cal F$ assumes the form
\begin{eqnarray*}
   \langle {\cal F}^{\mathrm{pow.}} \rangle & = & 
   a_{\cal F} \frac{4 C_F}{\pi p} \cdot
   \left( \frac{\mu_I}{\sqrt{s}}\right)^p \cdot
   \ln\! ^r\left(\frac{\sqrt{s}}{\mu_I}\right) \cdot
                                                  \nonumber \\
    & & 
   \cdot
   \left[
   \bar{\alpha}_{p-1}(\mu_I)-\as(\sqrt{s}) 
         - \frac{\beta_0}{2\pi}
  \left(\ln\frac{\sqrt{s}}{\mu_I}+\frac{K}{\beta_0}
                                  +\frac{1}{p}\right)\as^2(\sqrt{s})
   \right]\ ,
\end{eqnarray*}
where $C_F = 4/3$. The factor 
$\beta_0 = (11 C_A-2 N_f)/3$ stems from the QCD $\beta$-function of the
renormalisation group equation. It depends on the number of colours, 
$C_A = 3$, and number of active quark flavours $N_f$, for which we used 
$N_f=5$
throughout the analysis. The term $K = (67/18-\pi^2/6) C_A - 5/9 \cdot N_f$
originates from the choice of the $\overline{\mathrm{MS}}$ renormalisation
scheme.
The remaining coefficient $a_{\cal F}$ and the parameters $p$ and $r$ depend 
on the event shape observable. For completeness, these coefficients and 
parameters obtained in Reference~\cite{bib-webber} are also listed in 
Table~\ref{tab-powcor}. 

\subsection{ Determination of \as\ using power corrections} 
We determined $\as(\mz)$ by \chisq\ fits of the expression
\begin{displaymath} 
 \langle {\cal F}\rangle = \langle {\cal F}^{\mathrm{pert.}} \rangle + 
                           \langle {\cal F}^{\mathrm{pow.}}  \rangle . 
\end{displaymath} 
to the mean values of the five observables investigated in this
analysis\footnote{Our results for $\langle M_H^2/s\rangle$ are $0.0745 \pm 0.0011$ 
and $0.0679 \pm 0.0008$ at $\sqrt{s} = 35$~GeV and $44$~GeV respectively. The      
errors are the statistical and systematical uncertainties added in quadrature.}    
including the measured mean values obtained by other
experiments at different centre-of-mass 
energies~\cite{bib-L3alphas,bib-OPALNLLA,bib-meanvalues,bib-DELPHI-powcor}.
For the 
central values of \as\ from the fits we chose a renormalisation 
scale factor of $\xmu=1$ and an infrared scale of $\mu_I=2$~GeV.
The \chisqd\ of all fits were between $0.8$ ($\langle M_H^2/s\rangle$)
and $4.2$ ($\langle B_T\rangle$).
We estimated the systematic uncertainties by varying \xmu\ from
$0.5$ to $2$ and $\mu_I$ from $1$ to $3$~GeV. Since the precision 
of the coefficient $a_{\cal F}$ as given in Reference~\cite{bib-webber}
for the heavy jet mass $M_H$ and the two jet broadening measures, 
$B_T$ and $B_W$ is only $\pm 50\%$, we assigned an additional
uncertainty to \as\ due to the variation of these coefficients by
this amount.

In the case of $\langle y_{23}\rangle$ 
no coefficient $a_{\cal F}$
is given in Reference \cite{bib-webber}.
We investigated the size of
$a_{\cal F}$ 
by fitting with $\as$ fixed to
the world average~\cite{bib-world-alphas-sb,bib-world-alphas-ms}
$\as^{\mathrm{w.a.}}(\mz)=0.118$. All fits with $p=1$ or $2$,
and $r=0$ or $1$
resulted in very small values of
$a_{\cal F}$ compatible with  zero.
From this we conclude that 
power corrections to the perturbative prediction for $\langle y_{23}\rangle$ 
can be neglected for the energy range considered. Therefore, we used the 
perturbative prediction only, for which we obtained a good fit with
$\chisqd=1$. 

\begin{figure}[!htb]
\vspace*{-7mm}
\begin{center}
\resizebox{74mm}{!}{\includegraphics{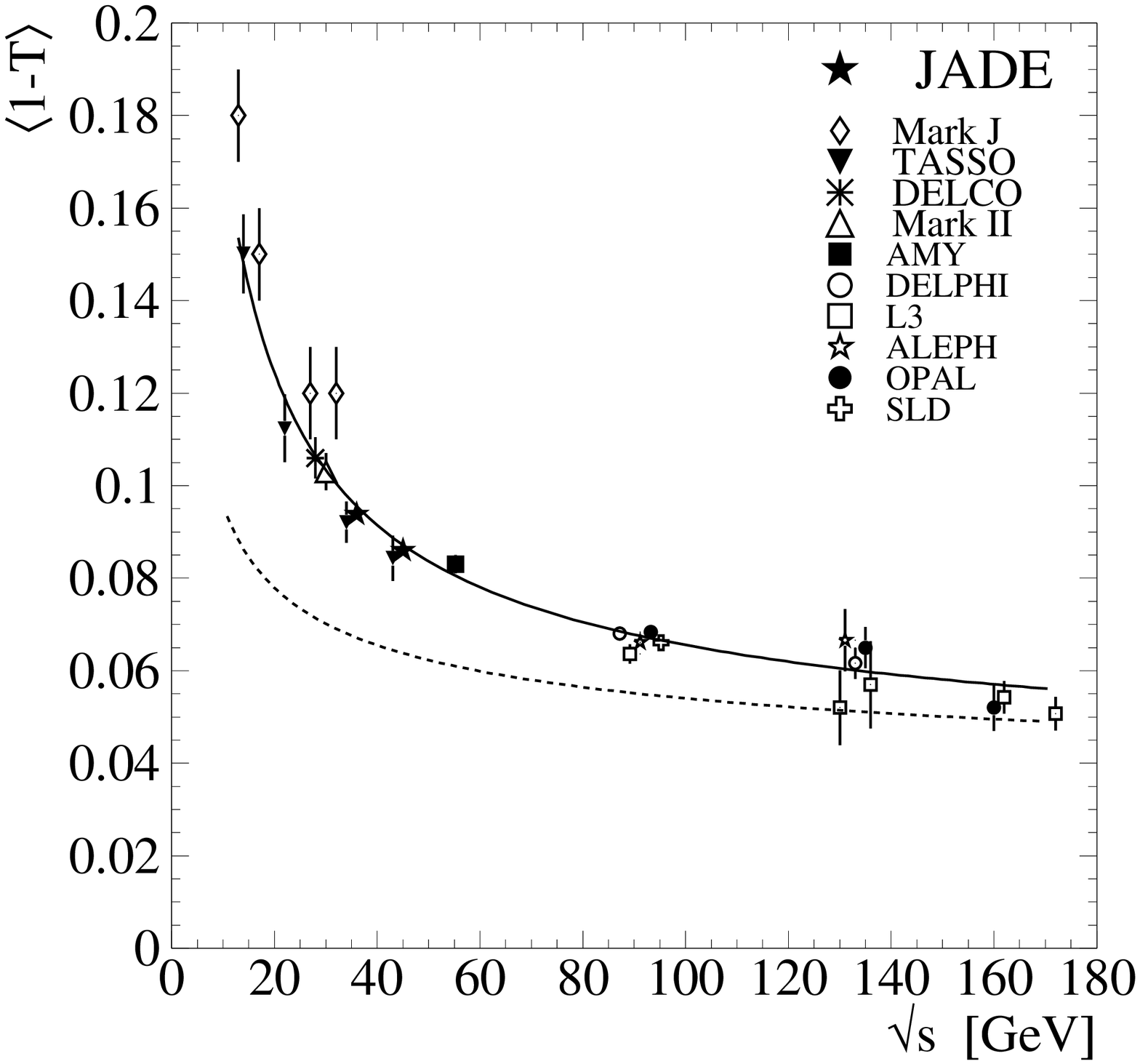}}
\resizebox{74mm}{!}{\includegraphics{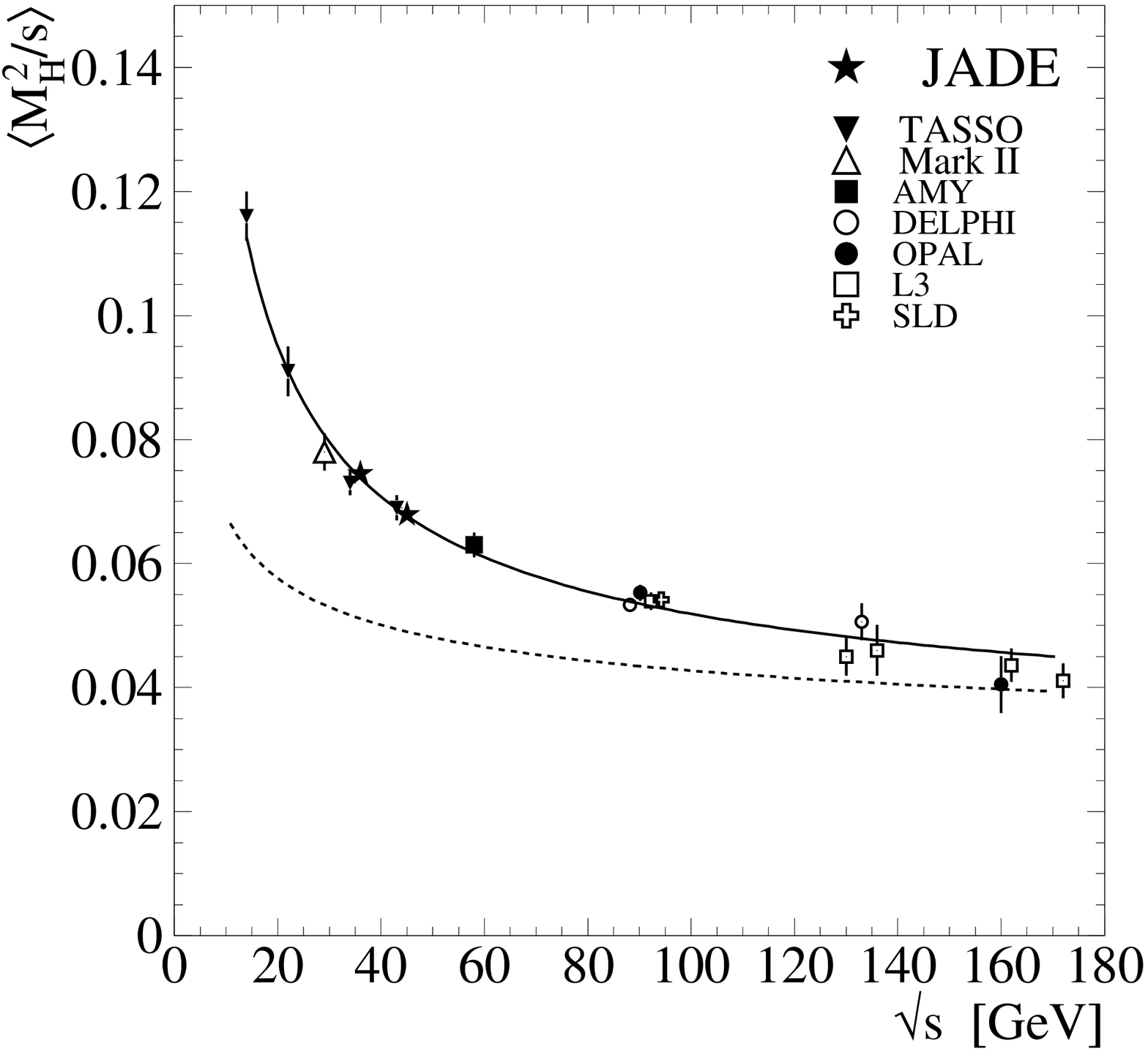}}\\
\resizebox{74mm}{!}{\includegraphics{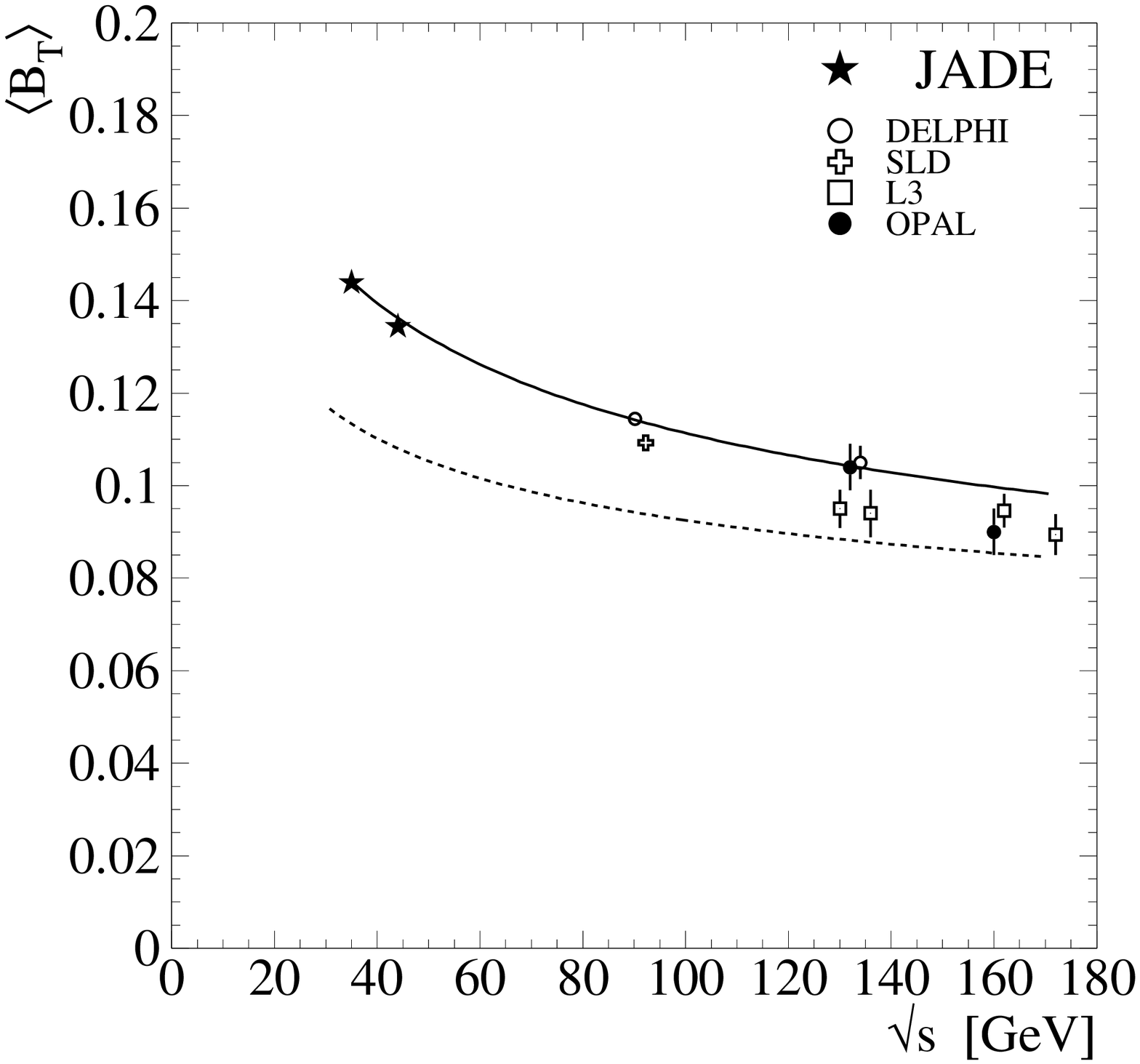}}
\resizebox{74mm}{!}{\includegraphics{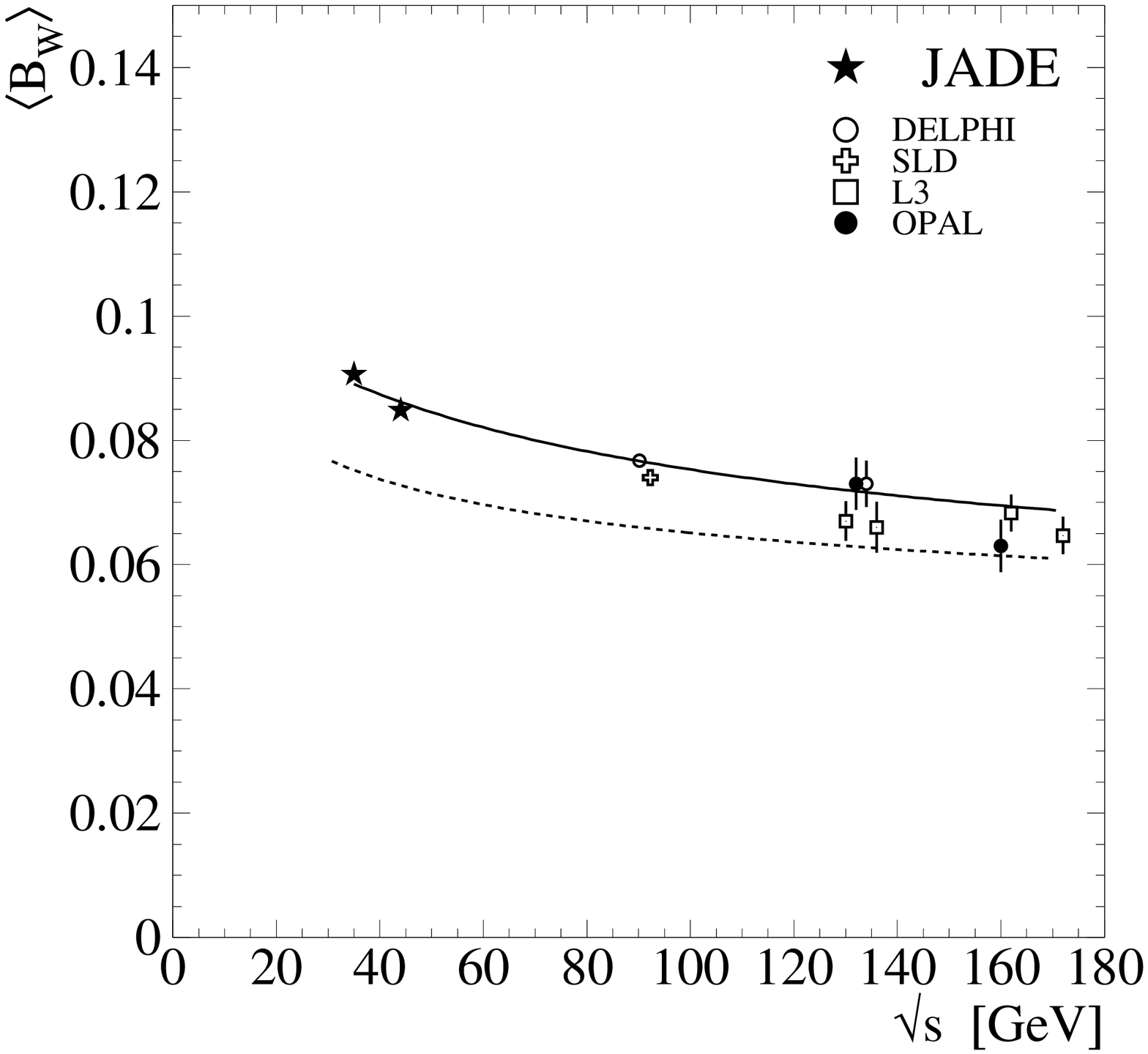}}\\
\resizebox{74mm}{!}{\includegraphics{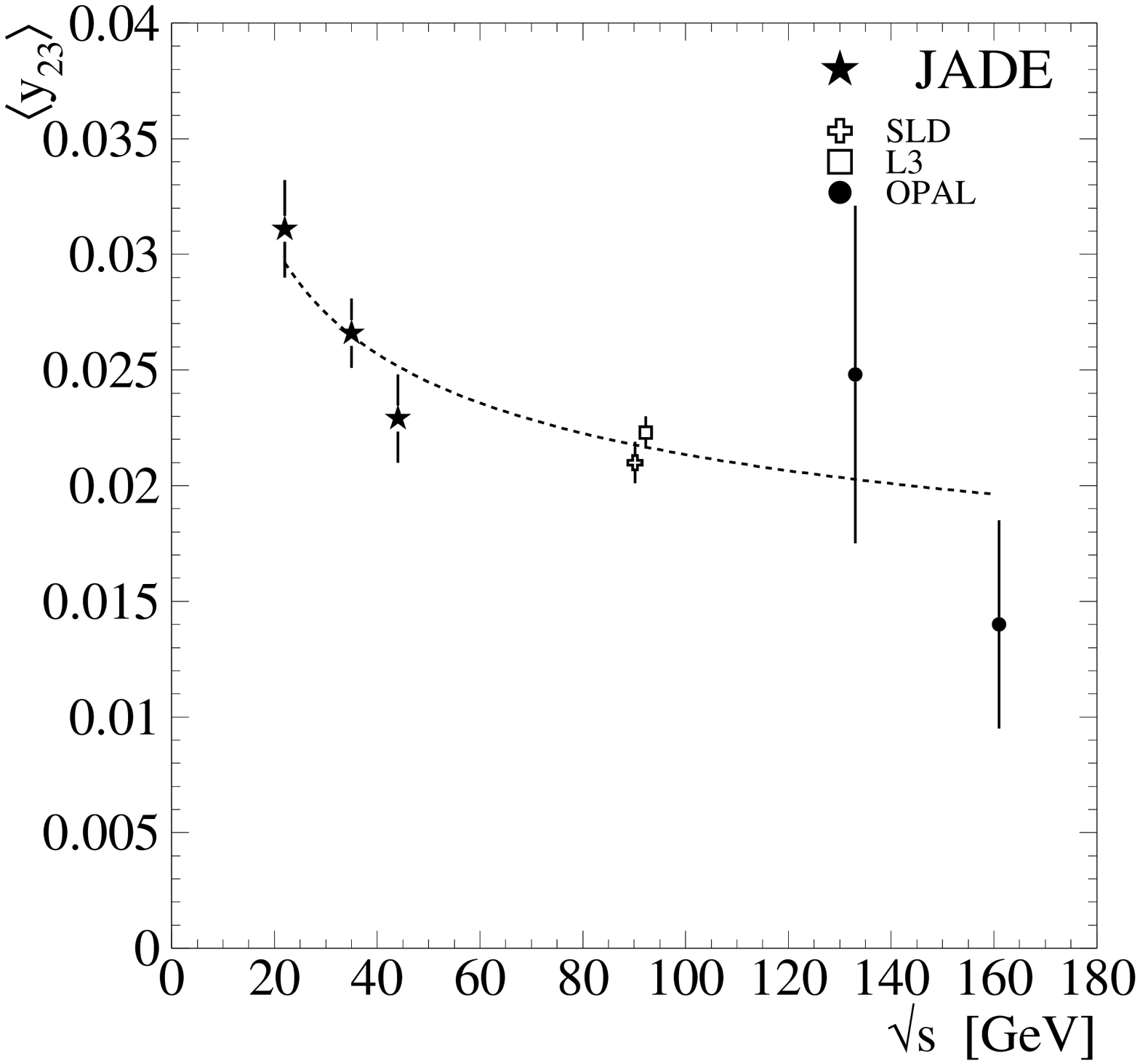}}
\end{center}
\vspace*{-5mm}
\caption{\label{fig-as-powcor}
Energy dependence of the mean values of thrust $\langle 1-T\rangle$,
heavy jet mass $\langle M_H^2/s\rangle$, total $\langle B_T\rangle$
and wide jet broadening $\langle B_W\rangle$, and of the differential
2-jet rate $\langle y_{23}\rangle$ are 
shown~\protect\cite{bib-L3alphas,bib-OPALNLLA,bib-meanvalues,bib-DELPHI-powcor}. 
The solid curve is the result
of the fit using perturbative calculations plus power corrections while
the dashed line is the perturbative prediction for the same value of
\asmz.} 
\end{figure}

The results of the fits are shown in Figure~\ref{fig-as-powcor}
and the numeric values are listed in Table~\ref{tab-as-powcor}.
It presents the values for \as\ and for $\bar{\alpha}_0$, the
experimental errors and systematic uncertainties of the fit
results.
We consider these results based on power corrections as a test of this      
theoretical prediction. It should be noted that the theoretically expected  
universality of $\bar{\alpha}_0$ is not observed. The issue of universality 
is further addressed in~\cite{bib-dokshitzer}.                              

Employing the procedure used in Section~\ref{sec-alphas} to combine the 
individual \as\ values, we obtained
\begin{displaymath}
   \as(\mz) = 0.1155\ ^{+0.0062}_{-0.0045}
\end{displaymath}
where the error is the experimental uncertainty ($\pm 0.0013$), the
renormalisation scale uncertainty ($^{+0.0045}_{-0.0033}$),
the uncertainty due to the choice of the infrared scale 
($^{+0.0029}_{-0.0019}$) and the uncertainties of the 
non-perturbative coefficients $a_{\cal F}$ ($^{+0.0028}_{-0.0020}$), all
combined in quadrature. This result is in good agreement with the
world average value~\cite{bib-world-alphas-sb} of 
$\as^{\mathrm{w.a.}}(\mz)=0.118\pm 0.006$.
Our value is also in agreement with the results of similar studies 
for different sets of observables by
the DELPHI Collaboration~\cite{bib-DELPHI-powcor} and by the
H1 Collaboration~\cite{bib-H1-powcor}.

\section{Summary and Conclusions }
\label{sec-conclusions}
Data recorded by the JADE experiment at centre-of-mass
energies around $22$, $35$, and $44$~GeV were analysed in terms
of event shape distributions and differential 2-jet rates.
For most of the observables 
no experimental results have previously been presented, 
because the total and wide jet broadening, $B_T$ and $B_W$, as
well as the Durham jet finding scheme were proposed only 
after the shutdown of the experiments at the PETRA accelerator.

The measured distributions were corrected for detector 
and initial state photon radiation effects using original 
Monte Carlo simulation data 
for $35$ and $44$~GeV. 
The simulated data are based
on the JETSET parton shower generator version~6.3. 
The same event 
generator was also employed to correct the data for hadronisation 
effects in order to determine the strong coupling constant \as.

Our measurements of \as\ are based on the most complete theoretical 
calculations available to date. For all observables 
theoretical calculations exist in \oaa\ %
and in the next-to-leading log approximation.
These two calculations were combined using the ln($R$)-matching scheme.

The final values of \as\ 
at the three different centre-of-mass energies are
\begin{eqnarray*}
\as(44\ {\mathrm{GeV}}) & = & 0.137 
                         \ ^{+0.010}
                           _{-0.007} \\
\as(35\ {\mathrm{GeV}}) & = & 0.143 
                         \ ^{+0.011}
                           _{-0.007} \\
\as(22\ {\mathrm{GeV}}) & = & 0.161 
                         \ ^{+0.016}
                           _{-0.011}\ ,
\end{eqnarray*}
where the errors are statistical, experimental systematics, Monte Carlo 
modelling and higher order QCD uncertainties added in quadrature. The 
dominant contributions to the total error came from the choice of the
renormalisation scale and from uncertainties due to quark mass
effects.

The \as\ result at $22$~GeV was obtained from the differential 2-jet 
rate only.
Note, however, that for $35$ and $44$~GeV the 
\as\ value obtained from the differential 2-jet rate 
has the smallest total error and 
is very close to 
the weighted average as can be inferred from Figure~\ref{fig-asresult-numbers}. 
We therefore consider the \as\ value 
obtained   
at $22$~GeV 
a good 
approximation of the projected result of a more    
comprehensive study at this energy.                

The fits for \as\ were also performed using the \oaa\ %
calculation alone. 
All results were found to be consistent with each other. 

These results agree well with those which are available from previous
measurements of \as\ in the PETRA and PEP energy range; see e.g.
\cite{bib-naroska,bib-sb-budapest} for reviews of that time.
Our results, however, include more detailed systematic studies, are
based on more observables and use more advanced theoretical calculations;
nevertheless they exhibit 
smaller total errors. 

Similarities between the main components of the JADE
detector~\cite{bib-JADEdet}
at PETRA and the OPAL detector~\cite{bib-OPALdetector} at LEP, as well as
between this analysis and studies performed by the OPAL
Collaboration~\cite{bib-eventshapes,bib-OPALresummed,bib-OPALNLLA} 
at $\sqrt{s} = 91.2$,
$133$, and $161$~GeV, suggest the energy dependence of 
\as\ in the centre-of-mass energy range of $\sqrt{s} = 22$-$161$~GeV
can be reliably tested, 
because the systematic uncertainties of these measurements are 
partly correlated. 

\begin{figure}[!htb]
\resizebox{\textwidth}{!}{\includegraphics{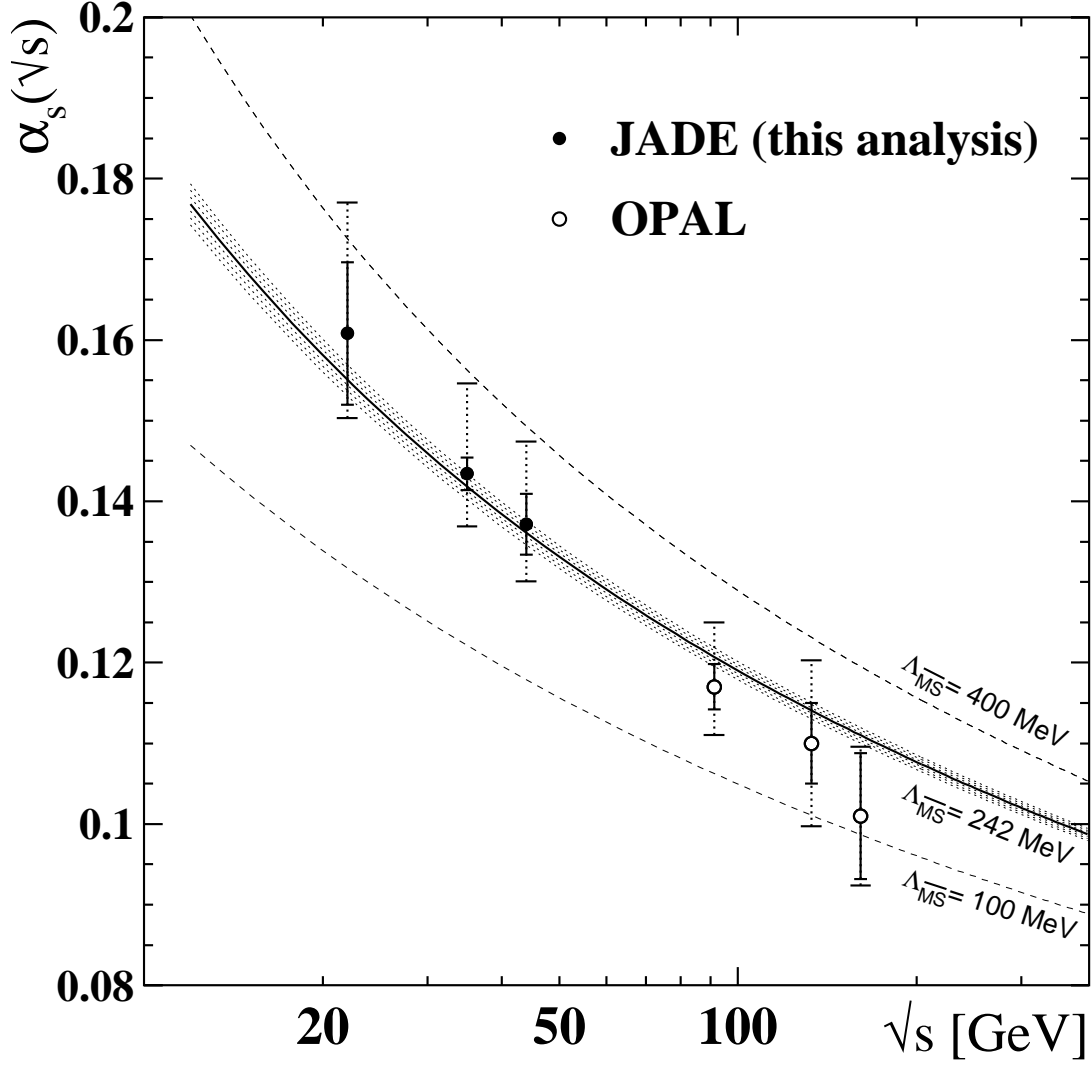}}
\caption{\label{fig-world-NLLA-as}
Values of \as\ from \oaa +NLLA fits, as a function of centre-of-mass 
energy. The solid error bars are the
statistical and experimental uncertainties added in quadrature, the
dotted error bars are the total errors. 
The results
from OPAL~\protect\cite{bib-eventshapes,bib-OPALNLLA}
for the same set of observables
are shown as representative for the LEP experiments because the
relevant detector subsystems of OPAL are similar to those of JADE.
The solid line and the shaded band represent the QCD prediction for 
$\asmz = 0.1207 \pm
0.0012$ corresponding to $\lmsbf = (242\pm 15)$~MeV
which was obtained from a \chisq\ fit to the data taking only
experimental errors into account.
For comparison the QCD predictions for $\lmsbf = 100$~MeV and
$400$~MeV are shown by the dashed curves.
}
\end{figure}

The \as\ results from OPAL and from this analysis are shown in 
Figure~\ref{fig-world-NLLA-as}. 
The result of 
a $\chisq$ fit of the \oaaa\ QCD prediction~\cite{bib-PDG} 
to the data is shown by the 
solid line. The fit resulted 
in $\asmz = 0.1207 \pm 0.0012$~\footnote{This value of \asmz\ corresponds 
to a QCD scale $\lmsbf = 242 \pm 15$~MeV for five active quark flavours.}
and $\chisqd = 4.9/5$,
taking into account only statistical and experimental uncertainties, which
are displayed in Fig.~\ref{fig-world-NLLA-as} as the solid, innermost error
bars.
The other systematic uncertainties, due to hadronisation and to unknown
higher order contributions, are assumed to be fully correlated at all
energies and thus are not considered in this test of the {\em energy
dependence} of \as.
A visible trend of the lower energy results all lying above and
the higher energy ones lying below the fitted QCD curve can be consistently
explained within the assigned experimental uncertainties which is indicated
by the value of $\chisqd = 1$.

A \chisq\ fit for the hypothesis of a constant value of \as\ gives 
$\as=0.1328 \pm 0.0014$ and $\chisqd = 101/5$,
which has a vanishing probability.
The energy dependence of \as\ is therefore significantly demonstrated by 
the
results from the combined JADE and OPAL data.

Evolving our \as\ measurements to $\sqrt{s} = \mz$ the results obtained at
$44$, $35$ and $22$~GeV transform to $0.122\,^{+0.008}_{-0.006}$, 
$0.122\,^{+0.008}_{-0.006}$, and $0.124\,^{+0.009}_{-0.007}$, respectively. 
The combination of these values gives $\asmz = 0.122\,^{+0.008}_{-0.005}$. 
This value is consistent with the direct measurement at $\sqrt{s} = \mz$ 
by the OPAL Collaboration of 
$\as ( \mz ) = 0.117\,^{+0.008}_{-0.006}$~\cite{bib-OPALresummed}, for 
the same subset of observables.

The energy dependence of the mean values of the distributions can be
directly compared with analytic QCD predictions plus power corrections
for hadronisation effects~\cite{bib-webber}.
Until recently, such studies were hardly possible since for most of the
observables no results were available at energies below the
$\znull$ mass scale. With the inclusion of the results presented in this
paper,  comprehensive fits of the analytic predictions to the data are now
possible. Our studies resulted in 
\begin{displaymath}
   \as(\mz) = 0.116\ ^{+0.006}
                     _{-0.005}
\end{displaymath}
which is
in good agreement with our results from the \oaa+NLLA fits, with measurements
at LEP~\cite{bib-DELPHI-powcor} and at HERA~\cite{bib-H1-powcor} and also
with the world average value. 

\par \par
In summary, new studies of hadronic final states of
$\epem$-annihilations in the PETRA energy range provided valuable 
information which was not available before. 
New results of \as, obtained in a similar manner as those from the
experiments at LEP, provide a significant test of the running of \as\
and thus of the non-abelian nature of QCD. 
Evolved to the $\znull$ mass scale, the results are in good agreement
with those obtained at LEP, and are of similar precision.
A direct comparison of the energy dependence of the mean values of the
measured distributions with analytic QCD calculations plus power
corrections provide alternative ways to test QCD, without
the need to rely on phenomenological hadronisation models.

Work has been started to further decrease the overall uncertainties of
the results presented in this paper, and to study more aspects of QCD
using the JADE data samples.
This can be achieved by the use of more recent event generators
and the JADE detector simulation software.
This will provide the possibility to study the data at the
lowest PETRA energies, around $\sqrt{s} = 14$ and 22~GeV, in more detail,
i.e. for energy scales at which the variation of \as\ is strongest.
In addition, the significance of results from data at PETRA energies
will increase from a better and more fundamental treatment of the
b-quark mass, in theory~\cite{bib-NLO-quark-mass} as well as in experiment.

\medskip
\bigskip\bigskip\bigskip
\appendix
\par
\section*{Acknowledgements}
\par
We are grateful to the members of the former JADE Collaboration 
for providing the possibility to further analyse their data. 
We thank the DESY computer centre for copying old IBM format tapes 
to modern data storage devices before the shutdown of the DESY-IBM. 
We are especially indepted to G.~Eckerlin, E.~Elsen and J.~Olsson for
their valuable help to recover the data files and for numerous
suggestions and comments on the analysis and on this manuscript.
We thank R.~Barlow for proof reading this manuscript.
We also acknowledge the 
effort of J.~von~Krogh, P.~Bock and many other colleagues to search for 
files and tapes containing JADE software, data and Monte Carlo simulation.
S.K. and O.B. are also grateful to S.~Catani and M.H.~Seymour for providing
the program EVENT2.

%
\clearpage
\section*{ Tables }

\begin{table}[!htb]
\begin{center}
%
%
\begin{tabular}{|c||r@{ $\pm$ }l@{ $\pm$ }l|}   
\hline
$1-T$           & 
      \multicolumn{3}{c|}{$1/\sigma \cdot {\mathrm{d}}\sigma/{\mathrm{d}}(1-T)$} \\
\hline\hline
$ 0.00 $-$0.02  $&$   0.989 $&$     0.061 $&$     0.192   $\\
$ 0.02 $-$0.04  $&$   8.73  $&$     0.26  $&$     0.89    $\\
$ 0.04 $-$0.06  $&$  12.85  $&$     0.35  $&$     0.92    $\\
$ 0.06 $-$0.08  $&$   8.58  $&$     0.28  $&$     0.84    $\\
$ 0.08 $-$0.10  $&$   4.97  $&$     0.20  $&$     0.18    $\\
$ 0.10 $-$0.12  $&$   3.84  $&$     0.18  $&$     0.33    $\\
$ 0.12 $-$0.14  $&$   2.54  $&$     0.14  $&$     0.17    $\\
$ 0.14 $-$0.16  $&$   1.88  $&$     0.12  $&$     0.23    $\\
$ 0.16 $-$0.18  $&$   1.47  $&$     0.10  $&$     0.18    $\\
$ 0.18 $-$0.20  $&$   1.141 $&$     0.091 $&$     0.104   $\\
$ 0.20 $-$0.23  $&$   0.808 $&$     0.062 $&$     0.139   $\\
$ 0.23 $-$0.26  $&$   0.486 $&$     0.047 $&$     0.066   $\\
$ 0.26 $-$0.30  $&$   0.326 $&$     0.035 $&$     0.047   $\\
$ 0.30 $-$0.35  $&$   0.239 $&$     0.027 $&$     0.036   $\\
$ 0.35 $-$0.40  $&$   0.047 $&$     0.012 $&$     0.019   $\\
$ 0.40 $-$0.50  $&$   0.001 $&$     0.001 $&$     0.002   $\\
\hline\hline
mean value       &$  0.0860 $&$     0.0008 $&$    0.0011 $\\
\hline
\end{tabular}
%
%
\begin{tabular}{|c||r@{ $\pm$ }l@{ $\pm$ }l|}   
\hline
$M_H/\sqrt{s}$           & 
      \multicolumn{3}{c|}{$1/\sigma \cdot {\mathrm{d}}
                            \sigma/{\mathrm{d}}(M_H/\sqrt{s})$} \\
\hline\hline
$ 0.00 $-$0.06  $&$   0.002 $&$     0.000 $&$     0.003   $\\
$ 0.06 $-$0.10  $&$   0.022 $&$     0.002 $&$     0.007   $\\
$ 0.10 $-$0.14  $&$   0.576 $&$     0.025 $&$     0.077   $\\
$ 0.14 $-$0.18  $&$   4.06  $&$     0.11  $&$     0.37    $\\
$ 0.18 $-$0.22  $&$   6.94  $&$     0.19  $&$     0.29    $\\
$ 0.22 $-$0.26  $&$   5.00  $&$     0.16  $&$     0.20    $\\
$ 0.26 $-$0.30  $&$   3.30  $&$     0.13  $&$     0.23    $\\
$ 0.30 $-$0.34  $&$   2.146 $&$     0.096 $&$     0.116   $\\
$ 0.34 $-$0.38  $&$   1.222 $&$     0.074 $&$     0.228   $\\
$ 0.38 $-$0.42  $&$   0.936 $&$     0.065 $&$     0.101   $\\
$ 0.42 $-$0.46  $&$   0.567 $&$     0.048 $&$     0.153   $\\
$ 0.46 $-$0.50  $&$   0.376 $&$     0.042 $&$     0.097   $\\
$ 0.50 $-$0.55  $&$   0.118 $&$     0.019 $&$     0.020   $\\
$ 0.55 $-$0.60  $&$   0.015 $&$     0.005 $&$     0.005   $\\
\hline\hline
mean value       &$  0.2470 $&$     0.0010 $&$    0.0008 $\\
\hline
\end{tabular}

\vspace*{5mm}
%
%
\begin{tabular}{|c||r@{ $\pm$ }l@{ $\pm$ }l|}   
\hline
$B_T$           & 
      \multicolumn{3}{c|}{$1/\sigma \cdot {\mathrm{d}}\sigma/{\mathrm{d}}B_T$} \\
\hline\hline
$ 0.00 $-$0.03  $&$   0.012 $&$     0.002 $&$     0.010   $\\
$ 0.03 $-$0.06  $&$   0.637 $&$     0.041 $&$     0.137   $\\
$ 0.06 $-$0.08  $&$   5.76  $&$     0.22  $&$     1.00    $\\
$ 0.08 $-$0.10  $&$   9.39  $&$     0.30  $&$     0.31    $\\
$ 0.10 $-$0.12  $&$   8.44  $&$     0.28  $&$     0.94    $\\
$ 0.12 $-$0.14  $&$   7.17  $&$     0.26  $&$     0.39    $\\
$ 0.14 $-$0.16  $&$   5.31  $&$     0.21  $&$     0.27    $\\
$ 0.16 $-$0.18  $&$   3.55  $&$     0.16  $&$     0.18    $\\
$ 0.18 $-$0.20  $&$   2.84  $&$     0.15  $&$     0.16    $\\
$ 0.20 $-$0.22  $&$   2.01  $&$     0.12  $&$     0.12    $\\
$ 0.22 $-$0.24  $&$   1.65  $&$     0.11  $&$     0.15    $\\
$ 0.24 $-$0.27  $&$   0.998 $&$     0.065 $&$     0.139   $\\
$ 0.27 $-$0.30  $&$   0.563 $&$     0.049 $&$     0.071   $\\
$ 0.30 $-$0.35  $&$   0.253 $&$     0.024 $&$     0.043   $\\
$ 0.35 $-$0.40  $&$   0.018 $&$     0.006 $&$     0.013   $\\
\hline\hline
mean value       &$  0.1344 $&$     0.0007 $&$    0.0015 $\\
\hline
\end{tabular}
%
%
\begin{tabular}{|c||r@{ $\pm$ }l@{ $\pm$ }l|}   
\hline
$B_W$           & 
      \multicolumn{3}{c|}{$1/\sigma \cdot {\mathrm{d}}\sigma/{\mathrm{d}}B_W$} \\
\hline\hline
$ 0.00 $-$0.02  $&$   0.034 $&$     0.006 $&$     0.032   $\\
$ 0.02 $-$0.04  $&$   3.06  $&$     0.14  $&$     0.44    $\\
$ 0.04 $-$0.06  $&$  13.94  $&$     0.37  $&$     0.78    $\\
$ 0.06 $-$0.08  $&$  12.56  $&$     0.34  $&$     0.37    $\\
$ 0.08 $-$0.10  $&$   6.88  $&$     0.23  $&$     0.46    $\\
$ 0.10 $-$0.12  $&$   4.62  $&$     0.18  $&$     0.40    $\\
$ 0.12 $-$0.14  $&$   3.27  $&$     0.15  $&$     0.30    $\\
$ 0.14 $-$0.16  $&$   1.95  $&$     0.12  $&$     0.26    $\\
$ 0.16 $-$0.18  $&$   1.367 $&$     0.096 $&$     0.194   $\\
$ 0.18 $-$0.20  $&$   1.038 $&$     0.084 $&$     0.108   $\\
$ 0.20 $-$0.23  $&$   0.605 $&$     0.052 $&$     0.032   $\\
$ 0.23 $-$0.26  $&$   0.261 $&$     0.037 $&$     0.055   $\\
$ 0.26 $-$0.30  $&$   0.020 $&$     0.005 $&$     0.007   $\\
\hline\hline
mean value       &$  0.0848 $&$     0.0006 $&$    0.0004 $\\
\hline
\end{tabular}
\end{center}
\caption[dummy]{\label{tab-eventshapes-44GeV}
Event shape data at $\protect\sqrt{s}=44$~GeV for the observables 
described in the text. The values were corrected for detector 
and for initial state radiation effects. 
The first errors denote the statistical and the second 
the experimental systematic uncertainties.
}
\end{table}

\newpage

\begin{table}[!htb]
\vspace*{-7mm}
\begin{center}
%
%
\begin{tabular}{|c||r@{ $\pm$ }l@{ $\pm$ }l|}   
\hline
$1-T$           & 
      \multicolumn{3}{c|}{$1/\sigma \cdot {\mathrm{d}}\sigma/{\mathrm{d}}(1-T)$} \\
\hline\hline
$ 0.00 $-$0.02  $&$  0.638 $&$  0.028 $&$  0.181 $\\
$ 0.02 $-$0.04  $&$  6.43  $&$  0.12  $&$  0.23  $\\
$ 0.04 $-$0.06  $&$ 11.00  $&$  0.17  $&$  0.25  $\\
$ 0.06 $-$0.08  $&$  9.47  $&$  0.16  $&$  0.43  $\\
$ 0.08 $-$0.10  $&$  6.43  $&$  0.13  $&$  0.21  $\\
$ 0.10 $-$0.12  $&$  4.049 $&$  0.095 $&$  0.173 $\\
$ 0.12 $-$0.14  $&$  3.033 $&$  0.084 $&$  0.239 $\\
$ 0.14 $-$0.16  $&$  1.962 $&$  0.065 $&$  0.151 $\\
$ 0.16 $-$0.18  $&$  1.704 $&$  0.062 $&$  0.179 $\\
$ 0.18 $-$0.20  $&$  1.209 $&$  0.050 $&$  0.117 $\\
$ 0.20 $-$0.23  $&$  0.922 $&$  0.036 $&$  0.066 $\\
$ 0.23 $-$0.26  $&$  0.754 $&$  0.035 $&$  0.095 $\\
$ 0.26 $-$0.30  $&$  0.453 $&$  0.022 $&$  0.049 $\\
$ 0.30 $-$0.35  $&$  0.186 $&$  0.012 $&$  0.061 $\\
$ 0.35 $-$0.40  $&$  0.070 $&$  0.008 $&$  0.010 $\\
$ 0.40 $-$0.50  $&$  0.001 $&$  0.001 $&$  0.001 $\\
\hline\hline
mean value       &$  0.0938 $&$ 0.0004 $&$ 0.0015 $\\
\hline
\end{tabular}
%
%
\begin{tabular}{|c||r@{ $\pm$ }l@{ $\pm$ }l|}   
\hline
$M_H/\sqrt{s}$           & 
      \multicolumn{3}{c|}{$1/\sigma \cdot {\mathrm{d}}
                            \sigma/{\mathrm{d}}(M_H/\sqrt{s})$} \\
\hline\hline
$ 0.00 $-$0.06  $&$  0.002 $&$  0.000 $&$  0.004 $\\
$ 0.06 $-$0.10  $&$  0.017 $&$  0.001 $&$  0.009 $\\
$ 0.10 $-$0.14  $&$  0.288 $&$  0.009 $&$  0.041 $\\
$ 0.14 $-$0.18  $&$  2.566 $&$  0.043 $&$  0.095 $\\
$ 0.18 $-$0.22  $&$  6.278 $&$  0.090 $&$  0.361 $\\
$ 0.22 $-$0.26  $&$  5.463 $&$  0.088 $&$  0.319 $\\
$ 0.26 $-$0.30  $&$  3.823 $&$  0.073 $&$  0.173 $\\
$ 0.30 $-$0.34  $&$  2.390 $&$  0.056 $&$  0.084 $\\
$ 0.34 $-$0.38  $&$  1.643 $&$  0.047 $&$  0.088 $\\
$ 0.38 $-$0.42  $&$  1.008 $&$  0.037 $&$  0.039 $\\
$ 0.42 $-$0.46  $&$  0.626 $&$  0.030 $&$  0.040 $\\
$ 0.46 $-$0.50  $&$  0.427 $&$  0.025 $&$  0.067 $\\
$ 0.50 $-$0.55  $&$  0.153 $&$  0.013 $&$  0.034 $\\
$ 0.55 $-$0.60  $&$  0.020 $&$  0.004 $&$  0.005 $\\
\hline\hline
mean value       &$  0.2601 $&$ 0.0006 $&$ 0.0016 $\\
\hline
\end{tabular}

\vspace*{5mm}
%
%
\begin{tabular}{|c||r@{ $\pm$ }l@{ $\pm$ }l|}   
\hline
$B_T$           & 
      \multicolumn{3}{c|}{$1/\sigma \cdot {\mathrm{d}}\sigma/{\mathrm{d}}B_T$} \\
\hline\hline
$ 0.00 $-$0.03  $&$  0.018 $&$  0.003 $&$  0.028 $\\
$ 0.03 $-$0.06  $&$  0.381 $&$  0.019 $&$  0.103 $\\
$ 0.06 $-$0.08  $&$  3.371 $&$  0.089 $&$  0.279 $\\
$ 0.08 $-$0.10  $&$  8.02  $&$  0.15  $&$  0.82  $\\
$ 0.10 $-$0.12  $&$  8.50  $&$  0.15  $&$  0.16  $\\
$ 0.12 $-$0.14  $&$  7.38  $&$  0.14  $&$  0.31  $\\
$ 0.14 $-$0.16  $&$  6.27  $&$  0.12  $&$  0.33  $\\
$ 0.16 $-$0.18  $&$  4.52  $&$  0.10  $&$  0.13  $\\
$ 0.18 $-$0.20  $&$  3.267 $&$  0.084 $&$  0.149 $\\
$ 0.20 $-$0.22  $&$  2.429 $&$  0.072 $&$  0.202 $\\
$ 0.22 $-$0.24  $&$  1.748 $&$  0.060 $&$  0.149 $\\
$ 0.24 $-$0.27  $&$  1.277 $&$  0.042 $&$  0.090 $\\
$ 0.27 $-$0.30  $&$  0.811 $&$  0.033 $&$  0.061 $\\
$ 0.30 $-$0.35  $&$  0.262 $&$  0.013 $&$  0.047 $\\
$ 0.35 $-$0.40  $&$  0.020 $&$  0.003 $&$  0.006 $\\
\hline\hline
mean value       &$  0.1439 $&$ 0.0004 $&$ 0.0012 $\\
\hline
\end{tabular}
%
%
\begin{tabular}{|c||r@{ $\pm$ }l@{ $\pm$ }l|}   
\hline
$B_W$           & 
      \multicolumn{3}{c|}{$1/\sigma \cdot {\mathrm{d}}\sigma/{\mathrm{d}}B_W$} \\
\hline\hline
$ 0.00 $-$0.02  $&$  0.041 $&$  0.006 $&$  0.046 $\\
$ 0.02 $-$0.04  $&$  1.628 $&$  0.059 $&$  0.276 $\\
$ 0.04 $-$0.06  $&$ 11.79  $&$  0.18  $&$  0.46  $\\
$ 0.06 $-$0.08  $&$ 12.18  $&$  0.17  $&$  0.31  $\\
$ 0.08 $-$0.10  $&$  8.87  $&$  0.14  $&$  0.42  $\\
$ 0.10 $-$0.12  $&$  5.11  $&$  0.10  $&$  0.20  $\\
$ 0.12 $-$0.14  $&$  3.63  $&$  0.088 $&$  0.255 $\\
$ 0.14 $-$0.16  $&$  2.479 $&$  0.074 $&$  0.184 $\\
$ 0.16 $-$0.18  $&$  1.631 $&$  0.059 $&$  0.286 $\\
$ 0.18 $-$0.20  $&$  1.092 $&$  0.049 $&$  0.052 $\\
$ 0.20 $-$0.23  $&$  0.739 $&$  0.035 $&$  0.133 $\\
$ 0.23 $-$0.26  $&$  0.276 $&$  0.021 $&$  0.051 $\\
$ 0.26 $-$0.30  $&$  0.020 $&$  0.004 $&$  0.004 $\\
\hline\hline
mean value       &$  0.0906 $&$ 0.0003 $&$ 0.0009 $\\
\hline
\end{tabular}

\end{center}
\caption[dummy]{\label{tab-eventshapes-35GeV}
Event shape data as for Table~\protect\ref{tab-eventshapes-44GeV}
but measured at $\protect\sqrt{s}=35$~GeV.
}
\end{table}

\newpage

\begin{table}[!htb]
\vspace*{-7mm}
\begin{center}
%
%
%
%
\begin{tabular}{|r||r@{ $\pm$ }l@{ $\pm$ }l|}   
\hline
$44$~GeV & 
      \multicolumn{3}{c|}
       {$1/\sigma \cdot {\mathrm{d}}\sigma/{\mathrm{d}}y_{\mathrm{23}}$} \\
\hline\hline
$ 0.000$-$0.001 $&$  15.2   $&$     1.1   $&$     3.7     $\\
$ 0.001$-$0.002 $&$  67.4   $&$     3.1   $&$     9.0     $\\
$ 0.002$-$0.005 $&$  86.7   $&$     2.3   $&$     4.1     $\\
$ 0.005$-$0.010 $&$  49.4   $&$     1.4   $&$     5.9     $\\
$ 0.010$-$0.020 $&$  16.00  $&$     0.51  $&$     0.98    $\\
$ 0.020$-$0.030 $&$   7.26  $&$     0.32  $&$     0.97    $\\
$ 0.030$-$0.040 $&$   4.04  $&$     0.23  $&$     0.69    $\\
$ 0.040$-$0.050 $&$   2.35  $&$     0.18  $&$     0.48    $\\
$ 0.050$-$0.060 $&$   1.94  $&$     0.16  $&$     0.61    $\\
$ 0.060$-$0.080 $&$   1.302 $&$     0.094 $&$     0.078   $\\
$ 0.080$-$0.100 $&$   0.897 $&$     0.079 $&$     0.081   $\\
$ 0.100$-$0.130 $&$   0.601 $&$     0.053 $&$     0.054   $\\
$ 0.130$-$0.160 $&$   0.484 $&$     0.053 $&$     0.171   $\\
$ 0.160$-$0.200 $&$   0.276 $&$     0.035 $&$     0.034   $\\
$ 0.200$-$0.250 $&$   0.142 $&$     0.022 $&$     0.076   $\\
$ 0.250$-$0.400 $&$   0.024 $&$     0.006 $&$     0.023   $\\
\hline\hline
mean value       &$  0.0229 $&$     0.0005 $&$    0.0019 $\\
\hline
\end{tabular}
%
%
\begin{tabular}{|r||r@{ $\pm$ }l@{ $\pm$ }l|}   
\hline
$35$~GeV & 
      \multicolumn{3}{c|}
       {$1/\sigma \cdot {\mathrm{d}}\sigma/{\mathrm{d}}y_{\mathrm{23}}$} \\
\hline\hline
$ 0.000$-$0.001 $&$  8.45  $&$  0.47  $&$  3.23  $\\
$ 0.001$-$0.002 $&$ 46.5   $&$  1.4   $&$  3.6   $\\
$ 0.002$-$0.005 $&$ 73.5   $&$  1.1   $&$  2.7   $\\
$ 0.005$-$0.010 $&$ 45.74  $&$  0.67  $&$  2.13  $\\
$ 0.010$-$0.020 $&$ 19.32  $&$  0.30  $&$  0.66  $\\
$ 0.020$-$0.030 $&$  8.09  $&$  0.19  $&$  0.16  $\\
$ 0.030$-$0.040 $&$  4.75  $&$  0.15  $&$  0.30  $\\
$ 0.040$-$0.050 $&$  3.09  $&$  0.12  $&$  0.19  $\\
$ 0.050$-$0.060 $&$  2.39  $&$  0.11  $&$  0.14  $\\
$ 0.060$-$0.080 $&$  1.568 $&$  0.060 $&$  0.098 $\\
$ 0.080$-$0.100 $&$  1.096 $&$  0.052 $&$  0.109 $\\
$ 0.100$-$0.130 $&$  0.816 $&$  0.038 $&$  0.141 $\\
$ 0.130$-$0.160 $&$  0.449 $&$  0.027 $&$  0.048 $\\
$ 0.160$-$0.200 $&$  0.365 $&$  0.023 $&$  0.083 $\\
$ 0.200$-$0.250 $&$  0.170 $&$  0.014 $&$  0.039 $\\
$ 0.250$-$0.400 $&$  0.026 $&$  0.004 $&$  0.005 $\\
\hline\hline
mean value       &$  0.0266 $&$ 0.0003 $&$ 0.0015 $\\
\hline
\end{tabular}

\vspace*{5mm}
%
%
\begin{tabular}{|c||r@{ $\pm$ }l@{ $\pm$ }l|}   
\hline
$22$~GeV & 
      \multicolumn{3}{c|}
       {$1/\sigma \cdot {\mathrm{d}}\sigma/{\mathrm{d}}y_{\mathrm{23}}$} \\
\hline\hline
$ 0.000$-$0.001 $&$   1.91  $&$     0.87  $&$     2.84    $\\
$ 0.001$-$0.002 $&$  18.5   $&$     3.6   $&$     7.7     $\\
$ 0.002$-$0.005 $&$  37.4   $&$     3.2   $&$    10.7     $\\
$ 0.005$-$0.010 $&$  41.4   $&$     2.5   $&$     1.6     $\\
$ 0.010$-$0.020 $&$  24.4   $&$     1.4   $&$     3.5     $\\
$ 0.020$-$0.030 $&$  11.64  $&$     0.90  $&$     0.12    $\\
$ 0.030$-$0.040 $&$   7.61  $&$     0.77  $&$     1.57    $\\
$ 0.040$-$0.050 $&$   3.76  $&$     0.51  $&$     1.27    $\\
$ 0.050$-$0.060 $&$   3.206 $&$     0.490 $&$     0.086   $\\
$ 0.060$-$0.080 $&$   2.38  $&$     0.30  $&$     0.38    $\\
$ 0.080$-$0.100 $&$   1.17  $&$     0.21  $&$     0.81    $\\
$ 0.100$-$0.130 $&$   0.85  $&$     0.14  $&$     0.19    $\\
$ 0.130$-$0.160 $&$   0.49  $&$     0.11  $&$     0.11    $\\
$ 0.160$-$0.200 $&$   0.278 $&$     0.077 $&$     0.070   $\\
$ 0.200$-$0.250 $&$   0.205 $&$     0.062 $&$     0.066   $\\
$ 0.250$-$0.400 $&$   0.022 $&$     0.016 $&$     0.042   $\\
\hline\hline
mean value       &$  0.0311 $&$     0.0011 $&$    0.0018  $\\
\hline
\end{tabular}
\end{center}
\caption[dummy]{\label{tab-D2shapes}
Differential 2-jet rate $D_2$ at $\protect\sqrt{s}=44$~GeV,
at $35$~GeV and at 22~GeV. The values were corrected for detector 
and for initial state radiation effects. 
The first errors denote the statistical and the second the 
experimental systematic uncertainties.
}
\end{table}

\newpage

\begin{table}[!htb]
\vspace*{-7mm}
\begin{center}
\begin{tabular}{|r||r|r|r|r|r|}   \hline
 &\multicolumn{1}{c|}{\thr} &\multicolumn{1}{c|}{\mh}
 &\multicolumn{1}{c|}{\bt}  &\multicolumn{1}{c|}{\bw}
 &\multicolumn{1}{c|}{\dtwo}     \\
\hline\hline
\as($44$~GeV)     &\bf 0.1457  &\bf 0.1423  &\bf 0.1417  &\bf 0.1278  &\bf 0.1344   \\
\hline\hline
fit range         &$0.08$-$0.3$ 
                               & $0.22$-$0.46$ 
                                            & $0.080$-$0.27$ 
                                                         & $0.06$-$0.16$
                                                                  & $0.005$-$0.200$ \\
\hline\hline
\chisqd           & $9.4/8$    & $31.0/5$   & $27.1/8$   & $48.8/4$   & $37.8/10$    \\
\hline\hline
Statistical error &$\pm 0.0017$&$\pm 0.0017$&$\pm 0.0014$&$\pm 0.0016$&$\pm 0.0019$
\\ \hline\hline
tracks only       &  $-0.0011$ &  $-0.0012$ &  $-0.0027$ &  $-0.0004$ &  $-0.0022$  \\
\hline
clusters only     &  $+0.0027$ &  $+0.0036$ &  $+0.0009$ &  $+0.0015$ &  $+0.0016$  \\
\hline
$\cos\theta_T $ 
                  &$\pm 0.0001$&$\pm 0.0009$&$\pm 0.0005$&$\pm 0.0008$&$\pm 0.0006$  \\
\hline
$p_{\mathrm{miss}} $ 
                  &$\pm 0.0006$&$\pm 0.0008$&$\pm 0.0001$&$\pm 0.0002$&$\pm 0.0005$  \\
\hline
$p_{\mathrm{bal}} $ 
                  &$\pm 0.0004$&$\pm 0.0004$&$\pm 0.0003$&$\pm 0.0004$&$\pm 0.0002$  \\
\hline
$N_{\mathrm{ch}} $ 
                  &  $+ 0.0002$&  $+ 0.0004$&  $+ 0.0003$&  $+ 0.0003$&  $+ 0.0006$  \\
\hline
$E_{\mathrm{vis}} $ 
                  &$\pm 0.0003$&$\pm 0.0001$&$\pm 0.0003$&$\pm 0.0002$&$\pm 0.0002$  \\
\hline
fit range 
                  &$\pm 0.0015$&$\pm 0.0032$&$\pm 0.0023$&$\pm 0.0048$&$\pm 0.0032$  \\
\hline\hline
Experimental syst.&$\pm 0.0032$&$\pm 0.0050$&$\pm 0.0037$&$\pm 0.0051$&$\pm 0.0040$ \\
\hline\hline
$a-0.225$         &  $+0.0019$ &  $+0.0018$ &  $+0.0017$ &  $+0.0010$ & $< 0.0001$  \\
\hline
$a+0.225$         &  $-0.0016$ &  $-0.0021$ &  $-0.0017$ &  $-0.0009$ &  $-0.0005$  \\
\hline
$\sigma_q-30$~MeV &  $+0.0010$ &  $+0.0001$ &  $+0.0009$ &  $+0.0007$ &  $+0.0004$  \\
\hline
$\sigma_q+30$~MeV &  $-0.0009$ &  $-0.0003$ &  $-0.0011$ &  $-0.0006$ &  $-0.0009$  \\
\hline
LUND symmetric    &  $+0.0012$ &  $+0.0009$ &  $+0.0017$ &  $+0.0015$ &  $+0.0005$  \\
\hline
$Q_0+500$~MeV     &  $-0.0008$ &  $+0.0011$ &  $-0.0007$ &  $+0.0012$ &  $+0.0026$  \\
\hline
$Q_0-500$~MeV     &  $+0.0003$ &  $-0.0007$ &  $+0.0002$ &  $-0.0002$ &  $-0.0015$  \\
\hline
$\Lambda-50$~MeV  &  $-0.0005$ &  $+0.0001$ &  $-0.0013$ &  $+0.0001$ &  $+0.0003$  \\
\hline
$\Lambda+50$~MeV  &  $+0.0008$ &  $-0.0005$ &  $+0.0008$ & $< 0.0001$ &  $-0.0011$  \\
\hline
udsc only         &  $+0.0040$ &  $+0.0007$ &  $+0.0064$ &  $+0.0047$ &  $+0.0049$  \\
\hline
MC statistics     &$\pm 0.0011$&$\pm 0.0011$&$\pm 0.0009$&$\pm 0.0011$&$\pm 0.0012$ \\
\hline\hline
\raisebox{2mm}{MC modelling}       
                  &$\stackrel{\textstyle +0.0050}{-0.0030}$
                        &$\stackrel{\textstyle +0.0033}{-0.0032}$
                              &$\stackrel{\textstyle +0.0072}{-0.0032}$
                                    &$\stackrel{\textstyle +0.0054}{-0.0027}$
                                         &$\stackrel{\textstyle +0.0067}{-0.0045}$ \\
\hline\hline
$\xmu=0.5$        &  $-0.0089$ &  $-0.0067$ &  $-0.0100$ &  $-0.0065$ &  $-0.0007$  \\
\hline
$\xmu=2.0$        &  $+0.0115$ &  $+0.0092$ &  $+0.0125$ &  $+0.0082$ &  $+0.0045$  \\
\hline
\hline\hline
\raisebox{2mm}{Total error}       
                  &$\stackrel{\textstyle +0.0131}{-0.0101}$
                        &$\stackrel{\textstyle +0.0111}{-0.0091}$
                              &$\stackrel{\textstyle +0.0150}{-0.0112}$
                                    &$\stackrel{\textstyle +0.0112}{-0.0088}$
                                          &$\stackrel{\textstyle +0.0091}{-0.0063}$ \\
\hline
\end{tabular}
\end{center}
\caption[dummy]{\label{tab-asresult-44GeV}
Values of \as($44$~GeV) derived using the \oaa+NLLA QCD calculations
with $\xmu=1$ and the $\ln(R)$-matching scheme, fit ranges and \chisqd\ %
values for each of the five event shape observables. In addition, the
statistical and systematic uncertainties are given. Where a signed value is
quoted, this indicates the direction in which \as($44$~GeV) changed with
respect to the standard analysis. The scale uncertainty and quark mass
effects are treated as asymmetric uncertainties of \as. 
}
\end{table}

\newpage

\begin{table}[!htb]
\vspace*{-7mm}
\begin{center}
\begin{tabular}{|r||r|r|r|r|r|}   \hline
 &\multicolumn{1}{c|}{\thr} &\multicolumn{1}{c|}{\mh}
 &\multicolumn{1}{c|}{\bt}  &\multicolumn{1}{c|}{\bw}
 &\multicolumn{1}{c|}{\dtwo}     \\
\hline\hline
\as($35$~GeV)     &\bf 0.1510  &\bf 0.1445  &\bf 0.1448  &\bf 0.1326  &\bf 0.1448   \\
\hline\hline
fit range         &$0.08$-$0.3$ 
                               & $0.22$-$0.46$ 
                                            & $0.080$-$0.27$ 
                                                         & $0.06$-$0.16$
                                                                  & $0.020$-$0.200$ \\
\hline\hline
\chisqd           & $25.2/8$   & $32.8/5$   & $23.7/8$   & $23.0/4$   & $20.3/8$    \\
\hline\hline
Statistical error &$\pm 0.0009$&$\pm 0.0009$&$\pm 0.0007$&$\pm 0.0009$&$\pm 0.0014$
\\ \hline\hline
tracks only       &  $-0.0016$ & $< 0.0001$ &  $-0.0010$ &  $-0.0006$ &  $-0.0019$  \\
\hline
clusters only     &  $+0.0012$ &  $+0.0015$ &  $-0.0009$ &  $-0.0018$ &  $-0.0006$  \\
\hline
$\cos\theta_T $ 
                  &$\pm 0.0004$&$\pm 0.0005$&$\pm 0.0001$&$\pm 0.0002$&$\pm 0.0012$  \\
\hline
$p_{\mathrm{miss}} $ 
                  &$\pm 0.0001$&$\pm 0.0001$&$\pm 0.0001$&$\pm 0.0003$&$\pm 0.0004$  \\
\hline
$p_{\mathrm{bal}} $ 
                  &$\pm 0.0006$&$\pm 0.0001$&$\pm 0.0002$&$\pm 0.0006$&$\pm 0.0004$  \\
\hline
$N_{\mathrm{ch}} $ 
                  &  $+ 0.0006$&  $+ 0.0005$&  $+ 0.0005$&  $+ 0.0006$&  $+ 0.0005$  \\
\hline
$E_{\mathrm{vis}} $ 
                  &$\pm 0.0001$&$\pm 0.0001$&$\pm 0.0001$&$\pm 0.0001$&$\pm 0.0002$  \\
\hline
fit range 
                  &$\pm 0.0009$&$\pm 0.0017$&$\pm 0.0008$&$\pm 0.0016$&$\pm 0.0017$  \\
\hline\hline
Experimental syst.&$\pm 0.0021$&$\pm 0.0024$&$\pm 0.0014$&$\pm 0.0026$&$\pm 0.0030$ \\
\hline\hline
$a-0.225$         &  $+0.0028$ &  $+0.0035$ &  $+0.0023$ &  $+0.0018$ &  $-0.0002$  \\
\hline
$a+0.225$         &  $-0.0027$ &  $-0.0033$ &  $-0.0021$ &  $-0.0020$ &  $+0.0002$  \\
\hline
$\sigma_q-30$~MeV &  $+0.0018$ &  $+0.0008$ &  $+0.0013$ &  $+0.0015$ &  $+0.0008$  \\
\hline
$\sigma_q+30$~MeV &  $-0.0015$ &  $-0.0006$ &  $-0.0012$ &  $-0.0013$ &  $-0.0004$  \\
\hline
LUND symmetric    &  $+0.0040$ &  $+0.0034$ &  $+0.0027$ &  $+0.0029$ &  $+0.0009$  \\
\hline
$Q_0+500$~MeV     &  $-0.0006$ &  $+0.0015$ &  $-0.0014$ &  $+0.0014$ &  $+0.0024$  \\
\hline
$Q_0-500$~MeV     &  $+0.0001$ &  $-0.0006$ &  $+0.0006$ &  $-0.0008$ &  $-0.0004$  \\
\hline
$\Lambda-50$~MeV  &  $-0.0008$ &  $-0.0006$ &  $-0.0021$ &  $-0.0003$ &  $+0.0009$  \\
\hline
$\Lambda+50$~MeV  &  $+0.0011$ &  $+0.0007$ &  $+0.0018$ &  $+0.0003$ &  $-0.0008$  \\
\hline
udsc only         &  $+0.0074$ &  $+0.0025$ &  $+0.0086$ &  $+0.0077$ &  $+0.0055$  \\
\hline
MC statistics     &$\pm 0.0008$&$\pm 0.0008$&$\pm 0.0007$&$\pm 0.0008$&$\pm 0.0013$ \\
\hline\hline
\raisebox{2mm}{MC modelling}       
                  &$\stackrel{\textstyle +0.0092}{-0.0054}$
                        &$\stackrel{\textstyle +0.0060}{-0.0055}$
                              &$\stackrel{\textstyle +0.0099}{-0.0048}$
                                    &$\stackrel{\textstyle +0.0089}{-0.0045}$
                                         &$\stackrel{\textstyle +0.0065}{-0.0034}$ \\
\hline\hline
$\xmu=0.5$        &  $-0.0100$ &  $-0.0077$ &  $-0.0107$ &  $-0.0078$ &  $-0.0008$  \\
\hline
$\xmu=2.0$        &  $+0.0129$ &  $+0.0103$ &  $+0.0134$ &  $+0.0097$ &  $+0.0055$  \\
\hline
\hline\hline
\raisebox{2mm}{Total error}       
                  &$\stackrel{\textstyle +0.0160}{-0.0116}$
                        &$\stackrel{\textstyle +0.0122}{-0.0098}$
                              &$\stackrel{\textstyle +0.0167}{-0.0118}$
                                    &$\stackrel{\textstyle +0.0134}{-0.0094}$
                                          &$\stackrel{\textstyle +0.0091}{-0.0048}$ \\
\hline
\end{tabular}
\end{center}
\caption[dummy]{\label{tab-asresult-35GeV}
Values of \as($35$~GeV) derived as in Table~\protect\ref{tab-asresult-44GeV}
but at $35$~GeV.
}
\end{table}

\begin{table}[!htb]
\vspace*{-7mm}
\begin{center}
\begin{tabular}{|r||r|}   \hline
 &\multicolumn{1}{c|}{\dtwo}     \\
\hline\hline
\as($22$~GeV)     &\bf 0.1607   \\
\hline\hline
fit range         & $0.060$-$0.200$ \\
\hline\hline
\chisqd           & $1.7/4$     \\
\hline\hline
Statistical error &$\pm 0.0083$
\\ \hline\hline
tracks only       &  $+0.0023$  \\
\hline
clusters only     &  $-0.0030$  \\
\hline\hline
Experimental syst.&$\pm 0.0030$ \\
\hline\hline
$a-0.225$         &  $-0.0015$  \\
\hline
$a+0.225$         &  $+0.0010$  \\
\hline
$\sigma_q-30$~MeV &  $+0.0010$  \\
\hline
$\sigma_q+30$~MeV &  $-0.0004$  \\
\hline
LUND symmetric    &  $+0.0034$  \\
\hline
$Q_0+500$~MeV     &  $+0.0031$  \\
\hline
$Q_0-500$~MeV     &  $+0.0001$  \\
\hline
$\Lambda-50$~MeV  &  $+0.0011$  \\
\hline
$\Lambda+50$~MeV  &  $-0.0009$  \\
\hline
udsc only         &  $+0.0105$  \\
\hline
MC statistics     &$\pm 0.0025$  \\
\hline\hline
\raisebox{2mm}{MC modelling}       
                  &$\stackrel{\textstyle +0.0119}{-0.0056}$ \\ 
\hline\hline
$\xmu=0.5$        & $< 0.0001$  \\
\hline
$\xmu=2.0$        &  $+0.0066$  \\
\hline
\hline\hline
\raisebox{2mm}{Total error}       
                  &$\stackrel{\textstyle +0.0162}{-0.0105}$ \\
\hline
\end{tabular}
\end{center}
\caption[dummy]{\label{tab-asresult-22GeV}
Value of \as($22$~GeV) derived as in Table~\protect\ref{tab-asresult-44GeV}
but only for the differential 2-jet rate $D_2$ at $22$~GeV.
}
\end{table}
\newpage

\begin{table}[!htb]
\begin{center}
\begin{tabular}{|r||r|r|r|r|r||r|}   \hline
  \multicolumn{1}{|c||}{$44$~GeV}
 &\multicolumn{1}{c|}{\thr} &\multicolumn{1}{c|}{\mh}
 &\multicolumn{1}{c|}{\bt}  &\multicolumn{1}{c|}{\bw}
 &\multicolumn{1}{c|}{\dtwo}&\multicolumn{1}{|c|}{averaged}     \\
\hline\hline
\as($44$~GeV)  
     &\bf 0.1510  &\bf 0.1532  &\bf 0.1681  &\bf 0.1406  &\bf 0.1302  &\bf 0.1442 \\
\hline\hline
fit range         
     &$0.12$-$0.35$ 
                  & $0.26$-$0.50$ 
                               & $0.16$-$0.35$ 
                                            & $0.10$-$0.20$
                                                         & $0.01$-$0.20$ 
                                                                      &             \\
\hline\hline
\chisqd\ ($\xmu=1$)
     &$   3.0   $ &$    2.3  $ &$    3.7  $ &$    2.2  $ &$    1.1  $ &            \\
\hline\hline
\xmu\ fitted
     &$    0.056$ &$    0.132$ &$    0.600$ &$    0.070$ &$    0.080$ &            \\
\hline
\chisqd\ (\xmu\ free)
     &$   2.0   $ &$    2.0  $ &$    3.6  $ &$    2.0  $ &$    1.8  $ &            \\
\hline\hline
Statistical error 
     &$\pm 0.0028$&$\pm 0.0027$&$\pm 0.0025$&$\pm 0.0026$&$\pm 0.0021$&$\pm 0.0025$ \\ 
\hline
Experimental syst.
     &$\pm 0.0038$&$\pm 0.0051$&$\pm 0.0036$&$\pm 0.0034$&$\pm 0.0038$&$\pm 0.0029$ \\ 
\hline
\raisebox{2mm}{MC modelling}       
     &$\stackrel{\textstyle +0.0041}{-0.0027}$
            &$\stackrel{\textstyle +0.0034}{-0.0033}$
                   &$\stackrel{\textstyle +0.0074}{-0.0032}$
                          &$\stackrel{\textstyle +0.0040}{-0.0031}$
                                 &$\stackrel{\textstyle +0.0057}{-0.0036}$ 
                                         &$\stackrel{\textstyle +0.0048}{-0.0029}$ \\
\hline
Higher orders
     &$\pm 0.0241$&$\pm 0.0117$&$\pm 0.0072$&$\pm 0.0086$&$\pm 0.0036$&$\pm 0.0072$ \\ 
\hline\hline
\raisebox{2mm}{Total error}       
     &$\stackrel{\textstyle +0.0249}{-0.0247}$
            &$\stackrel{\textstyle +0.0135}{-0.0135}$
                   &$\stackrel{\textstyle +0.0112}{-0.0090}$
                          &$\stackrel{\textstyle +0.0104}{-0.0101}$
                                 &$\stackrel{\textstyle +0.0080}{-0.0067}$ 
                                        &$\stackrel{\textstyle +0.0095}{-0.0086}$ \\
\hline
\end{tabular}
\end{center}
\caption[dummy]{\label{tab-oaa-results-44GeV}
Values of \as($44$~GeV) derived using the \oaa\ QCD calculations with 
fixed $\xmu=1$ and \xmu\ fitted. The statistical and systematic 
uncertainties are also given. 
}
\end{table}

\begin{table}[!htb]
\vspace*{-7mm}
\begin{center}
\begin{tabular}{|r||r|r|r|r|r||r|}   \hline
  \multicolumn{1}{|c||}{$35$~GeV}
 &\multicolumn{1}{c|}{\thr} &\multicolumn{1}{c|}{\mh}
 &\multicolumn{1}{c|}{\bt}  &\multicolumn{1}{c|}{\bw}
 &\multicolumn{1}{c|}{\dtwo}&\multicolumn{1}{|c|}{averaged}     \\
\hline\hline
\as($35$~GeV)  
     &\bf 0.1560  &\bf 0.1654  &\bf 0.1699  &\bf 0.1508  &\bf 0.1485  &\bf 0.1560 \\
\hline\hline
fit range         
     &$0.12$-$0.30$ 
                  & $0.30$-$0.50$ 
                               & $0.16$-$0.30$ 
                                            & $0.10$-$0.20$
                                                         & $0.04$-$0.20$ 
                                                                      &             \\
\hline\hline
\chisqd\ ($\xmu=1$)
     &$   5.8   $ &$    2.3  $ &$    1.6  $ &$    3.4  $ &$    2.9  $ &            \\
\hline\hline
\xmu\ fitted
     &$    0.040$ &$    0.342$ &$    0.367$ &$    0.056$ &$    0.074$ &            \\
\hline
\chisqd\ (\xmu\ free)
     &$   2.4   $ &$    2.2  $ &$    0.2  $ &$    1.4  $ &$    3.1  $ &            \\
\hline\hline
Statistical error 
     &$\pm 0.0015$&$\pm 0.0019$&$\pm 0.0013$&$\pm 0.0014$&$\pm 0.0020$&$\pm 0.0016$ \\ 
\hline
Experimental syst.
     &$\pm 0.0034$&$\pm 0.0034$&$\pm 0.0025$&$\pm 0.0030$&$\pm 0.0052$&$\pm 0.0026$ \\ 
\hline
\raisebox{2mm}{MC modelling}       
     &$\stackrel{\textstyle +0.0065}{-0.0036}$
            &$\stackrel{\textstyle +0.0048}{-0.0048}$
                   &$\stackrel{\textstyle +0.0134}{-0.0064}$
                           &$\stackrel{\textstyle +0.0079}{-0.0047}$
                                  &$\stackrel{\textstyle +0.0033}{-0.0032}$ 
                                         &$\stackrel{\textstyle +0.0045}{-0.0029}$ \\
\hline
Higher orders
     &$\pm 0.0279$&$\pm 0.0089$&$\pm 0.0140$&$\pm 0.0085$&$\pm 0.0068$&$\pm 0.0093$ \\ 
\hline\hline
\raisebox{2mm}{Total error}       
     &$\stackrel{\textstyle +0.0289}{-0.0284}$
            &$\stackrel{\textstyle +0.0109}{-0.0108}$
                   &$\stackrel{\textstyle +0.0196}{-0.0157}$
                          &$\stackrel{\textstyle +0.0121}{-0.0103}$
                                 &$\stackrel{\textstyle +0.0094}{-0.0094}$ 
                                        &$\stackrel{\textstyle +0.0108}{-0.0102}$ \\
\hline
\end{tabular}
\end{center}
\caption[dummy]{\label{tab-oaa-results-35GeV}
Values of \as($35$~GeV) derived as in Table~\protect\ref{tab-oaa-results-44GeV}
but at $35$~GeV.
}
\end{table}

\begin{table}[!htb]
\vspace*{-7mm}
\begin{center}
\begin{tabular}{|r||r|r|r|r|r||r|}   \hline
  \multicolumn{1}{|c||}{(a)} 
 &\multicolumn{1}{c|}{$\langle 1-T \rangle$} 
 &\multicolumn{1}{c|}{$\langle M_H^2/s \rangle$}
 &\multicolumn{1}{c|}{$\langle B_T \rangle$}  
 &\multicolumn{1}{c|}{$\langle B_W \rangle$}
 &\multicolumn{1}{c||}{$\langle y_{23} \rangle$}
 &\multicolumn{1}{|c|}{average}     \\
\hline\hline
\asmz  
    &\bf 0.1204  &\bf 0.1118  &\bf 0.1158  &\bf 0.1105  &\bf 0.1232  &\bf 0.1155 \\
\hline\hline
$Q$ range [GeV]
    & $13$-$172$ & $14$-$172$ & $35$-$161$ & $35$-$161$ & $22$-$161$ & \\
\hline\hline
\chisqd       
    & $42.6/24$  & $10.9/14$  & $37.4/9$   & $21.1/9$   & $5.0/6$    & \\
\hline\hline
experimental
    &$\pm 0.0013$&$\pm 0.0010$&$\pm 0.0018$&$\pm 0.0015$&$\pm 0.0020$&$\pm 0.0013$ \\
\hline\hline
$\xmu=0.5$    
    &  $-0.0050$ &  $-0.0027$ &  $-0.0039$ &  $-0.0012$ &  $-0.0043$ &  $-0.0033$  \\
\hline
$\xmu=2.0$     
    &  $+0.0061$ &  $+0.0037$ &  $+0.0048$ &  $+0.0020$ &  $+0.0057$ &  $+0.0045$  \\
\hline\hline
$\mu_I=1$~GeV   
    &  $+0.0023$ &  $+0.0021$ &  $+0.0056$ &  $+0.0054$ & 
\multicolumn{1}{|c||}{---}       
                                                                     &  $+0.0029$  \\
\hline
$\mu_I=3$~GeV 
    &  $-0.0018$ &  $-0.0016$ &  $-0.0043$ &  $-0.0037$ & 
\multicolumn{1}{|c||}{---}       
                                                                     &  $-0.0019$  \\
\hline\hline
\raisebox{2mm}{$a_{\cal F} \pm 50\%$}
    &  \multicolumn{1}{c|}{\raisebox{2mm}{---}}
    &  $\stackrel{\textstyle -0.0022}{+0.0025}$ 
    &  $\stackrel{\textstyle -0.0047}{+0.0064}$ 
    &  $\stackrel{\textstyle -0.0052}{+0.0073}$ 
    &  \multicolumn{1}{|c||}{\raisebox{2mm}{---}}
    &  $\stackrel{\textstyle -0.0020}{+0.0028}$  \\
\hline\hline
\raisebox{2mm}{Total error}       
    &$\stackrel{\textstyle +0.0066}{-0.0055}$
           &$\stackrel{\textstyle +0.0050}{-0.0040}$
                  &$\stackrel{\textstyle +0.0099}{-0.0077}$
                         &$\stackrel{\textstyle +0.0094}{-0.0067}$
                                &$\stackrel{\textstyle +0.0060}{-0.0046}$ 
                                       &$\stackrel{\textstyle +0.0062}{-0.0045}$ \\
\hline
\multicolumn{7}{c}{\vspace*{7mm}} \\
  \cline{1-5}
  \multicolumn{1}{|c||}{(b)} 
 &\multicolumn{1}{c|}{$\langle 1-T \rangle$} 
 &\multicolumn{1}{c|}{$\langle M_H^2/s \rangle$}
 &\multicolumn{1}{c|}{$\langle B_T \rangle$}  
 &\multicolumn{1}{c|}{$\langle B_W \rangle$}
 &\multicolumn{2}{c} {}                                 \\
  \cline{1-5} \cline{1-5}
$\bar{\alpha}_0$
    &\bf 0.543   &\bf 0.457   &\bf 0.342   &\bf 0.264   
 &\multicolumn{2}{c} {}                                 \\
  \cline{1-5} \cline{1-5}
experimental
    &$\pm 0.014 $&$\pm 0.009 $&$\pm 0.007 $&$\pm 0.002 $ 
 &\multicolumn{2}{c} {}                                 \\
  \cline{1-5} \cline{1-5}
$\xmu=0.5$    
    &  $+0.002 $ &  $+0.013 $ &  $+0.009 $ &  $+0.030 $ 
 &\multicolumn{2}{c} {}                                 \\
  \cline{1-5}
$\xmu=2.0$     
    &  $-0.001 $ &  $-0.008 $ &  $-0.006 $ &  $-0.019 $ 
 &\multicolumn{2}{c} {}                                 \\
  \cline{1-5} \cline{1-5}
\raisebox{2mm}{$a_{\cal F} \pm 50\%$}
    &  \multicolumn{1}{c|}{\raisebox{2mm}{---}}
    &  $\stackrel{\textstyle -0.076 }{+0.212 }$ 
    &  $\stackrel{\textstyle -0.036 }{+0.063 }$ 
    &  $\stackrel{\textstyle -0.024 }{+0.037 }$  
 &\multicolumn{2}{c} {}                                 \\
  \cline{1-5} \cline{1-5}
\raisebox{2mm}{Total error}       
    &$\stackrel{\textstyle +0.015 }{-0.014 }$
           &$\stackrel{\textstyle +0.212 }{-0.077 }$
                  &$\stackrel{\textstyle +0.064 }{-0.038 }$
                         &$\stackrel{\textstyle +0.048 }{-0.031 }$ 
 &\multicolumn{2}{c} {}                                 \\
  \cline{1-5}
\end{tabular}
\end{center}
\caption{\label{tab-as-powcor}
Values of 
\asmz (a) 
and $\bar{\alpha}_0$ (b) derived using 
the \oaa\ calculations and power corrections with $\mu_I=2$~GeV and $\xmu=1$. 
Fit ranges and 
\chisqd\ values for each of the five event shape observables are included. 
In addition, the
statistical and systematic uncertainties are given. Where a signed value is
quoted, this indicates the direction in which 
\asmz\ and 
$\bar{\alpha}_0$ changed 
with respect to the standard analysis. The renormalisation and infrared scale 
uncertainties and the uncertainties due to the $a_{\cal F}$ coefficients are 
treated as an asymmetric uncertainty on 
\asmz. 
These uncertainties are treated 
similarly for $\bar{\alpha}_0$ but exclude the infrared scale uncertainty. 
}
\end{table}

%
\clearpage
\section*{ Figures }

\end{document}